\documentclass[usegraphicx,usenatbib]{mn2e}
\usepackage{color}
\usepackage{verbatim}
\usepackage{amssymb}
\usepackage{amsmath}
\usepackage{breqn}
\usepackage{epsf,multirow}
\usepackage{aas_macros}

\newcommand{\beq}{\begin{equation}}
\newcommand{\eeq}{\end{equation}}

\newcommand{\hmpc}{\,$h^{-1}$Mpc }
\newcommand{\hmpcii}{\,$h^{-1}$Mpc}

\newcommand{\hgpcii}{\,\ensuremath{h^{-3}{\rm Gpc}^3}}
\newcommand{\baf}{\,baryonic acoustic feature }
\newcommand{\bafii}{\,baryonic acoustic feature}
\newcommand{\Epaper}{\,\cite{eisenstein05b} }

\newcommand{\GpaperaltA}{\,\citealt{gaztanaga08iv}}

\newcommand{\PWpaper}{\,\cite{padmanabhan08a} }
\newcommand{\PWpaperii}{\,\cite{padmanabhan08a}}
\newcommand{\PWpaperalt}{\,\citealt{padmanabhan08a}}

\newcommand{\AP}{\,AP }
\newcommand{\APii}{\,AP}

\newcommand{\avg}[1]{\ensuremath{\langle{#1}\rangle}}
\newcommand{\dM}{\,D_{\rm A}}
\newcommand{\dA}{\,D_{\rm A}}

\newcommand{\SH}{\,{\mathcal S}}
\newcommand{\DIST}{\,{\mathcal D}}
\newcommand{\TRUE}{\,{\mathcal T}}

\begin{document}

\title[Improving measurements of $H(z)$ and $D_{\rm A}(z)$ by analyzing clustering anisotropies]
{
Improving measurements of $H(z)$ and $D_{\rm A}(z)$ by analyzing clustering anisotropies
}
\author[Kazin E., S\'anchez A. \& Blanton M.]
{\parbox[t]{\textwidth}{
Eyal A. Kazin$^{1}$\thanks{E-mail: eyalkazin@gmail.com},
Ariel G. S\'anchez$^{2}$,
Michael R. Blanton$^{1}$,
}
\vspace*{6pt} \\ 
$^{1}$ Center for Cosmology and Particle Physics, New York University, 4 Washington Place, New York, NY 10003, USA.\\
$^{2}$ Max-Planck-Institut f\"ur Extraterrestrische Physik, Giessenbachstra\ss e, 85748 Garching, Germany.\\
\\
}
\date{Submitted to MNRAS}
\maketitle

\begin{abstract}

The baryonic acoustic feature in galaxy clustering is a
promising tool for constraining the nature of the cosmic
acceleration, through measurements of expansion rates
$H$ and angular diameter distances $\dA$.  Angle-averaged
measurements of clustering yield constraints on the
quantity $\dA^2/H$. However, to break the degeneracy
between these two parameters one must measure
the anisotropic correlation function as a function of both
line-of-sight (radial) and transverse separations.  Here we investigate how
to most effectively to do so,  using analytic techniques and
mock catalogues.  In particular, we examine multipole expansions
of the correlation function as well as ``clustering wedges"
$\xi(\Delta \mu, s)$, where $\mu=s_{||}/s$ and  $s_{||}$ is
the radial component of separation $\vec{s}$.  Both techniques
allow strong constraints on $H$ and $\dA$, as
expected.  The radial wedges strongly depend on $H$
and the transverse wedges are sensitive to $\dA$.
Analyses around the region of the acoustic peak constrain
$H$ $\sim 20\%$ better when using the wedge statistics
than when using the monopole-quadrupole combination.
However, we show that the hexadecapole allows substantially
stronger constraints than the monopole and quadrupole alone, 
as well as analyzing the full shape of $\xi$. 
Our findings here demonstrate that wedge statistics provide
a practical alternative technique to multipoles, that should be
useful to test systematics and will provide comparable or better
constraints.  Finally, we predict the constraints from galaxy
clustering that will be possible with a completed version of the
ongoing Baryonic Oscillation Spectroscopic Survey.

\end{abstract}


\section{Introduction}
\label{sec:intro}

The clustering of matter is a powerful tool 
to probe the evolution of the Universe. In recent years,   
the {\it baryonic acoustic feature} has  been detected in the clustering of galaxies.  
Originating from pre-recombination plasma waves at redshifts $z>1100$, this
feature is strongly detected in the temperature fluctuations of the Cosmic Microwave
Background radiation (CMB). Its detection at low redshifts serves as an important
confirmation of the cold dark matter (CDM) paradigm, and serves as a link between
the late and the early Universe (\citealt{peebles70a}, \citealt{meiksin99a}). 

The baryonic acoustic feature is also a practical cosmic {\it standard ruler}  
(\citealt{eisenstein98a}, \citealt{eisenstein98b}, \citealt{eisenstein99a}), 
because the plasma waves left a distinct imprint at the {\it surface of last scattering} ($z_*\sim 1100$).  
In the CMB temperature fluctuations $\Delta T/T$ this sound horizon scale of $\sim 150$ physical kpc 
corresponds to $\sim 1^\circ$ in the sky and is known to an accuracy of $<1.5\%$
($1\sigma$ level; \citealt{komatsu09a}). Hence it can be used as a calibrated scale of
the angular diameter distance $\dA(z_*)$. By measuring a related signature imprinted in
the distribution of matter at late times, one can perform geometrical 
measurements to determine angular diameter distances $\dA$ 
at different $z$, as well as measure expansion rates $H$. 
The importance of this statistical tool is amplified considering the 
recent discovery of the acceleration of the expansion of the Universe 
($z<1$;  \citealt{riess98}, \citealt{perlmutter99a}).  
Although both $\dA$ and $H$ are important cosmic measurements, measuring 
$H$ directly at various $z$ puts strong constraints on understanding the 
nature of the apparent acceleration of the observed Universe, 
e.g through the so-called ``dark energy" equation of state $w$   
(\citealt{blake03}, \citealt{seo03}, \citealt{linder03a}, \citealt{glazebrook05a}).

In the matter distribution, this signature appears as oscillations at $k>0.1\,h\,{\rm Mpc}^{-1}$   
in the power-spectrum, $P(k)$, corresponding to a bump of excess overdensity in 
the two-point correlation function, $\xi(s)$, at the characteristic comoving scale 
of $\sim 100$ \hmpcii. 
This scale, although not equal to the sound-horizon, is closely related  
to it, with well-understood differences 
(\citealt{meiksin99a}, \citealt{smith08}, \citealt{crocce08}, 
\citealt{sanchez08}).

The baryonic feature is imprinted in $\xi(\mu,s)$ and can be used to measure $H$ and $D_{\rm A}$. 
For purely geometric reasons, radial clustering (i.e, clustering in the line-of-sight of the observer; $\mu=1$) 
contains $H(z)$ information, where the transverse direction ($\mu=0$) yields $\dA(z)$, 
where $\mu$ is the cosine of the angle between the total separation 
of the galaxies and the radial direction. 
 
 For S/N reasons, most studies have focused 
on the angle averaged signal (monopole) 
$\xi_0(s)$
(\citealt{eisenstein05b}, 
\citealt{martinez08},
\citealt{cabre09i}, 
\citealt{labini09}, 
\citealt{sanchez09a}, 
\citealt{kazin10a}, 
Beutler et al. (in prep)), 
$P_0(k)$,   
(\citealt{cole05a}, 
\citealt{tegmark06a}, 
\citealt{hutsi06a},
\citealt{percival07a}, 
\citealt{percival09b}, 
\citealt{reid09a},
\citealt{blake11a}) 
and wavelets 
(\citealt{arnalte11a}) 
in clustering of galaxies and galaxy-clusters of the SDSS  
(\citealt{york00a}), the Two 
Degree Field Galaxy Redshift Survey (\citealt{colless03a}), 
the WiggleZ (\citealt{{drinkwater10a}}) survey 
and the 6dF Galaxy Survey (\citealt{jones09a}) 
galaxy samples.
The feature in the projected two-point 
function of photo-$z$ samples 
has been analysed by
(\citealt{padmanabhan07b}, 
\citealt{blake07}, 
\citealt{estrada09a},
\citealt{crocce11a}).

The \baf in the monopole has been shown to constrain the cosmological information in 
combination $\dA^2/H$. The degeneracy of this combination limits its constraining power on 
 expansion models. In this study we discuss techniques to use anisotropic 
clustering to break this degeneracy. 

Radial clustering measurements have been attempted on the Sloan Digital Sky Survey (SDSS) 
Luminous Red Galaxy sample (LRGs; \GpaperaltA), as well as on the much smaller volume MAIN
sample \citep{tian10a}. Interestingly, both studies show  strong 
clustering measurements, relative to the monopole and $\Lambda$CDM predictions, 
near where the baryonic acoustic feature is expected. 
However, \citet{kazin10b} suggested a 
different interpretation of these measurements indicating that the measurement could be the result 
of sample variance due to the limited volume.

A few studies have investigated using the information from the full $P(\mu,k)$ 
plane to constrain dark energy by using geometric redshift distortions.
\cite{alcock79} describe how an intrinsically spherical system appears anisotropic 
due to geometric distortions. They point out that by reconstructing the original 
spherical shape, the true cosmology can be obtained. This effect is manifested in clustering
measurements, which are assumed to be isotropic (the cosmic principle). When converting the observed
redshifts to comoving distances, the observer is required to assume a {\it fiducial} cosmology.
Choosing an incorrect cosmology causes geometric redshift-distortions 
(in addition to the dynamical
distortions, due to peculiar velocities of the galaxies).
For this reason the observed clustering signal provides an opportunity to apply the Alcock-Paczynski
test and constrain the true underlying cosmology, assuming dynamical effects are understood 
(\citealt{kaiser87}). In practice, the \baf plays an important role as it breaks degeneracies  
between amplitude uncertainties ($\sigma_8$, bias of tracer to underlying matter over-densities) 
and geometric shifts. 

\cite{hu03a} suggested disentangling  $H$ from $\dA$ by analyzing  baryonic acoustic rings 
in the two-dimensional power spectrum. Focusing on phase shifts, they found that this technique, 
based only on geometric effects, can constrain the expansion rate of the Universe
when applied to galaxy and galaxy-clusters samples at intermediate redshifts $z<0.5$, 
combined with CMB priors. \cite{wagner08a} used mock catalogues at $z=1,3$ to demonstrate
the usefulness of the technique, and show that light-cone effects do not have a significant
impact on the results. \cite{shoji09a} argue that $H$ and $\dA$ information is encoded in the 
full 2D shape, and present a generic algorithm that attempts to take into account dynamic
distortions on all scales, assuming non-linear effects are understood. 
For a first attempt to 
use the  $\xi(s_\perp,s_{||})$ plane of the SDSS LRGs
to constrain cosmology, see \cite{okumura08}.  
 
These studies, though, do not take into account the complexity of constructing a realistic,
reliable (and invertible) covariance matrix for the full 2D measurement of the power spectrum or
correlation function. Also, one short-term concern is the fact that near future 
surveys will have low S/N in the 2D plane. 

\PWpaper investigated a more practical approach in which the 2D results are projected 
into one dimensional statistics. They proposed to break the $H-\dA$ degeneracy by combining
\AP analysis of the monopole of the correlation function, $\xi_0$, with the quadrupole, $\xi_2$. 
They argue that these measurements can constrain the {\it warping} in the 2D correlation function
which is sensitive to $\dA\cdot H$, breaking the $\dA^2/H$ degeneracy obtained when probing dilations
in the monopole.

Here we follow up on their analysis by: 
\begin{enumerate}
\item{Investigating the effects of higher order multipoles};
\item{Introducing a new alternative projection statistic in the form of {\it clustering wedges}};
\item{Performing the analysis, for the first time, in configuration space ($\xi(\mu,s)$).}
\end{enumerate}
 
We propose to use wide clustering wedges $\xi(\Delta\mu,s)$ to yield constraints on $H$ 
and $\dA$. A similar concept has been suggested by \citet{kazin10b} (e.g., see their Figure 7).
Tests performed here convincingly show that even a wide ``transverse" wedge of 
$0<\mu<0.5$ strongly depends on $\dA$ and a wide  ``radial" wedge of $0.5<\mu<1$ constrains $H$. 
We show that these $\Delta\mu=1/2$ wedges do have intermixing terms, but these can be corrected for 
and are reduced with decreasing $\Delta\mu$.

The outline of the paper is as follows: 
in \S\ref{methods_section} we describe the statistical
methods and the mock galaxy catalogues used in our analysis. 
In \S\ref{theory_section} we describe 
various theoretical aspects of $z-$distortions (\S \ref{zdistortions_section}), 
the implicatoins of the AP effect on basic cosmological parameters  
via determinations of $H$ and $\dA$
(\S \ref{hzplane_section}), the different clustering statistics used in our analysis 
(\S  \ref{wedge_introduction}), and the extraction of $H$ and $\dA$ 
information through the \AP test (\S  \ref{dialtionwarping_section}). 
In \S \ref{poc_section} we run two proof of concept tests. 
Using mock catalogues, we show that the geometrically distorted signal can be retrieved from the true signal 
through the \AP test (\S \ref{wedgeshifting_section}), 
and by doing so retrieve unbiased constraints on $H$ and $\dA$. 
We start with the ideal case in which all effects are known except for the \AP effect in \S \ref{reproducing_hda_section}, 
and then, in \S \ref{practical_section}, we gradually add amplitude effects. In 
\S  \ref{multipolesorwedges_section} we investigate the uncertainties of $H$ and $\dA$ 
as a function of the separation range used and compare the results obtained by the various 1D projection
combinations. 
In \S \ref{boss_section} we present predictions for BOSS. 
Finally, \S \ref{discussion_section} presents a discussion, 
and \S\ref{conclusion_section} our main conclusions

Unless otherwise stated, we use the standard flat $\Lambda$CDM cosmology ($\Omega_{\rm K}=0$, $w=-1$) 
with $[\Omega_{\rm M0},\Omega_{b0},h,n_s,\sigma_8]=[0.25,0.04,0.7,1,0.8]$. 
We test geometric distortions by varying only the equation of state of dark energy $w$, 
when converting $z$ to comoving distances. Our choices of distortions are explained in \S\ref{hzplane_section}
compared to expected degeneracies for various choices of curvature $\Omega_{\rm K}$, and matter
density $\Omega_{\rm M0}$. Unless stated otherwise, all distances hereon are comoving. 

To avoid semantic confusion, we briefly explain here the terminology of the different spaces we explore. 
First, all analyses are based on two-point correlation functions, 
which we refer to as 
{\it configuration-space}, as opposed to $k-$space. 
Second, because geometric redshift distortions 
and dynamic redshift distortions are different in nature, 
we minimize the use of the generic term for both, ``redshift distortions", 
and call them ``geometric distortions" and ``dynamical distortions". 
Because we analyse geometric distortions with 
and without dynamical effects, hereon we avoid using the common expression  {\it redshift-space} 
($z-space$). Instead, when dynamical effects are applied we refer to it as {\it velocity-space},
and when they are not we refer to it as {\it real-space}.

\section{Methods of Analysis}
\label{methods_section}

\subsection{Statistical tools}
\label{stats_section}

In our analyses we use the \cite{landy93a} $\xi$ estimator.
For details of usage please see Appendix \ref{xi_estimators}. 

When constraining parameters, we use the standard $\chi^2$ technique, where
\beq
\chi^2(\Phi)= \sum_{i,j}\left(M_i\left(\Phi\right)-D_i\right)C_{ij}^{-1}(M_j(\Phi)-D_j), 
\eeq
where $i,j$ are the bins tested. For reasons described below,  
 the ``data"  $D$ is given by the {\it distorted} measurement, $\xi^{\mathcal D}$, meaning $\xi$ measured
from our mock catalogues when using the incorrect cosmology to convert redshifts to comoving
distances (the \AP effect). The base-template used for modelling is $\xi^{\mathcal T}$, meaning the actual true clustering signal. Hence, the models $M$ are given by the {\it shifted} measurements $\xi^{\mathcal S}$, that is 
the result of using the parameters $\Phi$ to shift the template $\xi^{\mathcal T}$.
 
 When limiting $\Phi$ to [$H,\dA$] we calculate $\chi^2$ using brute force on a 2D grid.  
When investigating a larger parameter space,  we apply  a Monte-Carlo-Markov-Chain (MCMC).
We verify that both methods yield similar results. 

The statistics we use have covariant uncertainties. 
(e.g., see Figure 4 in \citealt{taruya11a} for the correlation coefficient between $\xi_0$ and $\xi_2$.)
For this reason we define 
$D$ and $M$ in array format. 
For example, when analyzing monopole, quadrupole 
($\ell=0,2$) the data is defined as $D=[\xi_0,\xi_2]= \xi_{[\ell]}$.

We construct the covariance matrix $C_{ij}$ 
from the $N_{\rm mocks}=160$ mock true signals.
(For mock description see \S\ref{mocks_section}.)
When using multipoles (or wedges) combination $\xi_{[\ell]}$,
we define the covariance matrix as:
\beq
C_{ij}= \frac{1}{N_{\rm mocks}} \sum_{m,n=1}^{N_{\rm mocks}} \left(   (\xi_{[\ell]})_i^m - (\xi_{[\ell]})_j^n \right) \left(   (\xi_{[\ell]})_j^m - (\xi_{[\ell]})_i^n \right).
\eeq

\subsection{Mock galaxy catalogs}
\label{mocks_section}

To simulate the observer's point of view, we analyse the  mock galaxy catalogues from LasDamas  
and the Horizon Run. A similar version of these mock catalogues has been used 
in our previous analyses  
of the monopole (\citealt{kazin10a}) and radial clustering 
(\citealt{kazin10b}). 

The LasDamas simulations use a cosmology of [$\Omega_{M0},\Omega_{b0},n_s$,$h$,$\sigma_8$]=[0.25,0.04,1,0.7,0.8] 
and the Horizon Run uses [0.26,0.044,0.96,0.72,0.8], where $\Omega_{b0}$ is the present baryonic 
density and $n_s$ is the spectral index. Both these cosmologies are well motivated 
by constraints obtained by WMAP 5-year measurements of temperature fluctuations in the cosmic microwave 
background \citep{komatsu09a}. To understand effects of velocity-space 
we analyse all volumes both in velocity- and real-space.

The LasDamas collaboration provides  realistic LRG mock
catalogues\footnote{http://lss.phy.vanderbilt.edu/lasdamas/} by placing galaxies inside dark matter
halos using a Halo Occupation Distribution (HOD; \citealt{berlind02a}). HOD parameters were chosen
to reproduce the observed number density as well as the projected two-point correlation function
$w_p(r_p)$ of galaxies in the SDSS-LRG sample at separations $0.3<r_p<30$\hmpcii, 
below the scales considered here.  
 For more details see McBride et al. (in prep.).
We use a suit of 160 LRG volume-limited mock catalogues constructed from light cone samples
with a mean number density of $\bar{n}\sim 10^{-4}\,h^3\rm{Mpc}^{-3}$. Each mock catalogue covers the
redshift range $0.16<z<0.44$ and reproduces the SDSS angular mask,
giving a volume of $1.2\,{\rm Gpc}^3h^{-3}$.
The LasDamas  real-space catalogues are similar to the velocity-space catalogues in all aspects,
with the exception of the shift in $z$ due to peculiar velocities. 

The Horizon Run\footnote{http://astro.kias.re.kr/Horizon-Run/} 
provides an ensemble of $32$ BOSS volume realizations of mock LRG samples with a higher number
density than DR7, $\bar{n}\sim 3 \times 10^{-4}\,h^3\rm{Mpc}^{-3}$, as expected in BOSS.
LRG positions are determined by identifying physically self-bound dark matter sub-halos that
are not tidally disrupted by larger structures. For full details see \cite{kim09a}. 
We construct these mock catalogues by dividing each of the eight full sky samples of the Horizon
Run into four quadrants. We map real-space into velocity-space, and limit the samples to the
expected volume-limited region of the BOSS LRGs $(0.16<z<0.6)$.

%

\begin{figure*}
\includegraphics[width=0.45\textwidth]{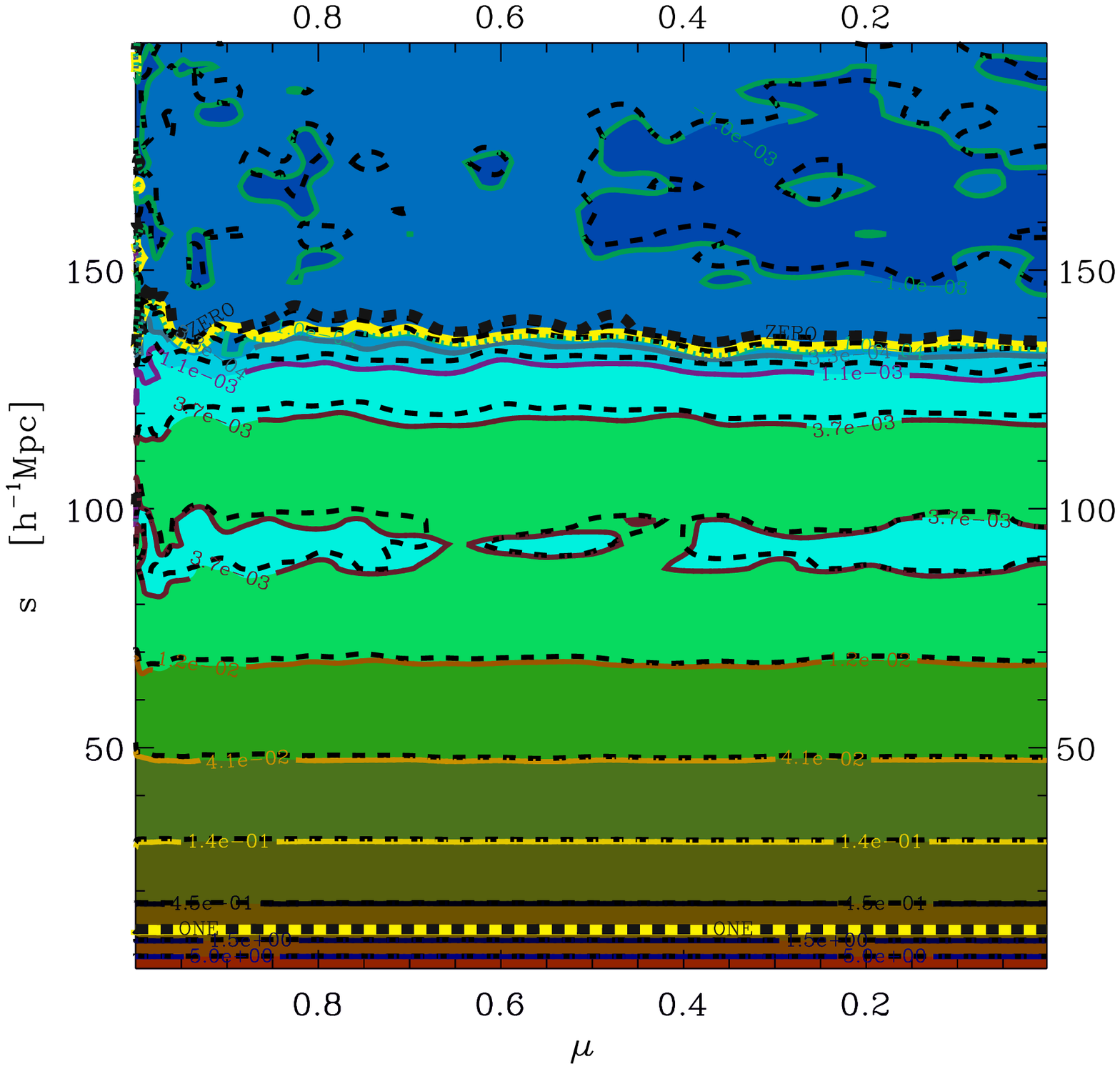}
\includegraphics[width=0.45\textwidth]{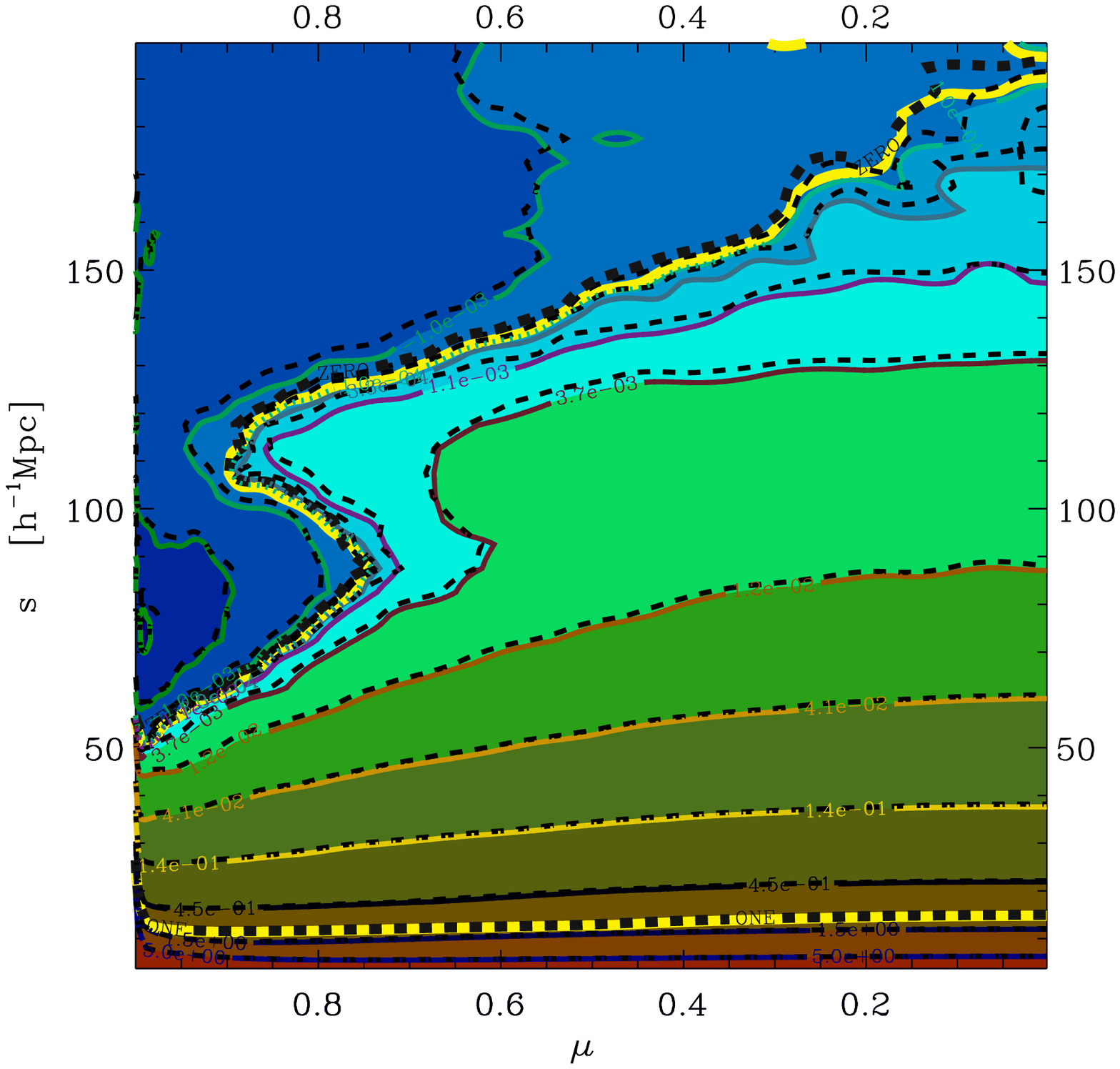}
\caption[$\xi(\mu,s)$ real- vs. velocity-space]{
Mean two-dimensional correlation functions $\xi(\mu,s)$ from the ensemble of mock 
catalogues in real- (left) and velocity-space (right). 
The contours are spaced logarithmically between 5 and $10^{-4}$ for positive values, 
and at  $-0.001,-0.005,-0.01$ for negative values.
The contours corresponding to the values $0,1$ are colored with thick yellow lines.
The solid contour lines, following the color scheme, correspond to the result obtained when 
using the correct cosmology when converting $z$ to comoving distances. The dashed lines show the geometrical distortions
obtained by assuming $w^{\mathcal D}=-1.1$ instead of the true value $w^{\mathcal T}=-1$. 
It can be clearly seen that dynamical effects dominate over the geometric. 
}
\label{a2pcf_ld}  
\end{figure*}

\section{Theory}
\label{theory_section}
\subsection{Redshift distortions: geometric vs. dynamic}
\label{zdistortions_section}

Redshift distortions arise due to two effects when converting the redshift $z_{\rm obs}$ of a
galaxy into a comoving distance:  
\beq\label{comoving_equation}
\chi=c\int_0^{z_{\rm obs}}{\frac{dz}{H(z)}}. 
\eeq 

 The first effect involves the assumption that the observed redshift is 
produced entirely by the expansion of the Universe $z_{\rm cos}$.  
This assumption is, of course, incorrect in the presence of peculiar velocities, which introduce an additional
Doppler component $z_{\rm pec}$ leading to radial shifts in the infered distances.
Although these shifts are small compared to the true distance $\chi(z_{\rm cos})$ 
(less than $1\%$ at $z\sim0.3$), they strongly affect clustering measurements 
which depend on separations between galaxies.  
We refer to these  as  {\it dynamical} distortions. 
 
Another, more subtle,  redshift distortion effect arises due to the 
conversion of redshift to distance using only approximately known cosmological parameters.
The conversion relies on the Hubble parameter (\citealt{friedman22a}) 
\begin{dmath}
\label{first_friedman}
H(z)^2=H_0^2\left(  \Omega_{\rm M0} \left( 1+z \right)^3+\Omega_{\rm K} \left( 1+z \right)^2+\Omega_{\rm {\rm DE}}  e^{\int_0^z{\frac{1+w\left( z' \right)}{1+z'}dz'}} \right),
\end{dmath}
where $\Omega_i$ are the standard cosmological density terms at present day for matter (M0),
curvature (K) and dark energy (DE). 
The {\it Hubble constant} $H_0\equiv H(0)$ (\citealt{hubble31}) 
factors out trivially and we thus express comoving distance in units of \hmpcii, 
where $h\equiv H_0/(100\,{\rm km}\,{\rm s}^{-1}{\rm Mpc}^{-1}$). 
The rest of the parameters have more important, and potentially measurable, effects. 
We refer to these \AP effects as {\it geometric} distortions.

One way of overcoming these effects is to recalculate 
clustering statistics for every set of parameters when 
determining cosmological constraints. 
However, that approach is currently not practical. 
Instead, we calculate  $\xi$ using a fixed 
fiducial set of parameters, 
and vary the result using linear equations. 
As we show below, this method is accurate enough. 

Figure  \ref{a2pcf_ld} illustrates  
dynamic and geometric distortions in the LasDamas mock catalogues 
using the anisotropic $\xi$ in the $\mu-s$ plane. The information
in this coordinate choice is similar to that in the commonly used $s_{||}-s_\perp$ plane. 
We define $\vec{s}$ to be the spatial separation vector 
with radial and transverse components $s_{||}$, $s_{\perp}$.
In real-space (left panel) the true signal corresponds to flat horizontal contour levels in $\xi(\mu,s)$,
shown as colored contours (solid lines)
A noticeable signature is the \baf around $s\sim 110$\hmpcii. 

The dashed lines show the result when we introduce 
geometric distortions by using $w=-1.1$ instead of the true value $w=-1$
in converting redshifts to comoving distance.  
These distortions are more noticeable at large scales, though 
they are also present on small scales.

The right-hand panel illustrates the equivalent measurements with the addition of 
dynamical distortions (velocity-space). 
It can be clearly seen that the dynamical distortions 
dominate over the geometric ones. 

Three noticeable features are worth mentioning here. 
First, the velocity dispersion effect  
is clearly seen in the clustering signal along the line of sight ($\mu=0$). 
Although commonly regarded as a small scale effect, 
it is still present on scales of $60$\hmpcii, 
as discussed by \cite{scoccimarro04a}.

Second,the {\it negative sea} along the radial direction
 is apparent at $s\sim60$\hmpc in this cosmology.  
Notice that in real-space (left plot) $\xi$ turns negative only at $\sim 135$\hmpcii.  

Third, the \bafii, which appears as a positive stripe in real-space clustering, appears here as
ridges which decrease strongly in amplitude towards the line-of-sight.  
Here the radial baryonic acoustic peak is negative, but can be positive depending 
on the value of the squashing parameter $\beta\equiv f/b$.

\cite{kaiser87} originally describes 
linear dynamical  distortions 
by coupling the logarithmic rate of change of the growth of 
structure $f$ to $\mu$. By doing so he  
related the underlying real-space isotropic $P(k)$ 
to the apparent anisotropic velocity-space one. 
The bias $b$ is introduced 
when relating to matter tracers. 
In this study we focus on the more subtle geometric  
distortions, and refer the reader to \cite{hamilton98a} for 
a review of dynamical distortions. 


\subsection{The cosmological power of the \AP effect}
\label{hzplane_section}


\begin{figure*}
\includegraphics[width=0.9\textwidth]{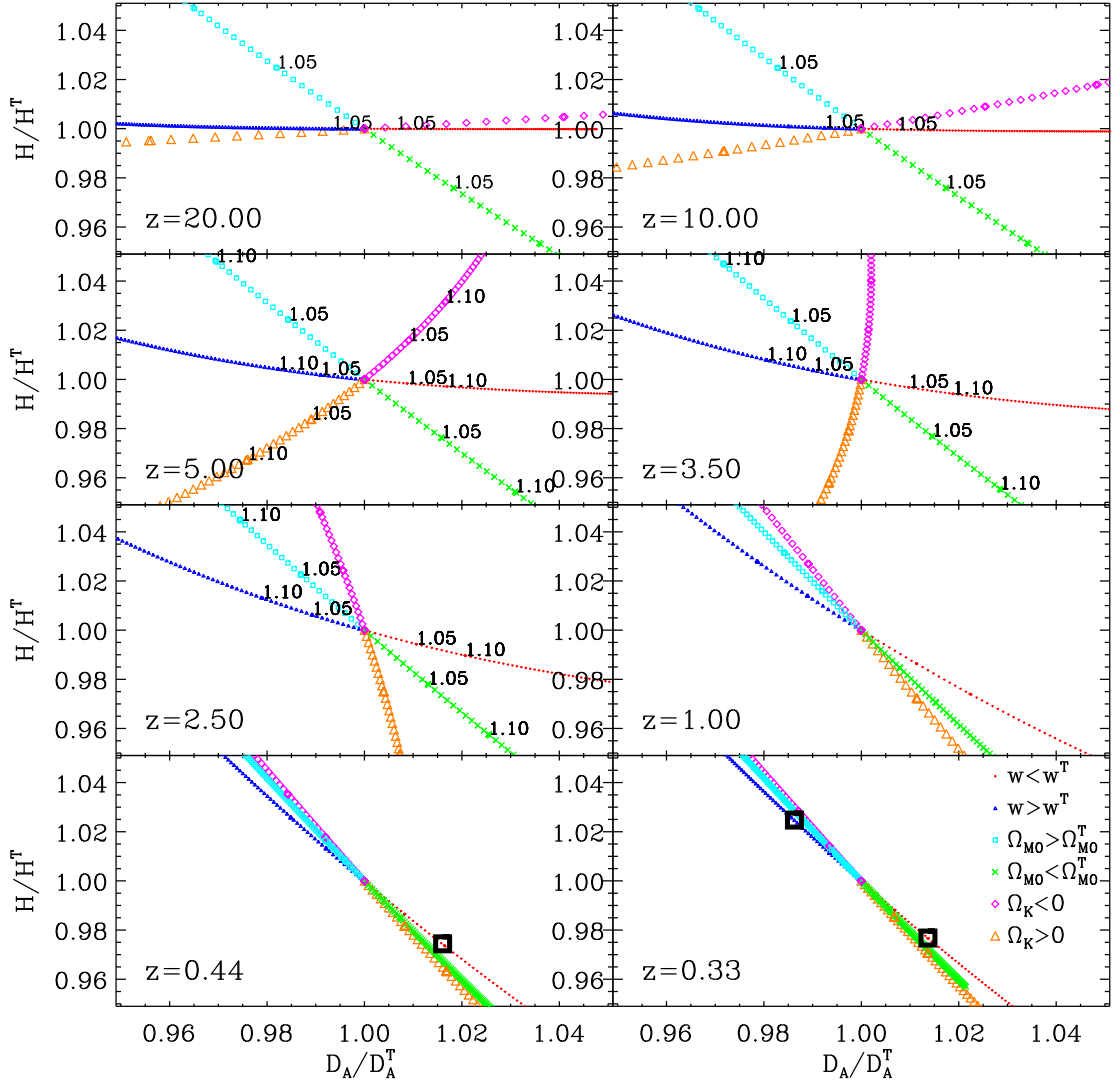}
\caption[Parameter degeneracy in $H-\dA$ plane]{
Each panel displays the degeneracies between 
$\Phi$=[$w$, $\Omega_{\rm M0}$, $\Omega_{\rm K}$]  
in the $H-\dM$ plane as a function of $z$. 
The axes, are in units of a ``true" cosmology $\Phi=[-1, \ 0.25, 0]$. 
The ``false" values are given in fractions in increments 
of $\Delta(frac)=0.005$. 
The legend reads such that: 
$w<w^{\TRUE}$ means $w=w^{\TRUE}\cdot frac$, 
$w>w^{\TRUE}$ means $w=w^{\TRUE}/ frac$, 
$\Omega_{\rm M0}>\Omega_{\rm M0}^{\TRUE}$ means $\Omega_{\rm M0}=\Omega_{\rm M0}^{\TRUE}\cdot frac$, 
$\Omega_{\rm M0}<\Omega_{\rm M0}^{\TRUE}$ means $\Omega_{\rm M0}=\Omega_{\rm M0}^{\TRUE}/ frac$, 
$\Omega_{\rm K}>0$ means $frac-1$ and 
$\Omega_{\rm K}<0$ means $1-frac$. 
In some high $z$ panels we highlight $5,10\%$ deviations in $\Phi$.  
The thick boxes at low $z$ indicate the geometric effects we test in this study. 
The figure clearly shows that at (cosmologically) 
low redshifts there is a large degeneracy between the parameters. 
This is relaxed at higher redshifts where distortions 
in $\Omega_{\rm K}$ and $w$ effect mostly $\dM$, and distortions 
in $\Omega_{\rm M0}$ yield similar results to low $z$.
}
\label{hzplane_parameters}  
\end{figure*}


Throughout this study we explore techniques to break the geometric $H-\dM$ degeneracy with
clustering. Here we examine how these constraints are related to 
fundamental cosmological parameters  
assuming a $\Lambda$CDM model. 

Following \cite{hogg99cosm} we define the 
{\it transverse comoving distance} as:   
\beq\label{dm_equation}
D_{\rm M} = \left\{ \begin{array}{lll}
D_{\rm H} \frac{1}{\sqrt{\Omega_{\rm K}}} \sinh\left(   \sqrt{\Omega_{\rm K}} \frac{\chi}{D_{\rm H}}  \right)  & \mbox{for $\Omega_{\rm K}>0$};\\
\chi & \mbox{for $\Omega_{\rm K}=0$};\\  
  D_{\rm H} \frac{1}{\sqrt{-\Omega_{\rm K}}} \sin\left(   \sqrt{-\Omega_{\rm K}} \frac{\chi}{D_{\rm H}}  \right)  & \mbox{for $\Omega_{\rm K}<0$}, \end{array} \right.
\eeq
where $D_{\rm H}$ is the Hubble distance $c/H_0$, and is related to the angular diameter distance by 
$D_{\rm M}=\dA(1+z)$. As we explain in \S   \ref{dialtionwarping_section},  
the \AP effect can be quantified by the dilation and warping  parameters $\alpha$ and $\epsilon$, 
which depend, in turn, on both $H$ and $\dM$ (see Equations  \ref{eq_dilation},  \ref{eq_warp}). 
These quantities depend in a non-straightforward fashion 
on the density parameters $\Omega_i$, and the dark energy equation of state $w$. 
$H$ is given by Equation (\ref{first_friedman}), while $\dM$ depends on $H$ and $\Omega_{\rm K}$
according to Equations (\ref{comoving_equation}) and (\ref{dm_equation}) . 

Figure \ref{hzplane_parameters} shows how $H$ and $\dA$ depend on cosmological parameters for a
number of redshifts. In each panel (each redshift) we hold two of the three 
parameters $\Omega_{\rm M0}$,  $\Omega_{\rm K}$, and $w$ 
(where $\Omega_\Lambda\equiv 1- \Omega_{\rm M0} -  \Omega_{\rm K}$) 
fixed to a ``true" value and modify the third 
from its fiducial according to the fraction indicated in the legend,
between $1$ to $1.5$.
We clearly see that at low redshifts 
$H$ and $\dA$ yield degenerate constraints on $w$, $\Omega_{\rm K}$ and $\Omega_{\rm M}$ 
that can be broken as $z$ increases. 
We notice that the dependence on $\Omega_{\rm M0}$ does 
not vary much as a function of redshift, 
where both $\Omega_{\rm K}$ and $w$ align with 
the $\dM$ axis at high $z$, meaning $H$ is not sensitive to these parameters. 

This plot demonstrates 
that the $\Omega_i, \ w$ degeneracy 
can be broken when 
applying the \AP effect 
at high redshift ($z>2$).

In this study we examine \AP effects 
when varying $w$ at mock mean redshifts 
$\avg{z}=0.33$, and 0.44 as indicated by thick 
boxes in bottom panels of Figure  \ref{hzplane_parameters}. 
Figure  \ref{hzplane_parameters} demonstrates that our 
results are similar to those we would 
have obtained by choosing
to vary $\Omega_{\rm K}$ or $\Omega_{\rm M0}$. 


\subsection{One-dimensional projections of $\xi(\mu,s)$: introducing clustering wedges}
\label{wedge_introduction}

We define clustering wedges as 
\beq\label{wedge_definition}
\xi(\Delta\mu,s)\equiv \frac{\int^{\mu_{\rm max}}_{\mu_{\rm min}}{\xi(\mu',s)}{{\rm d}\mu'}}{\int^{\mu_{\rm max}}_{\mu_{\rm min}}{{\rm d}\mu'}}, 
\eeq
where $\mu$ is the cosine of the angle between the total separation 
of the galaxies $\vec{s}$ and the line of sight.
We assume  here the {\it plane-parallel}, or {\it small angle}, approximation 
according to which two galaxies at the same distance from
the observer yield $\mu=0$ irrespective of their angular distance. 
We note that the \baf scale at $z=0.3$ corresponds to $\sim7^\circ$ in the sky, 
and is smaller at larger redshifts. \cite{samushia11a} discuss observer angle  effects. 
Due to our methods of building templates, our models incorporate  
large angle effects, and we do not test for them. 

Using spherical harmonics, the anisotropic $\xi(\mu,s)$ may be written as: 
\beq\label{eq_expansion}
\xi(\mu,s)=\sum_{{\rm even} \   \ell} {\mathcal P}_\ell(\mu)\xi_\ell(s),
\eeq
where ${\mathcal P}_\ell$ are Ledgendre polynomials 
(e.g, ${\mathcal P}_0=1$, ${\mathcal P}_2=\frac{1}{2}(\mu^2-1)$, ${\mathcal P}_4=\frac{1}{8}(35\mu^2-30\mu+3)$) and 
\beq\label{eq_extract_multipole}
\xi_\ell\equiv \frac{2\ell+1}{2}\int^1_{-1}{\mathcal P}_\ell(\mu)\xi(\mu,s){\rm d}\mu.
\eeq
 
Equations (\ref{wedge_definition}) and (\ref{eq_expansion}) can be used to 
find the relation between the clustering wedges and the multipoles. Discarding 
contributions from multipoles with $\ell>2$ this relation is given by:
\beq\label{wedge_equation}
\xi(\Delta\mu,s)= \xi_0 +\frac{1}{2} \left(   \frac{\mu_{\rm max}^3 -  \mu_{\rm min}^3   }{\mu_{\rm max} -  \mu_{\rm min} } -1 \right)\xi_2.
\eeq

A hexadecapole term would mean an additional term given by 
\beq\label{hex_term}
\frac{1}{8} \left(  \frac{ 7 \left(\mu_{\rm max}^5 -  \mu_{\rm min}^5 \right) - 10 \left(\mu_{\rm max}^3 -  \mu_{\rm min}^3 \right)  }{\mu_{\rm max} -  \mu_{\rm min} } +3 \right)\xi_4
\eeq
on the right hand side of Equation (\ref{wedge_equation}), and higher multipoles can 
be calculated in a similar manner.  

For simplicity, in this study we focus on clustering wedges defined by a width of $\Delta \mu=1/2$.
Of course, this analysis can be generalized to various wedge widths. 
We discuss the results obtained with various values of $\Delta\mu$ in Appendix \ref{wedgescorrestion_section}. 

Defining the radial wedge $\xi_{||}$ as that given by $0.5<\mu<1$ 
and the transverse $\xi_{\perp}$ as $0<\mu<0.5$, Equation (\ref{wedge_equation})
yields:
\beq\label{wedgemonoquad_equality}
\left( \begin{array}{c} \xi_{||} \\  \xi_{\perp}  \end{array} \right)  =
\left( \begin{array}{c c} 1 &  \frac{3}{8} \\   1 & -\frac{3}{8}  \end{array} \right)
\left( \begin{array}{c} \xi_{0} \\  \xi_{2}  \end{array} \right),  
\eeq

or

\beq
\left( \begin{array}{c} \xi_{0} \\  \xi_{2}  \end{array} \right)  =
\left( \begin{array}{c c} \frac{1}{2} &  \frac{1}{2} \\   \frac{4}{3} & -\frac{4}{3} \end{array} \right)
\left( \begin{array}{c} \xi_{||} \\  \xi_{\perp}  \end{array} \right).  
\eeq

The hexadecapole term would add a third column in the matrix on the right side of
Equation (\ref{wedgemonoquad_equality}) with absolute values of $15/128\sim 0.12$.

If $\xi(\mu,s)$ consisted only of $\ell=0,2$ terms, 
the two $\Delta \mu=0.5$ wedges would form a complementary basis to that of the multipoles. 
In the more generic case, 
these wide clustering wedges comprise an alternative, but not totally
complementary basis. It is easy to see that given any combination of even $\ell$s, 
the monopole is always the average of the $\Delta\mu=0.5$ wedges, 
but the quadrupole is 
combined with higher order multipole terms in a complicated fashion. 
This means that given non-zero $\xi_{\ell>2}$ terms, these wide wedges do not contain exactly the same 
information as [$\xi_0,\xi_2$], and hence form an alternative, non-complementary 
basis. To have a fully complementary basis to $\xi$ which contains $N$ multipoles would, of course,  
require the same number of wedges (or any other projection).

In Appendix \ref{projectionspractice_section}  
we test the relationships between the clustering 
wedges and multipoles. 
We find that the two wide clustering wedges ($\Delta\mu=0.5$) are defined fairly well by
the monopole and quadrupole in velocity-space (and monopole only in real-space), and hence may
be used as an alternative basis to these multipoles to project most of the information
contained in $\xi(\mu,s)$. 
In the next section we utilize this fact to show the effectiveness of the wedges 
to understand geometric distortions, and use them to constrain $H$ and $\dA$. 

\subsection{Dilation and warping in clustering: a treatment of multipoles and wedges}
\label{dialtionwarping_section}

Here we show that radial clustering wedges are, 
as expected, mostly sensitive to $H$ while the transverse 
ones are most sensitive to $\dA$, 
even for two wide $\Delta\mu=0.5$ clustering wedges.

\PWpaper parameterize geometric distortions in clustering. We make use of their Equations
$(2)-(4)$, and introduce them here in configuration space. 
We define $\vec{s}$ to be the true spatial separation  vector 
with radial and transverse components $s_{||}$, $s_{\perp}$. 
The geometrically distorted separations 
are indicated by a ${\mathcal D}$ superscript. 

As shown by \PWpaperii, distortions to the components of the separation   
can be parameterized by a factor $\alpha$ which causes 
isotropic {\it dilation} and  
a parameter $\epsilon$ that causes anisotropic {\it warping}, such that:
\begin{eqnarray}
\label{pll_compenent}
s^{\mathcal D}_{||}&=&s_{||}\alpha(1+\epsilon)^2 \\
\label{per_component}
s^{\mathcal D}_{\perp}&=&s_{\perp}\alpha(1+\epsilon)^{-1}.
\end{eqnarray}

The Jacobian of transformation between the true volume element ${\mathrm d}^3s$ and the distorted
${\mathrm d}^3s^{\mathcal D}$ is $\alpha^3$. 
Given that the comoving separation 
${\mathrm d}\chi=c\,{\rm d}z/H(z)$, and that 
the physical angular diameter distance is 
$(1+z)\dA=\chi$,\footnote{assuming flatness, see \S\ref{hzplane_section} for a more generic treatment} 
it is easy to show that the dilation parameter is given by
\beq \label{eq_dilation}
\alpha= 
\left( \frac{H^{\mathcal D}}{H} \right)^{1/3}\left(\frac{\dA}{\dA^{\mathcal D}} \right)^{2/3}.
\eeq

Applying Equation (\ref{eq_dilation}) to Equations (\ref{pll_compenent})
and (\ref{per_component}) yields 
\beq \label{eq_warp}
1+\epsilon=\left( \frac{H^{\mathcal D}\dA^{\mathcal D}}{H\dA} \right)^{1/3}.
\eeq

The combination of Equations (\ref{pll_compenent}) and (\ref{per_component}) yields:
\begin{eqnarray}
\label{s_distorted_equation}
s^{\mathcal D}&=&\alpha\left(1+2\epsilon {\mathcal P}_2\left(\mu\right)\right)s,\\
\label{mu_distored_equation}
(\mu^{\mathcal D})^2&=&\mu^2 + 6\epsilon(\mu^2-\mu^4). 
\end{eqnarray}
Note the difference in signs between configuration 
space $\mu$ and $k-$space $\mu_k$ (Equation (3) of \PWpaperalt).

Substituting these last two equations 
into Equation (\ref{eq_expansion}) yields:
\begin{eqnarray}
\label{mono_equation}
\xi^{\mathcal D}_0(s)&=&\xi_0(\alpha s)+\epsilon \left( \frac{2}{5}\frac{d\xi_2(s)}{d\ln(s)}+\frac{6}{5} \xi_2(\alpha s)\right), \\
\label{quad_equation}
\xi^{\mathcal D}_2(s)&=&\left( 1+\frac{6}{7}\epsilon \right)\xi_2(\alpha s) + \frac{4}{7}\epsilon \frac{d\xi_2(s)}{d\ln(s)}+\nonumber\\ 
&&2\epsilon\frac{d\xi_0(s)}{d\ln(s)}. 
\end{eqnarray}
Here we neglect terms of order ${\mathcal O}(\epsilon^2)$.  
See Appendix \ref{quad_terms_appendix} for inclusion of hexadecapole terms.

As \PWpaper mention, the second and third terms on the right hand side of 
Equation (\ref{mono_equation}) effectively cancel each other out, leaving 
$\xi^{\mathcal D}_0(s) \approx \xi_0(\alpha s)$. 
\Epaper and \cite{sanchez09a} demonstrated that this relationship works very well 
on the SDSS DR3 and DR6 LRG samples respectively, showing that the monopole alone
constrains the degenerate combination in $\alpha$, meaning $\dA^2/H$. 

\PWpaper showed that the combined information of $\xi_0(s)$ and $\xi_2(s)$
can be used to measure simultaneously $\alpha$ and $\epsilon$.  
Because these parameters depend on different combinations of $H$ and $\dA$,
this can in turn break the degeneracy between these parameters obtained 
from an analysis based only on the monopole.
Here we present a similar concept based on clustering wedges.

By combining Equations (\ref{wedgemonoquad_equality})
with Equations (\ref{mono_equation}) and (\ref{quad_equation}) it is possible to 
quantify the effect of geometrical distortions on the clustering wedges:
\begin{eqnarray}
\label{los_distorted}
\xi_{||}^{\mathcal D}(s)&=&\xi_{||}\left( \frac{H^{\mathcal D}}{H}s\right) +{\mathcal C}_{||}(\epsilon),\\
\label{trv_distorted}
\xi_{\perp}^{\mathcal D}(s)&=&\xi_{\perp}\left( \frac{\dA}{\dA^{\mathcal D}}s\right) +{\mathcal C}_{\perp}(\epsilon),
\end{eqnarray}
where we have used the fact that for small $\epsilon$,  
$\alpha(1+2\epsilon)\approx H^{\mathcal D}/H$ and 
$\alpha(1-\epsilon)\approx \dA/\dA^{\mathcal D}$. 
These equations hold for clustering wedges in general, 
where for $\Delta\mu=0.5$ the correction terms are given by
\begin{eqnarray}\label{correct_los}
 {\mathcal C}_{||}(\epsilon) &=& \epsilon\left(   -\frac{5}{4}\frac{d\xi_0(s)}{d\ln(s)}  - \frac{19}{140}\frac{d\xi_2(s)}{d\ln(s)} + \frac{213}{140}\xi_2(\alpha s) \right) \nonumber \\ 
 &=&  \epsilon\left(   -\frac{677}{840}\frac{d\xi_{||}(s)}{d\ln(s)}  - \frac{373}{840}\frac{d\xi_{\perp}(s)}{d\ln(s)}  \right) + \nonumber\\ 
    && \epsilon\left(       \frac{71}{35} \left( \xi_{||}(\alpha s) - \xi_{\perp}(\alpha s)  \right) \right),
\end{eqnarray}
and                        
\begin{eqnarray}\label{correct_los2}
 {\mathcal C}_{\perp}(\epsilon) &=&   \epsilon\left(   \frac{1}{4}\frac{d\xi_0(s)}{d\ln(s)}  - \frac{53}{280}\frac{d\xi_2(s)}{d\ln(s)} +\frac{123}{140}\xi_2(\alpha s) \right) \nonumber\\ 
  &=&  \epsilon\left(   -\frac{107}{840}\frac{d\xi_{||}(s)}{d\ln(s)}  + \frac{317}{840}\frac{d\xi_{\perp}(s)}{d\ln(s)}  \right) + \nonumber\\ 
    && \epsilon\left(       \frac{41}{35} \left( \xi_{||}(\alpha s) - \xi_{\perp}(\alpha s)  \right) \right).
\end{eqnarray}
We neglect ${\mathcal O}(\epsilon^2)$ and higher contributions.
Equation 25 of \cite{taruya10a} 
gives a more generic treatment of the linear \AP effect in the $P(\mu,k)$ plane, 
whereas the equations presented here are 1D projections. 

In \S \ref{wedgeshifting_section} we show the validity of the equations presented here. 
In \S \ref{reproducing_hda_section} we demonstrate that correcting for the \AP effect 
yields $H$ and $\dA$ to high accuracy, and compare the wedge technique to multipoles.

\section{Projections in practice: testing the \AP effect}
\label{poc_section}

Here we demonstrate the applicability of Equations (\ref{los_distorted}) and (\ref{trv_distorted}) 
using analytic formulae and mock galaxy catalogues. We show that, as expected from 
these equations the wide $\Delta \mu=0.5$ ``radial" clustering wedge dominantly constrains $H$, 
and the wide ``transverse" one is sensitive to $\dM$. 
This means that the information from these clustering wedges 
breaks the degeneracy in the combination $\dM^2/H$ obtained from the monopole only, 
and that the underlying values of $H$ and $\dA$ can be obtained at high accuracy. 
Here we also compare the results obtained in this way with those recovered 
from the alternative multipole technique of \PWpaperii.

As described in \S\ref{stats_section}, 
we define the true $\xi^{\TRUE}$ signal to be 
the mock mean results obtained using the true simulation cosmology when converting 
redshifts to comoving distances $\chi$. 
The distored signal $\xi^{\DIST}$  is similar to 
$\xi^{\TRUE}$ except 
that we use a different cosmology to convert $z$ into $\chi$. 
We perform this \AP effect both in real- and velocity-space, and thus we apply geometrical 
distortions in both cases. Finally, we define the shifted signal $\xi^{\SH}$  to be our attempts to 
reconstruct $\xi^{\DIST}$ from $\xi^{\TRUE}$. 
Technically this means that we transform $\xi^{\TRUE}$ to 
$\xi^{\SH}$ using  
Equations (\ref{los_distorted}) and (\ref{trv_distorted}) 
for the wedges and Equations (\ref{mono_equation}), (\ref{quad_equation}) for multipoles.

The tests we perform are as following:
\begin{enumerate}
\item \S \ref{wedgeshifting_section}: We show a near perfect dependence of the radial wedge on $H$ 
and the transverse wedge on $\dM$ by shifting $\xi^{\TRUE}$ results to match a $\xi^{\DIST}$ signal.
\item \S \ref{reproducing_hda_section}: We use $H$ and $\dM$ as varying parameters when fitting 
a model constructed from a template (the $\xi^{\TRUE}$ signal) 
to match ``data points" (the $\xi^{\DIST}$ signal), 
and show that  the best fit constraints on  these parameters agree with the true values.
\end{enumerate}

Our mechanism is similar in concept to that used by \PWpaperii, 
with the difference that we simulate the observer's point of view by including large-angle
effects and explicitly use a wrong cosmology when converting 
redshifts to comoving distances (Equation \ref{comoving_equation}). 
\PWpaper  warped  distant boxes according to a given value of $\epsilon$. 
One main difference is that we focus on low values of the warping parameter 
because our derivations are valid for small $\epsilon$. 
Here we focus on results obtained by changing the dark energy equation of state 
from its true value $w=-1$ value to $-1.1$  (yielding $\alpha=0.9832$, $\epsilon=-0.0033$ at the mean 
redshifts of the mocks, $\avg{z}=0.33$)
as well as $-0.9$ (yielding $\alpha=1.0175$, $\epsilon=+0.0035$) to examine two directions of shift.
These choices are semi-arbitrary, as $w$ is known to an accuracy of 
$10\%$ (\citealt{komatsu09a}, \citealt{sanchez09a}, \citealt{percival09b}, \citealt{reid09a}).  
The distorted cosmologies analysed here 
correspond to the squares shown in Figure  \ref{hzplane_parameters} 
in the $z=0.33$ panel. 
For this low redshift these variations  
are highly degenerate 
with misestimating $\Omega_{\rm M0}$ or $\Omega_{\rm K}$. 
Lastly, the radial direction used here ($\mu=1$) is the bisecting vector of $\vec{s}$
originated from the observer.

\subsection{Analyzing geometric distortions: proof of concept using mock catalogs}
\label{wedgeshifting_section}

In this section we test the accuracy of Equations (\ref{mono_equation})--(\ref{correct_los2}) 
using the suit of 160 SDSS-II mock galaxy catalogues described in \S\ref{mocks_section}.

We measure the correlation functions using the true cosmology of the simulations 
($\xi^{\TRUE}$ measurements) and the incorrect value of $w=-1.1$ ($\xi^{\DIST}$ measurements), and we use these equations
to shift the $\xi^{\TRUE}$ measurements to match the $\xi^{\DIST}$ ones ($\xi^{\SH}$ measurements).


\begin{figure*}
\includegraphics[width=0.45\textwidth]{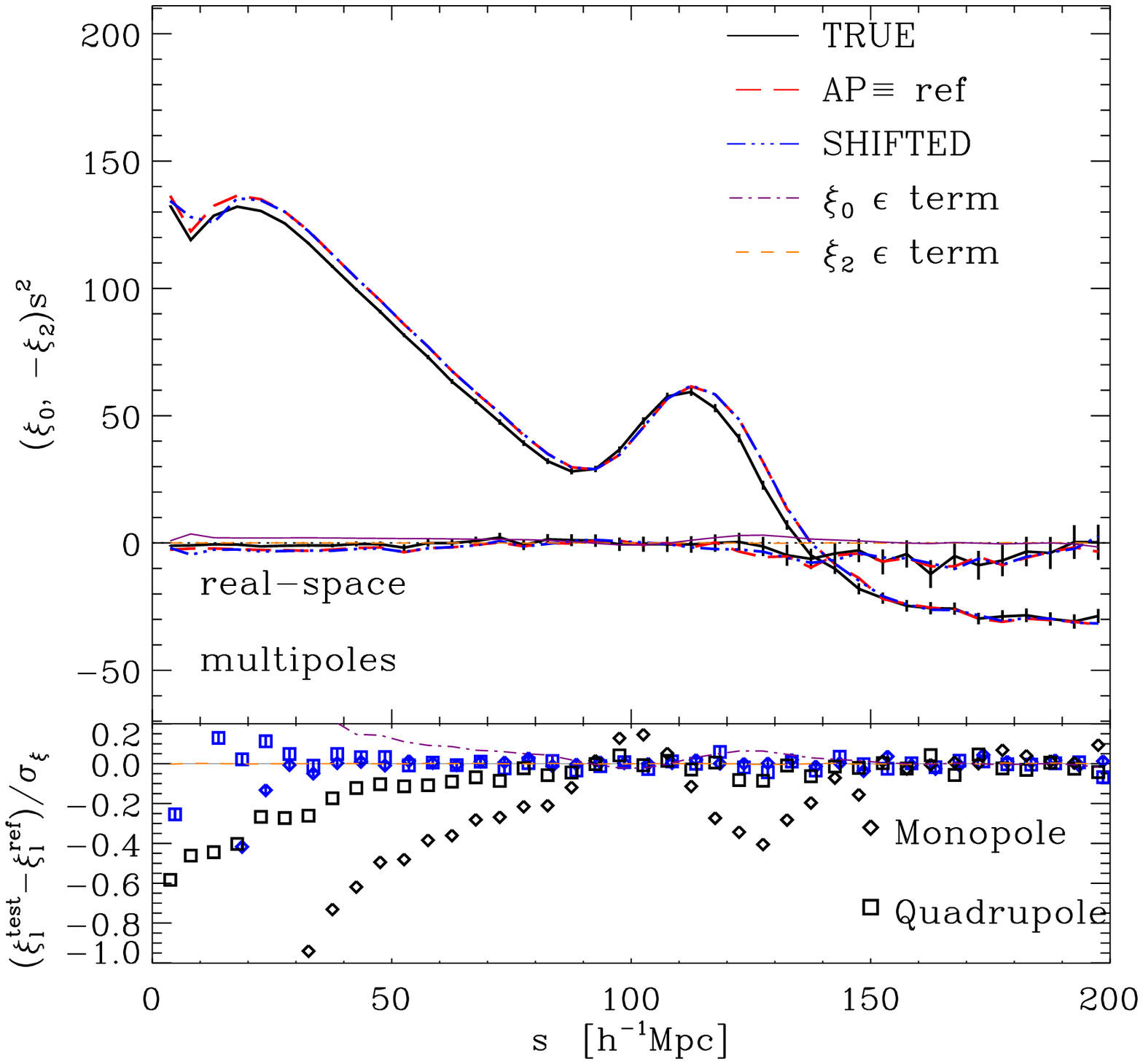}
\includegraphics[width=0.45\textwidth]{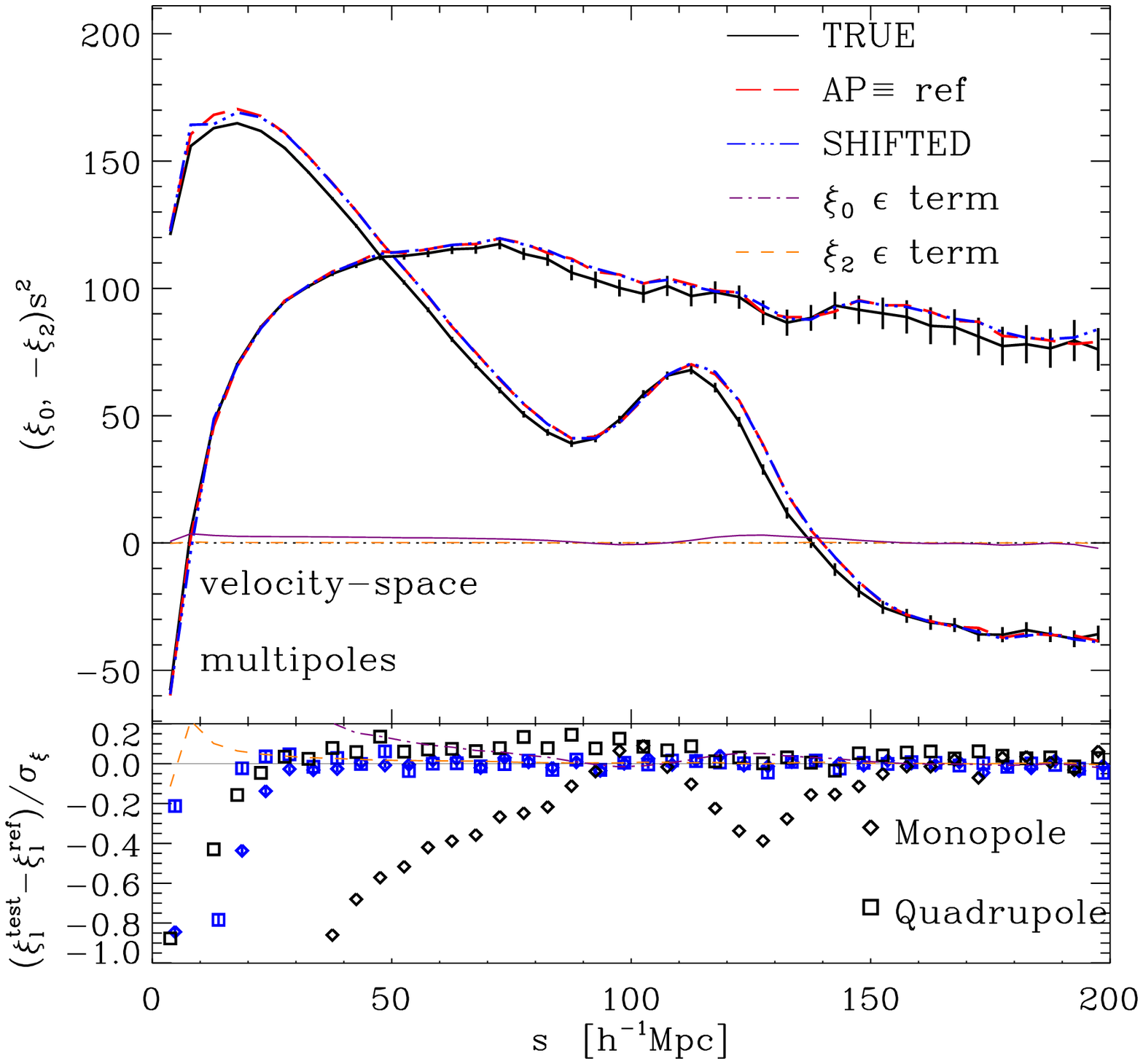}\\
\includegraphics[width=0.45\textwidth]{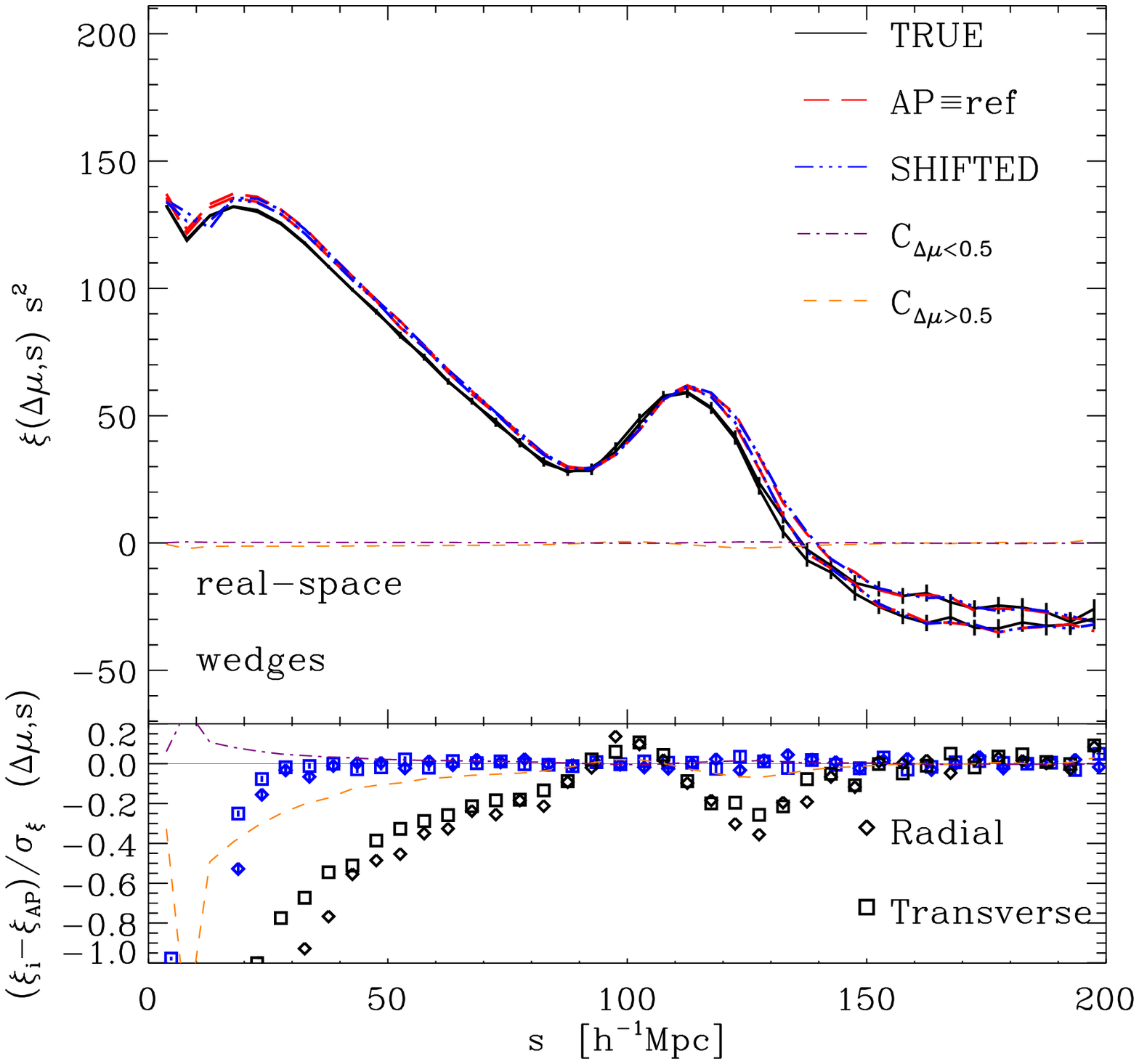}
\includegraphics[width=0.45\textwidth]{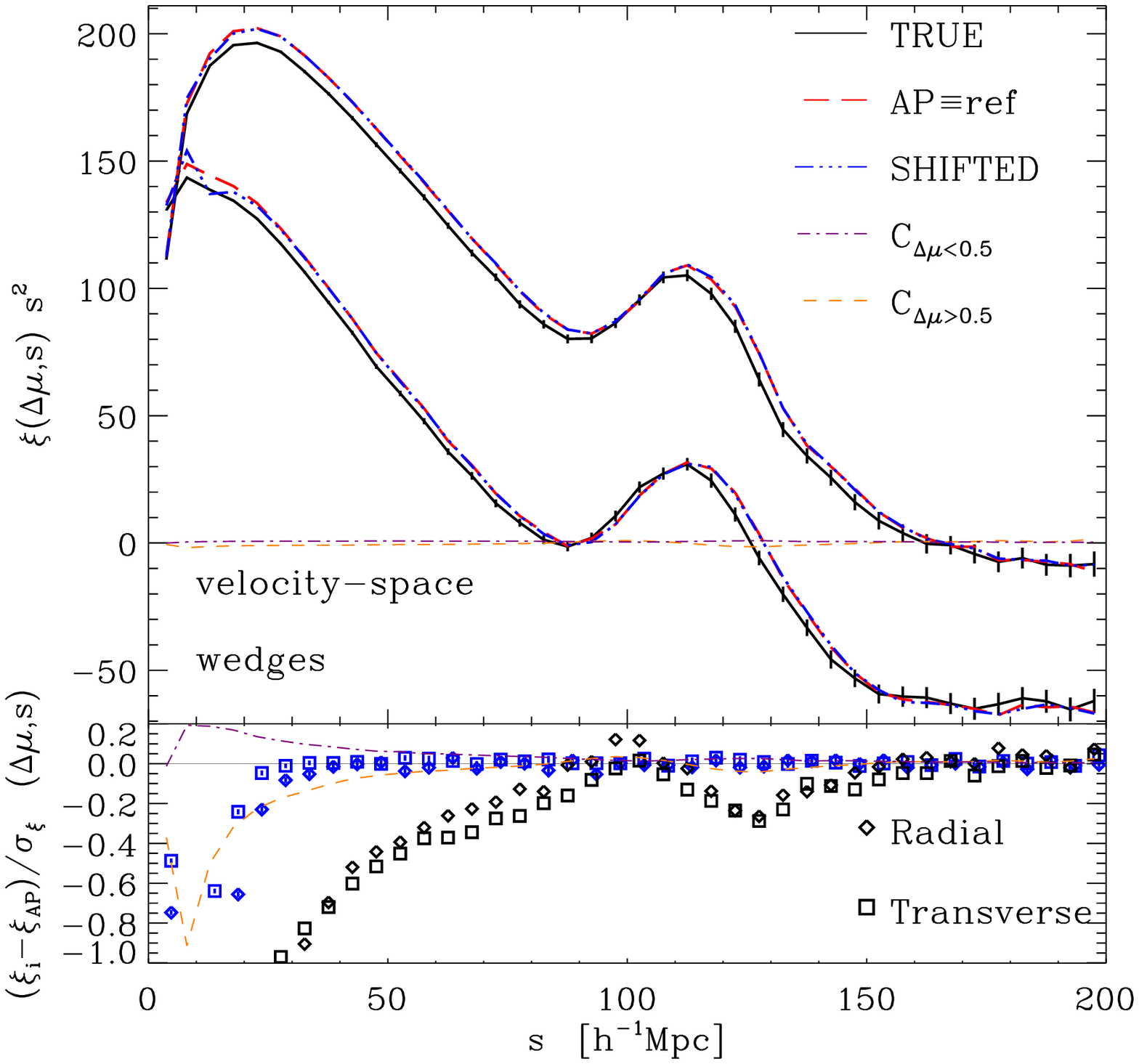}\\
\caption[Geometrical effects on 1D $\xi$ projections]{
Mock mean clustering projections (based 160 LasDamas realizations) 
with and without the \AP effect.
Left: real-space. Right: velocity-space. 
Top: multipoles (applying Equations \ref{mono_equation}, \ref{quad_equation}). 
Bottom: $\Delta\mu=1/2$ wedges (Equations \ref{los_distorted}-\ref{correct_los2}).
In the top panel of each plot are the 1D projections, 
and in the bottom panels the difference 
of each result with the DISTORTED signal (labeled as AP) 
normalized by the uncertainty of one realization. As indicated in legend, black solid lines/symbols 
are the true projection signals ($\TRUE$), the long-dashed red lines 
are the AP signals ($\DIST$). We apply the AP shift to the true signals to obtain 
the triple-dot-dashed blue lines/symbols ($\SH$). 
A perfect shift would yield a null result for
the blue symbols in the bottom panels. 
The dot-dashed purple and dashed orange lines, as indicated in the legend,
are the first order $\epsilon$ correction terms. The \AP distortion applied here is using
$w^{\rm FID}=-1.1$ instead of  the true value $-1$ when converting $z$ to comoving distances. 
}
\label{poc_ld}  
\end{figure*}

Our results are shown in Figure \ref{poc_ld}. The upper plots correspond to the 
results for the clustering multipoles and the bottom ones to the clustering wedges. 
The left plots show the measurements in real-space, and the right 
in velocity-space. Each plot consists of two panels. 
The top panels show $\xi\cdot s^2$
The $\xi^{\TRUE}$ results are shown by solid black lines, 
the $\xi^{\DIST}$ results (called AP) 
are dashed red lines, and the $\xi^{\SH}$ results are the dot-dashed lines. 
In the bottom panels, the $\xi^{\DIST}$ results form the reference 
to which we compare differences of $\xi^{\TRUE}$ (black) 
and $\xi^{\SH}$ (blue) in units of the uncertainty $\sigma_\xi$. 
In the multipoles the monopole 
results are in diamonds, and the quadrupole   
results are squares. 
In the wedges the radial wedge ($\Delta\mu>0.5$) 
results are in diamonds, and the transverse wedge ($\Delta\mu<0.5$)  
results are squares.

In the top left plot we see the \AP effect on the multipoles in real-space.  
We verify that the shift in the monopole from the $\xi^{\TRUE}_0$ signal (solid) to the 
$\xi^{\DIST}_0$ (dashed) is described very well at zeroth order in $\epsilon$, 
meaning by $\xi_0^{\mathcal D}(s)\sim\xi_0^{\TRUE}(\alpha s)$. Adding the warping $\epsilon$ terms 
adds little.  In velocity-space, however, we do notice improvements 
when adding the first order correction at $s<50$\hmpcii. 

In real-space we do not expect signal in higher order multipoles. 
In Figure  \ref{poc_ld}, though, there is a  slight detection of 
$\xi_2$ and even $\xi_{\ell>2}$ measurements.
These probably arise due to Poisson shot noise (either in the random points or data), or 
due to large angle effects. Nevertheless, we see that the \AP effect is understood for 
$\xi_2$ (Equation \ref{quad_equation}) in both real- and velocity-space. 
As expected in $\xi_2$, the $\alpha$ shift only is not sufficient, 
and adding the $\epsilon$ terms (blue) explain the \AP effect to  
high accuracy. We also test higher order corrections of $\xi_4$, d$\ln(\xi_s)/$d$s$ 
in the quadrupole and find them negligible.

As for the $\xi_4$ statistics, we find that in velocity-space  
our corrections work well at  $s>30$\hmpcii. 
We argue that it does not work at smaller scales 
because we do not use the expected $\xi_{\ell>4}$ terms, 
which are required due to  {\it leakage} of multipoles in the \AP effect.

We find similar trends for the clustering wedges statistics (bottom plots). 
In real-space (bottom left) we expect both $\xi^{\TRUE}$ wedges to coincide. 
We notice minute differences at $130$\hmpcii, 
(amplified in the plot by $s^2$ which make them visible). 
In velocity-space, as expected from the large quadrupole,  
there is a clear separation between the wedges, 
where the radial wedge is strongly suppressed and the transverse wedge elevated. 
The blue symbols indicate that the \AP is very well 
described by Equations  (\ref{los_distorted})-(\ref{correct_los2}). 
The first order correction terms ${\mathcal C}_{||,\perp}(\epsilon)$ 
(lines; see legend) are small ($\lesssim 1\%$ on most scales) with respect to the 
$\xi^{\TRUE}$ wedge signals. This indicates that even with our definition of 
wide ``radial" and ``transverse" wedges ($\Delta\mu=0.5$), 
the radial wedge is mostly sensitive to $H$, while the transverse wedge to $\dM$. 
That said, in Appendix  \ref{wedgescorrestion_section} 
we show that although setting ${\mathcal C}_{||,\perp}(\epsilon)=0$ 
yields fairly accurate $H,  \ \dA$ results, including ${\mathcal C}_{||,\perp}(\epsilon)$ 
can improve the results.

To summarize this test, we have proven here that \AP effects 
on the clustering multipoles and wedges 
are very well understood. We show that the dilation and warping terms shown 
by \PWpaper explain this effect very well 
in the monopole and quadrupole of the two-point correlation function. 
In addition we show an alternative approach to measuring the \AP effect by 
analyzing clustering wedges. 
We show here that even a wide ``radial" wedge is, as expected,
 mostly sensitive to $H$ and a wide ``transverse" wedge to $\dM$. 
In Appendix \ref{matsubara_section_ap} we perform similar 
tests on analytic formulae and obtain similar conclusions. 
In the next section we investigate the power of this method 
to obtain constraints on $H$ and $\dM$ from measurements of the
multipoles and wedges.

\subsection{Reproducing the true  $H$ and $\dA$}
\label{reproducing_hda_section}

Here we perform the \AP test on 
the  LRG SDSS-II mock catalogues described in \S\ref{mocks_section}
to measure $H$ and $\dM$.  
When quoting uncertainties in these parameters, 
we show what might be expected for a survey with the 
same number density as SDSS-II ($n\sim 10^{-4}\,h^{3}\rm{Mpc}^{-3}$)
but a volume twelve times larger, corresponding to the total Hubble volume 
(i.e, a sphere of radius $c/H_0$).

To simulate the observer's point of view, we assume the $\xi^{\DIST}$ measurements 
to be the ``data" points. We then find the best fitting 
models based on physical templates. Our first step is to perform 
an ideal test, where the template is the $\xi^{\TRUE}$ mock mean signal. 
In other words, we are not concerning ourselves, at this point, 
with uncertainties beyond the \AP effect. For example, 
this means we assume that we fully understand the amplitude ``bias", 
and dynamical $z$-distortions. We consider this merely as a 
``proof of concept" of the analysis, and in  \S  \ref{practical_section} take 
a more realistic approach by adding more {\it unknowns}. 

In \S \ref{stats_section} 
we describe the construction of the 
covariance matrix, 
in which we take into account covariances 
between the statistics. 
When manipulating 
the template, the covariance 
matrix is not varied, but 
rather fixed to the true  
cosmology. 
\cite{samushia11a} discuss 
the sensitivity of 
 $C_{ij}$ to amplitude parameters 
 and its insensitivity to shape parameters. 


\begin{figure*}
\includegraphics[width=0.45\textwidth]{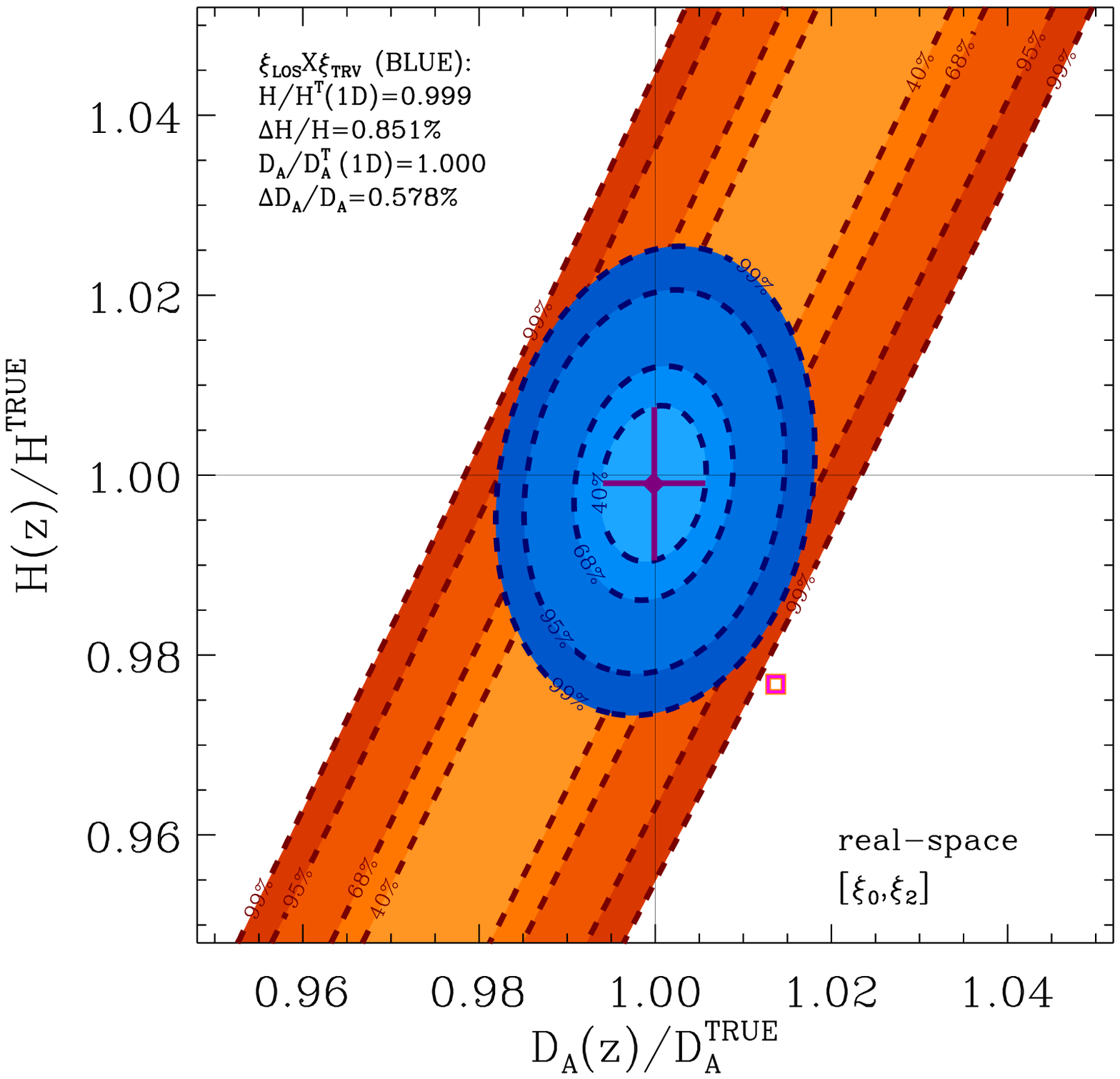}
\includegraphics[width=0.45\textwidth]{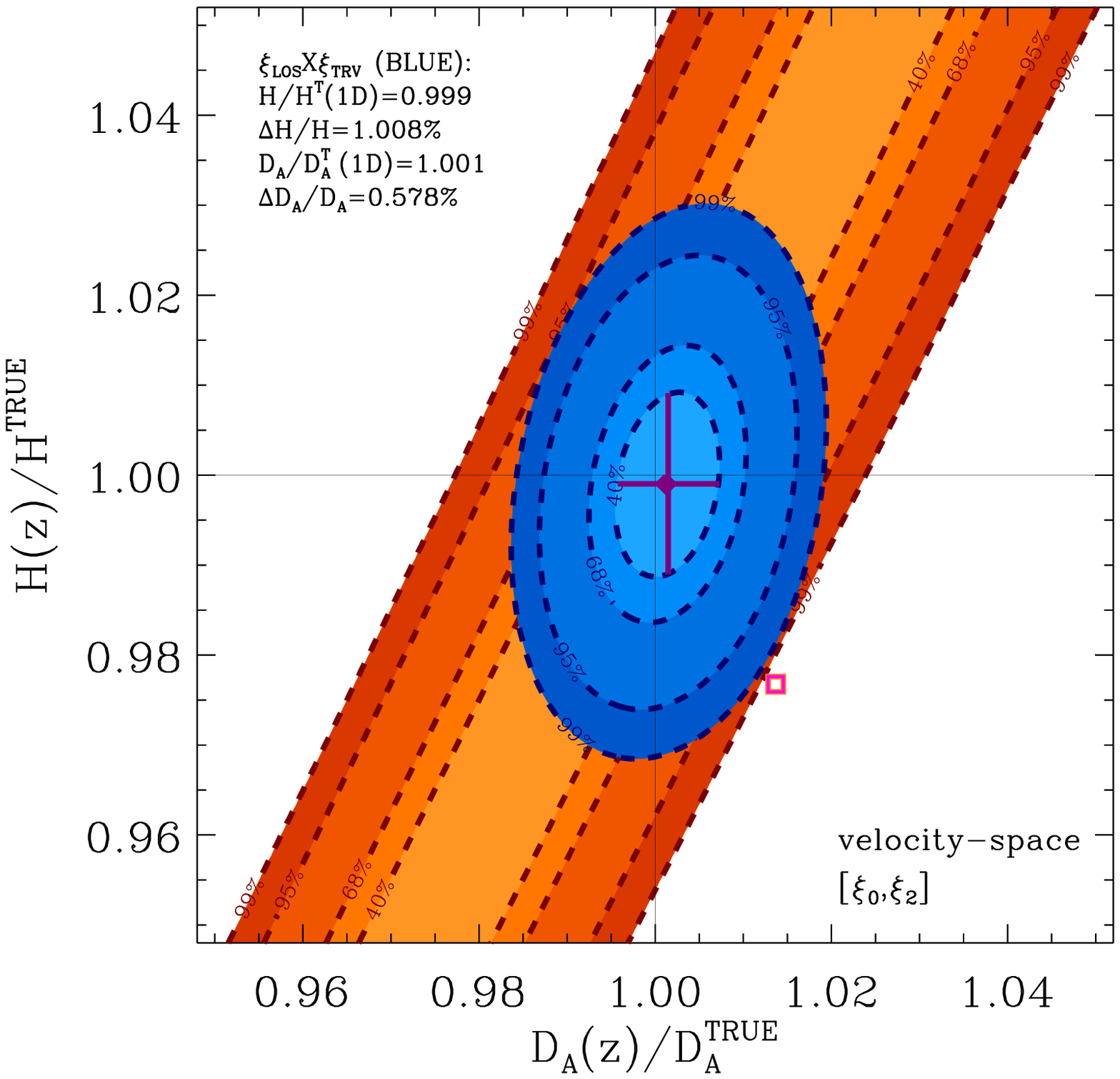}\\
\includegraphics[width=0.45\textwidth]{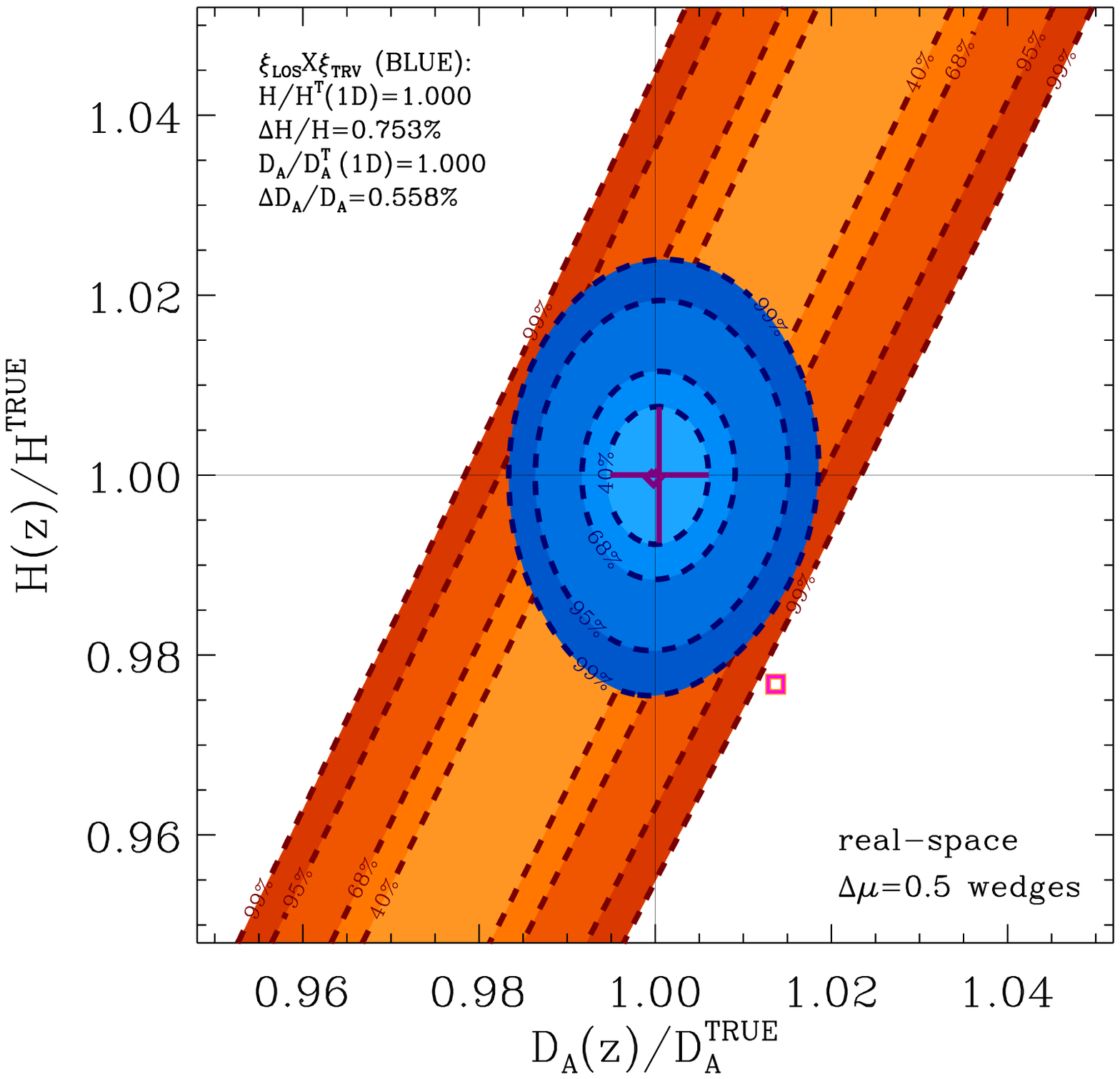}
\includegraphics[width=0.45\textwidth]{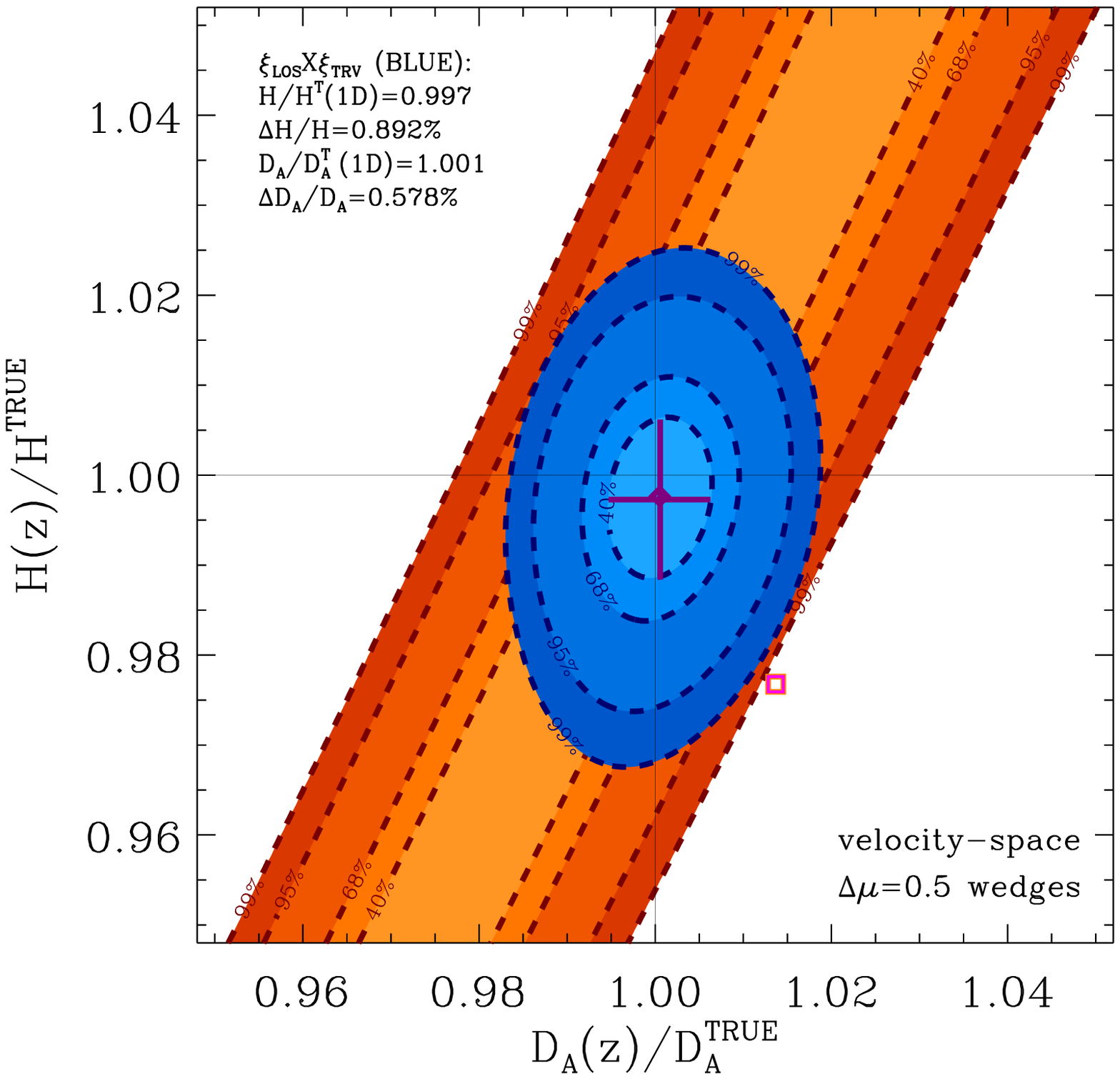}\\
\caption[Reproducing the true $H,\dA$: testing geometrical shifts]{
$H-\dA$ joint constraints from \AP shift tests 
when analyzing mock mean clustering 1D statistics.
Left panels: real-space. Right: velocity-space. 
The blue ellipses correspond to analysis 
of [$\xi_0,\xi_2]$ (top panels) and 
$\Delta\mu=1/2$ wedges (bottom). 
The orange bands are the results 
when analyzing only the $\xi_0$. 
The parameter space tested is $\Phi=$[$\dM$,$H$], 
where amplitude remains fixed (``\AP only").
The true cosmology of the simulations corresponds to [1,1],  
and the distortion cosmology tested 
when converting redshifts to comoving distances 
(pink box) corresponds to using $w^{\mathcal D}=-1.1$ 
instead of the true $-1$ value, and is the same for all tests.
The even contour likelihood mock mean values are for $40,\ 68.3,\ 95.4,\ 99 \%$ CL. 
The most likely 2D values are purple diamonds, and the 1D marginalized $1\sigma$ results are
the purple crosses. On the top of each plot we inscribe for each parameter $\Phi$ 
the deviation from the true value in the 1D marginalized results, as well as $\Delta \Phi/\Phi$. 
The constraints correspond roughly to a fictitious survey with 
a Hubble volume ($R\sim c/H_0$) and a galaxy density
of $n\sim 10^{-4} (h/{\rm Mpc})^{3}$. 
}
\label{hzdm_fitsi}
\end{figure*}


\begin{figure*}
\includegraphics[width=0.45\textwidth]{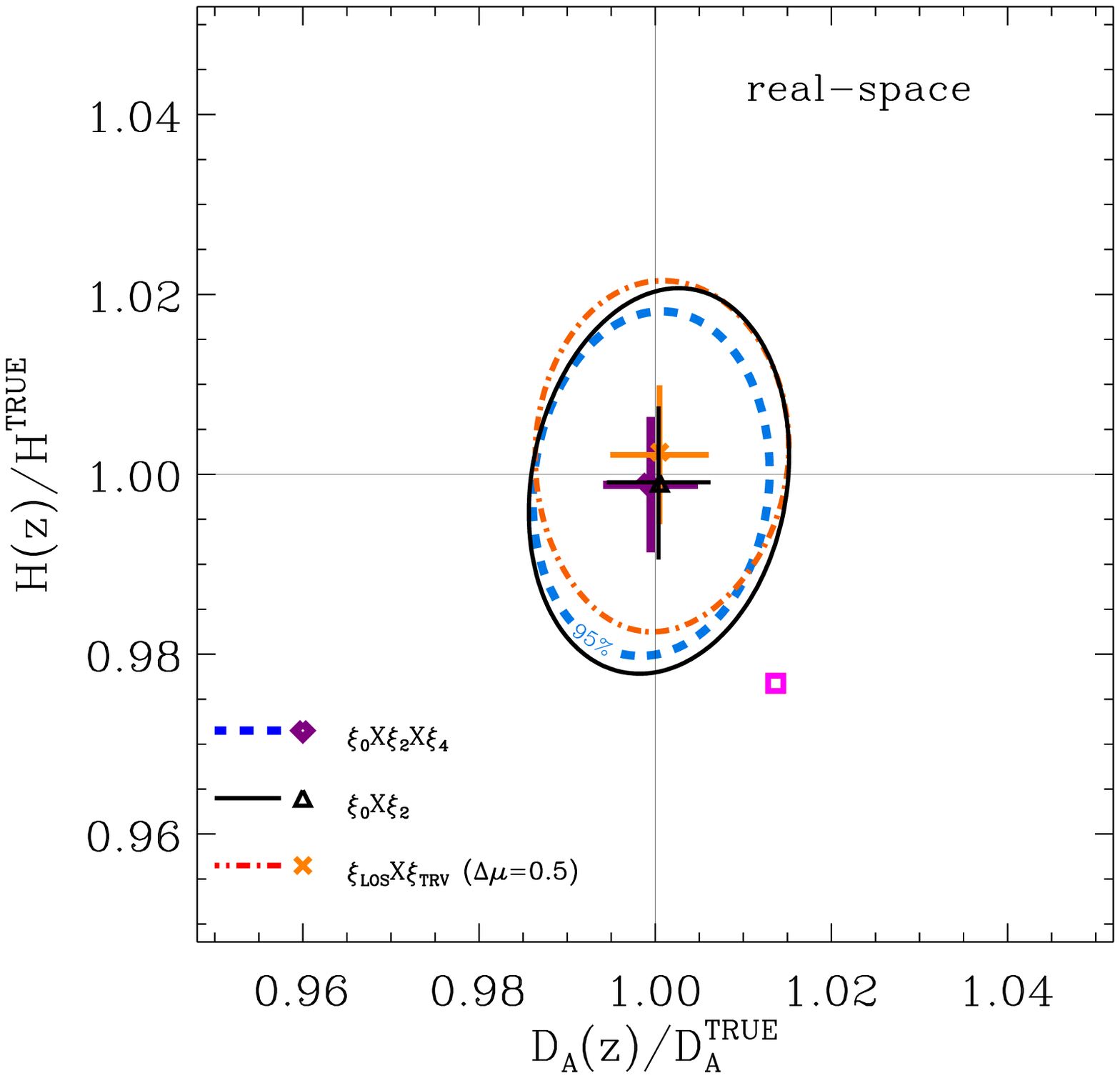}
\includegraphics[width=0.45\textwidth]{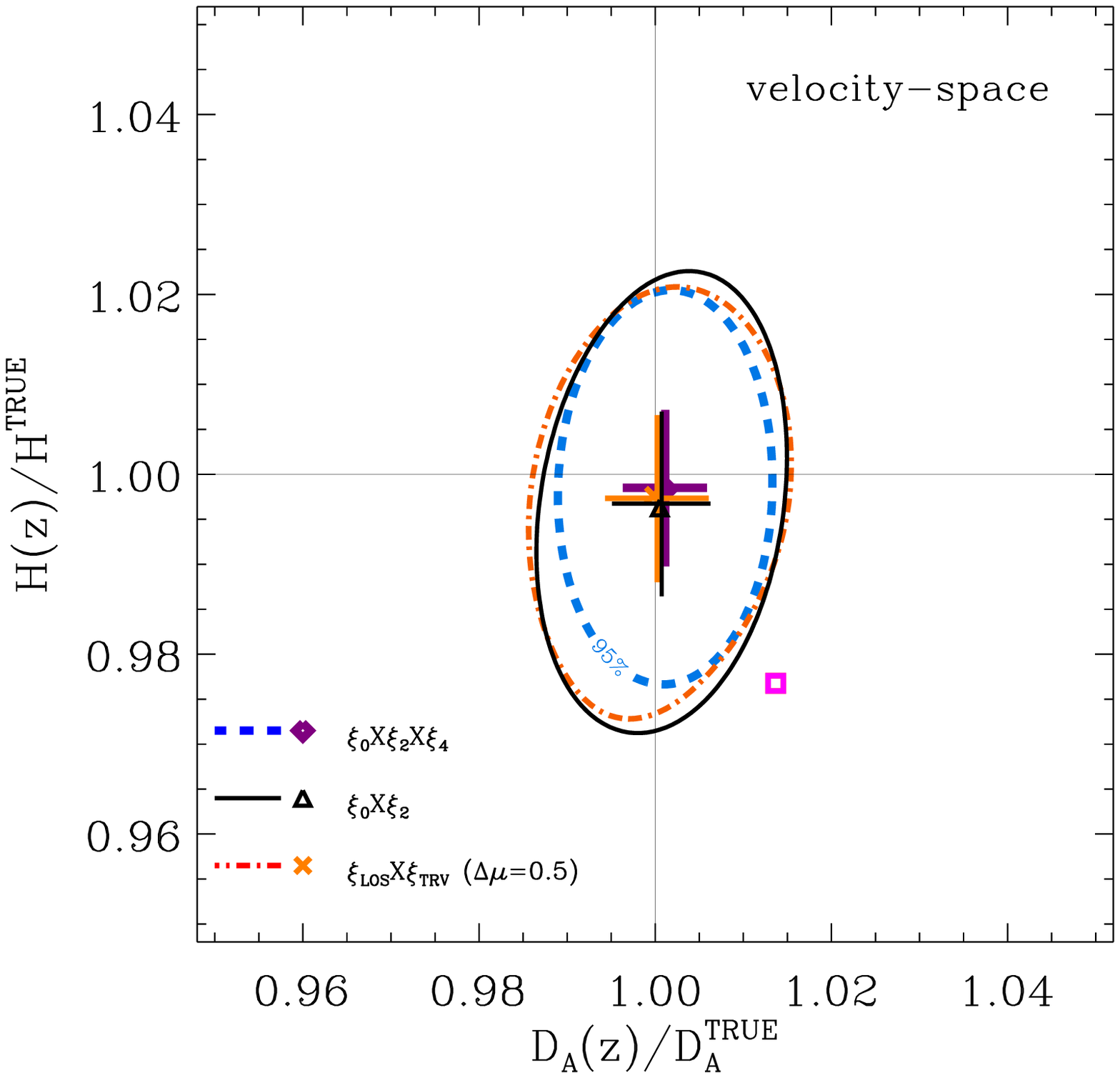}\\
\includegraphics[width=0.45\textwidth]{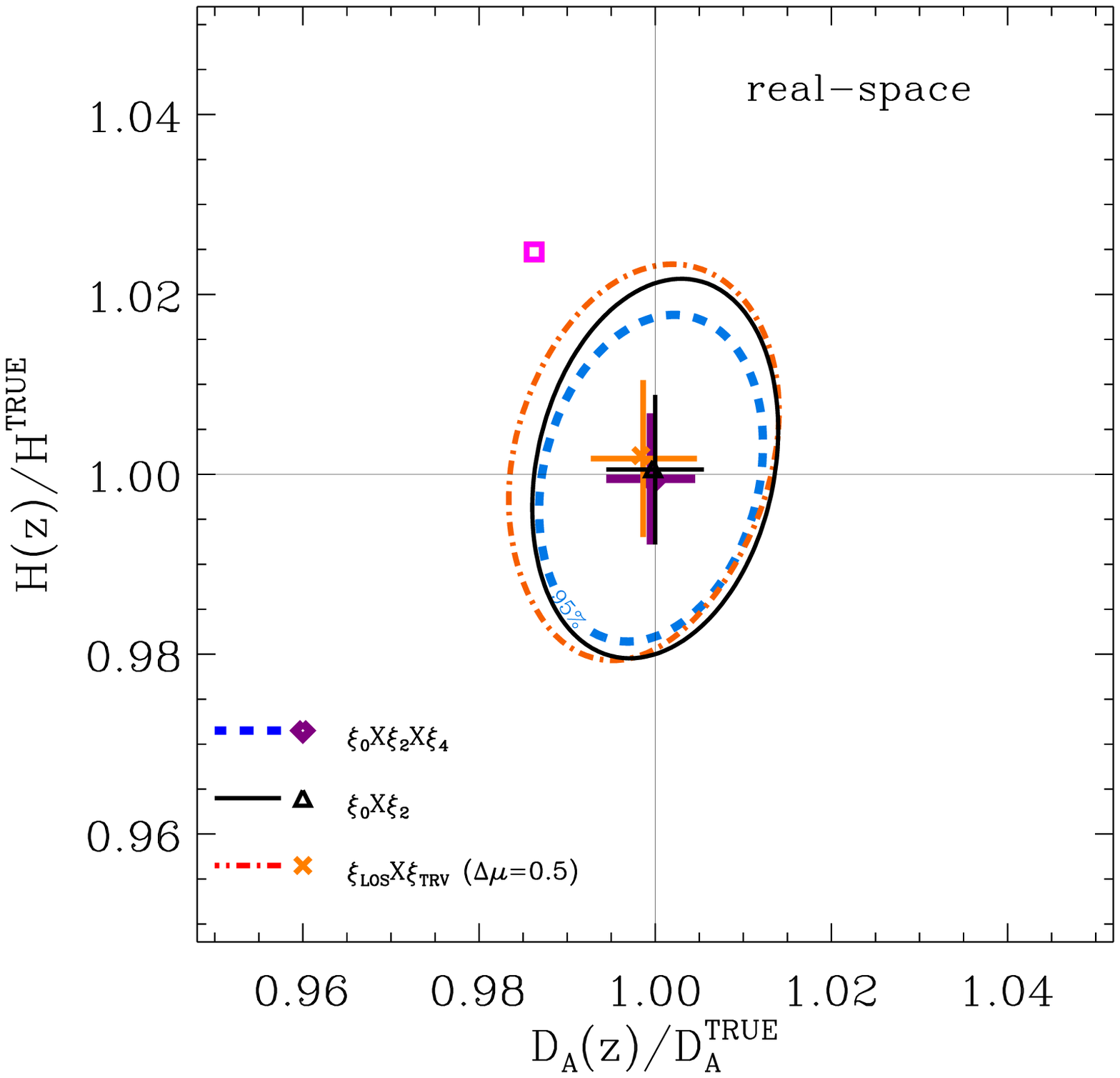}
\includegraphics[width=0.45\textwidth]{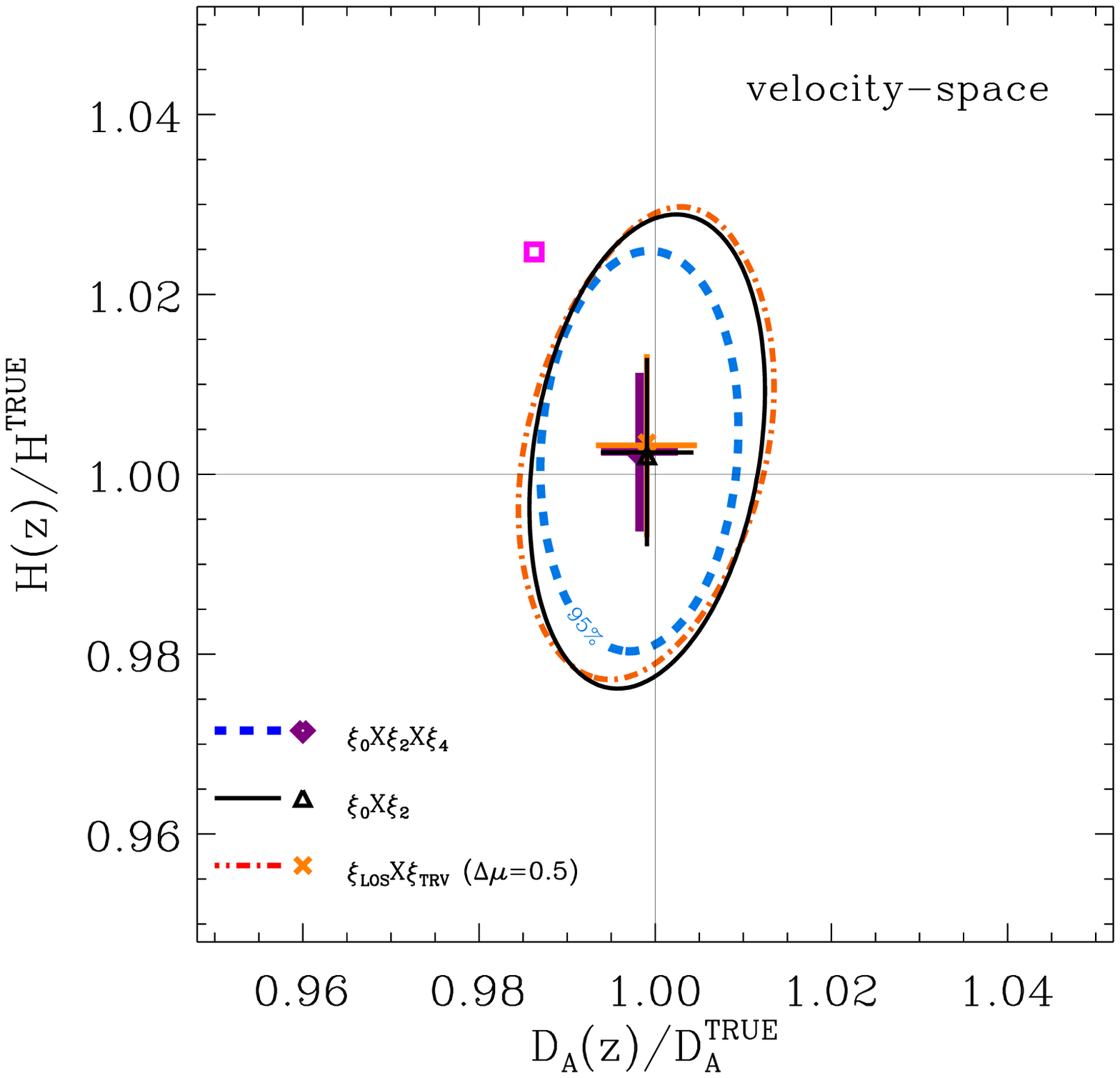}\\
\caption[Reproducing the true $H,\dA$: $\xi_4$ improvements]{
Joint $2\sigma$ constraints on the $H-\dA$ plane from \AP shift tests 
when analyzing mock mean clustering 1D statistics.
Left panels: real-space. Right: velocity-space. 
The \AP distortion tested is when using 
$w^{\mathcal D}$ different from the true value $w^{\mathcal T}=-1$  
when converting redshifts to comoving distances. 
Top panels show results for $w^{\mathcal D}=-1.1$,  
and bottom panels for $w^{\mathcal D}=-0.9$ 
(notice difference in input fiducial cosmology indicated by the magenta squares). 
The Parameter space analised is [$\dA$,$H$], where the amplitude remains fixed (``\AP only").
The dashed blue lines correspond to the constraints 
for the multipole combination [$\xi_0,\xi_2,\xi_4$], 
solid black lines to [$\xi_0,\xi_2$], 
and dot-dashed orange lines to $\Delta\mu=1/2$ wedges. 
Crosses indicate marginalized $1\sigma$ results according to color, and symbols most likely 2D value.
The $w^{\mathcal D}=-1.1$ clustering wedges and [$\xi_0,\xi_2$] results 
are the same as the blue ellipses in Figure  \ref{hzdm_fitsi}.  
Adding the hexadcapole information clearly improves constraints.
These constraints correspond roughly to a fictitious survey with a 
Hubble volume ($R\sim c/H_0$) and a galaxy density of $n\sim 10^{-4} (h/{\rm Mpc})^{3}$. 
}
\label{hzdm_hexcorrectionterm}
\end{figure*}

Our results are presented in Figures \ref{hzdm_fitsi}-\ref{hzdm_hexcorrectionterm}. 
Figure \ref{hzdm_fitsi} shows the two-dimensional marginalized constraints 
on the $\dA-H$ plane. The contours shown correspond to 40, 68, 95, and 99\% confidence levels. 
The best fit values of each parameter and their respective $1\sigma$ (68\%) CL regions 
after marginalizing over the other are indicated by the purple crosses. 
We find that the best fit model recovers, to  high accuracy,
the true values of these parameters. The pink box corresponds to the fiducial cosmology.
In the legend we display the calculated deviations from the true values 
in the 2D analysis (purple diamond). We also note the marginalized uncertainty  
for (i.e, the analog of the diagonal elements in Fisher-Matrix analysis). 
For the wedges we make use of first order correction terms 
${\mathcal C}_{\perp,||}(\epsilon)$, 
and in Appendix \ref{wedgescorrestion_section} discuss their importance.
For comparison, the wide orange contours 
display the constraining power of the monopole on its own.
The result is the well known $\dA^2/H$ degeneracy.

The fitting range chosen here is $40<s<150$\hmpc 
in bins of $\Delta s= 5$\hmpcii.
As discussed in  \cite{sanchez08} and \cite{shoji09a},  
we find that the extra information included when analyzing larger ranges yields 
tighter constraints.

We notice that in all cases the true parameters are recovered to  high accuracy, and 
the input incorrect cosmology is ruled out by over $3\sigma$ solely by the \AP effect on clustering.
That said, this level of precision is not expected when using this
technique in  near future galaxy samples, because marginalization over 
a larger parameter space would be required. 
We discuss this more detail in \S\ref{practical_section}. 

We do notice, though, differences in the results obtained by means of the
two projection techniques. We emphasize that these are fair tests, 
as every step along the analysis is equivalent as much as possible.
When analyzing real-space information (left plots), 
we see that $\dM$ is measured to similar accuracy with both techniques,
with the clustering wedges yielding slightly smaller uncertainties. 
Interestingly, the differences in the recovery of $H$ between the two 
methods are larger where the clustering wedges yield a more
accurate result, as well as smaller uncertainties. 
Throughout this study we compare these statistics, 
and find that the clustering wedges defined by $\Delta\mu=1/2$ 
perform better than $\xi_0,\xi_2$, for the most part, 
and motivate adding $\xi_4$ 
to improving constraints when using multipoles.

In the plots on the right we compare the performance of these   
statistics in velocity-space. Most notable is the fact that the uncertainties in $H$ 
increase substantially for both projection pairs compared 
to real-space results. Although the wedges yield tighter 
uncertainties in $H$, the multipoles generate slightly more accurate results.
As for $\dM$ they both yield similar uncertainties,
but clustering wedges yield more accurate results, 
and do not change much from real-space. 

When fitting for the $\Delta\mu=1/2$ clustering wedges, 
we take into account the intermixing terms ${\mathcal C}_{||,\perp}$ 
(Equations \ref{correct_los} and \ref{correct_los2}). 
In Appendix  \ref{wedgescorrestion_section} we explore their effectiveness 
and test results obtained by other wedge widths. We find that setting ${\mathcal C}_{||,\perp}=0$ 
(meaning radial wedge depends solely on $H$ and transverse wedges solely on $\dA$), 
yields results that are less accurate and uncertainties 
that are underestimated. We also find that  
the ${\mathcal C}_{||,\perp}$ terms are less important as one decreases $\Delta\mu$. 

In the top panels of Figure \ref{hzdm_hexcorrectionterm} 
we compare the results shown in Figure \ref{hzdm_fitsi} 
with ones obtained when adding $\xi_4$ to the multipole pair. 
The bottom plots corresponds to the same test but using a
different choice for the fiducial cosmology 
($w=-0.9$; notice the different location of the magenta box). 
The left plots are in real-space and the right in velocity-space. 

In all cases the true cosmology is recovered to  high accuracy. 
As expected, the $[\xi_0,\xi_2,\xi_4]$ combination (dash blue lines)
yields tighter constraints than when limiting the analysis to $[\xi_0,\xi_2]$
(solid black lines) or the $\Delta\mu=0.5$ wedges  (dot-dashed orange lines). 
Crosses indicate marginalized $1\sigma$ results according to color, and symbols most likely 2D value.
The real-space $w=-0.9$ result (bottom left plot) shows that the wedges 
do not generically yield tighter constraints than the monopole-quadrupole pair.  
In velocity-space, we notice that adding the $\xi_4$ information in the multipoles 
yields a smaller correlation coefficient between $H$ and $\dM$. 

Another oddity is the fact that in real-space $\xi_4$ is present at all, 
and can assist in improving constraints. 
We argue that it is probably an artifact 
of angular effects at large scales (see \S \ref{wedgeshifting_section}). 
We verify this by limiting our analysis to the range $s=[20,60]$\hmpcii, 
finding less of an improvement when using $\xi_4$, as its angular effect 
is negligible at these scales. 
These suggest that including $\xi_4$ is useful in breaking the $H-\dA$  
degeneracy in velocity-space. 

To summarize the results of this section, we find that in the ideal case analysed here 
the true parameters are recovered to  high accuracy with both
statistics. The input incorrect cosmology is ruled out by over $3\sigma$ 
solely by the \AP effect on clustering. We demonstrate here, for the first time, 
the power of the clustering wedges technique.
We find that using Equations (\ref{los_distorted}) and
(\ref{trv_distorted})  to describe the \AP distortions in
the clustering wedges it is possible to recover the true 
parameters as well as one can with the multipole expansion. 
This means that even a wide ``radial" wedge is most sensitive to $H$ 
while a ``transverse"  wedges is most sensitive to $\dA$. 
We also demonstrate that adding $\xi_4$ 
to the multipoles substantially improves constraints.

\subsection{Amplitude effects on uncertainties} \label{practical_section}
  
In the previous sections we simulate ``ideal" observational tests, in which 
the correlation function template from which the models are constructed, 
is fully understood except for the \AP effect. Here we address 
the fact that the observer does not have the luxury of {\it knowing} the
correct signal a priori. 


\begin{figure}
\includegraphics[width=0.47\textwidth]{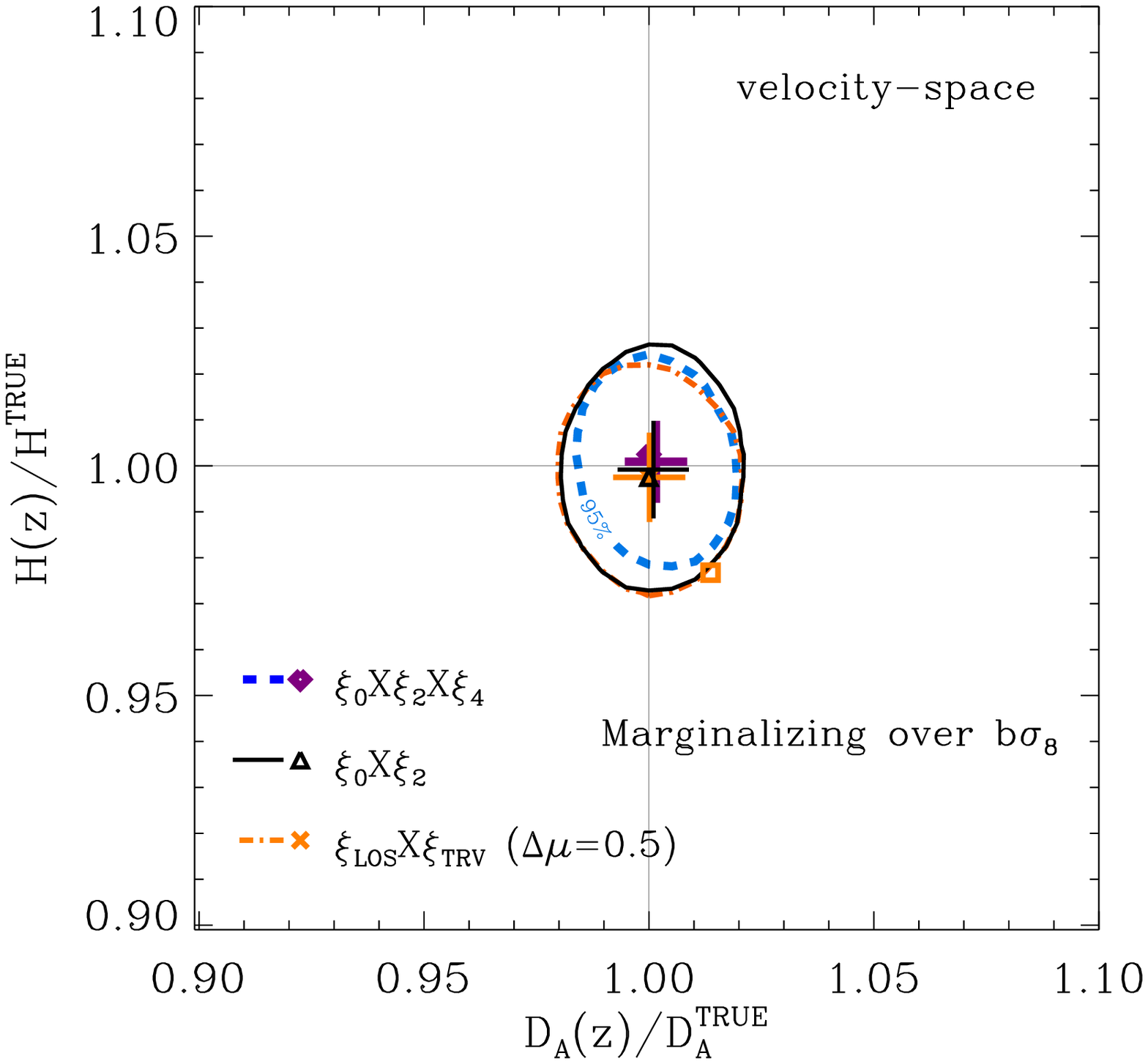}
\includegraphics[width=0.47\textwidth]{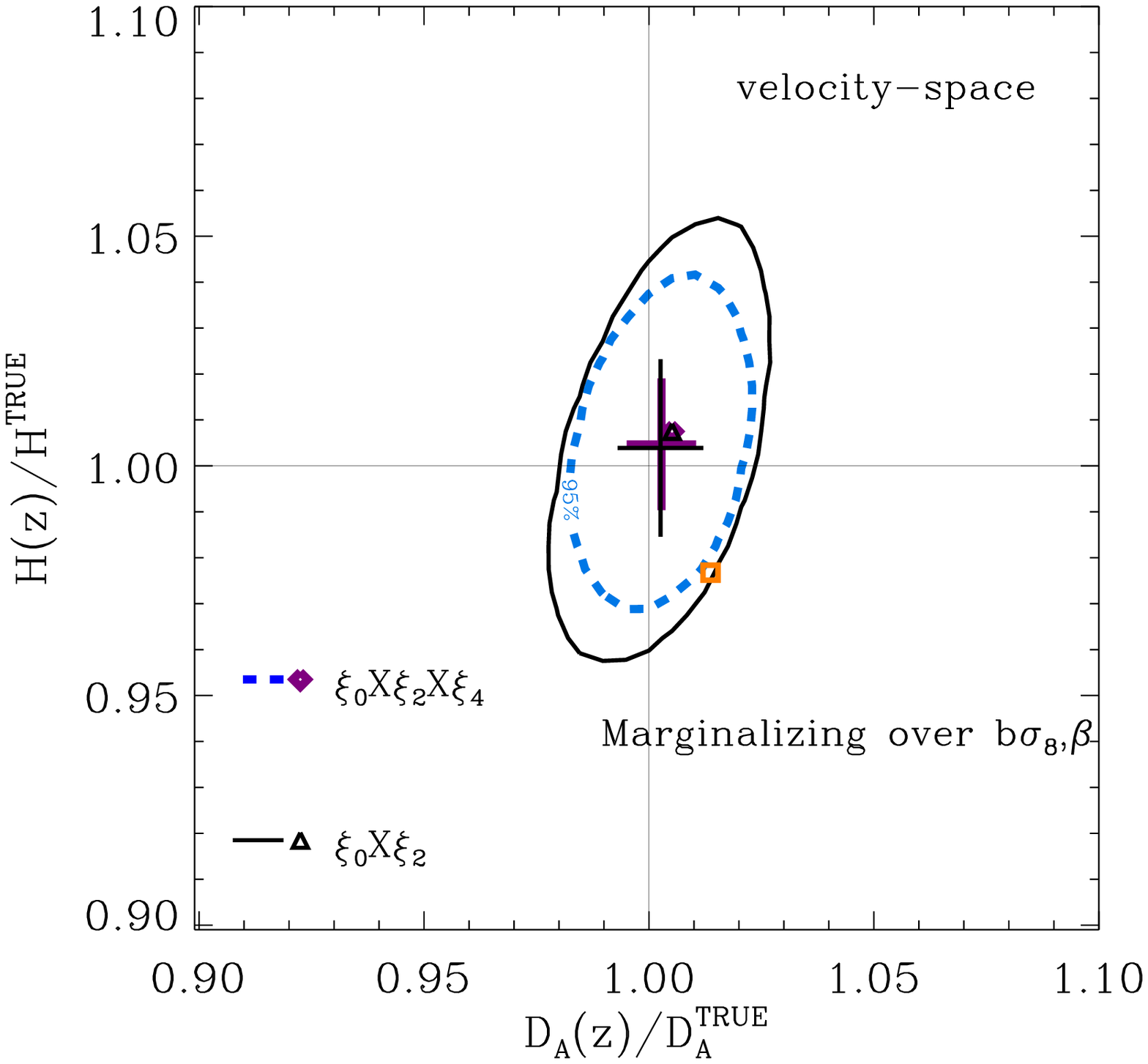}
\caption[Reproducing the true $H,\dA$: amplitude effects]{
Joint $2\sigma$ constraints on $H-\dA$ from \AP shift tests 
when analyzing mock mean velocity-space clustering 1D statistics.
The parameter space in the upper panel is $\Phi=[\dA,H,b\sigma_8]$.
and  $\Phi=[\dA,H,b\sigma_8,\beta]$ in the bottom. 
The analysis of the results shown here is similar 
to that performed when producing the results shown in Figure 
\ref{hzdm_hexcorrectionterm} only marginalizing over amplitude parameters.
$H-\dM$  joint constraints for the multipole combination [$\xi_0,\xi_2,\xi_4$] 
are shown in dashed blue lines, [$\xi_0,\xi_2$] are in solid black lines, 
and $\Delta\mu=1/2$ wedges are dot-dashed orange lines. 
Crosses indicate marginalized $1\sigma$ results according to color, 
and symbols most likely 2D value.
The \AP distortion applied here is using $w^{\rm FID}=-1.1$ (orange box) 
instead of  the true value $-1$ when converting $z$ to comoving distances.
These constraints roughly correspond to a fictitious survey with a Hubble volume 
($R\sim c/H_0$) and a galaxy density of $n\sim 10^{-4} (h/{\rm Mpc})^{3}$. 
}
\label{hzdm_fits_amplitude}
\end{figure}


To mention a few main concerns, one must understand the $\xi$
amplitude, shape, and various effects on the angular \bafii. 
The amplitude may be scale dependent for various reasons: 
non-linear tracer-matter bias, scale dependent dynamic distortion effects,  
observer angular effects, magnification bias (\citealt{hui07a}) 
and effects of non-gaussianities 
in initial conditions (\citealt{dalal08a}). 

Here we take into account the effect of uncertainties 
in the amplitudes of the various statistics and leave shape effects for a future study.
For this, we add amplitude parameters to the $H-\dM$ parameter space. 
We assume that the amplitudes of the multipoles and clustering wedges are given by 
 $A_{{\rm stat}}=B_{{\rm stat}} \, b^2\,\sigma_8^2$, where 
$b$ is the (scale-independent) bias parameter of the tracer with respect to the matter,  
$\sigma_8$ is the rms linear perturbation theory variance in spheres of radius $8$\hmpcii, 
and $B_{\rm stat}$ is the Kaiser squashing amplitude on large scales, which in real-space is unity. 
The subscript ``stat" signifies the fact that we test for various statistics such as the
wedges and multipoles.
  
For the multipoles in velocity-space $B_{\rm stat}$ 
is given by the usual \cite{kaiser87} prefixes  
\begin{eqnarray}
B_0&=&1+2/3\beta+1/5\beta^2, \\ 
B_2&=&4/3\beta + 4/7 \beta^2,\\
B_4&=& 8/35\beta^4,
\end{eqnarray}
of the squashing parameter $\beta(z)\equiv f/b \sim \Omega_{\rm M}(z)^{0.55}/b$, where $f\equiv d\ln(D_1)/d\ln(a)$, 
with $D_1$ the standard linear growth factor and and $a$ the scale factor.
Technically, as the templates used here correspond to the $\TRUE$ signal, 
when testing for the amplitude the models are based on 
the ratio of $A_{\rm stat}$ to its real value. 

When testing for $b\sigma_8$ we perform the analysis on wedges and multipoles. 
When testing for $\beta$, however, this is a non-trivial task for the wedges. 
This is due to the volume averaged monopole term which 
makes it non-trivial to account for $A_{\rm stat}$
in configuration space (see the terms involving $\overline{\xi}$ in equations 6-8 of
 \citealt{hamilton92}). In a $k-$space analysis this issue would be trivial, as it should be
when building a generic model, which we do not do here. For this reason, results shown 
marginalizing over $\beta$ are limited to multipoles. The fiducial $\beta$ tested is calculated through the
input $f(z)$ and the linear $b\sim 1.96$ which is inferred from matching linear theory to
the standard projected correlation function $w_p(s_\perp)$ of the mocks.

In Figure \ref{hzdm_fits_amplitude}  we show joint constraints 
on $H$ and $\dA$
using  the various statistics
when marginalizing over $b\sigma_8$ (top) and [$b\sigma_8,\beta$] (bottom). 
For the former case we assume perfect knowledge of the dynamic distortions, 
whereas for the latter we add the uncertainty of the Kaiser effect 
(disregarding velocity dispersion). 
The obtained constraints can be compared with those of Figure  \ref{hzdm_hexcorrectionterm}, 
and show a clear degradation of the uncertainties on $H$ nearly by a factor of two
and a slight degradation of the uncertainties on $\dM$
with respect to the results obtained without uncertainties in the amplitudes.  
In \S \ref{multipolesorwedges_section} we study the effect of varying the range of 
scales included in the analysis.
These plots show fair comparisons between [$\xi_0,\xi_2$], [$\xi_0,\xi_2,\xi_4$] and $\Delta\mu=0.5$
clustering wedges. 
  
We see that the true cosmology is recovered to high accuracy, 
but our choice of incorrect cosmology $w=-1.1$ is ruled out 
by [$\xi_0,\xi_2$] at only $2\sigma$ (top) and $\sim1.5\sigma$ (bottom), 
due to the marginalization over the amplitude parameters. 
The importance of the first finding is that multipole \AP Equations (\ref{mono_equation}) and 
(\ref{quad_equation}), hold to a very good degree. 
We obtain similar conclusions for Equations  (\ref{los_distorted}),
(\ref{trv_distorted}) and (\ref{correct_los}) regarding the clustering wedges. 
As in the test performed in \S \ref{reproducing_hda_section}, we 
find an increase in the uncertainties of $H$ at the $10\%$ level 
when going from real- to velocity-space. 

To conclude these tests we find that increasing the allowed range of 
the amplitude parameters degrades the constraints, but retains accuracy in recovering
the true values of $H$ and $\dM$. When comparing between the different 
statistics our conclusions of this test are similar to those obtained in 
\S \ref{reproducing_hda_section}. We demonstrate that $\xi_4$ improves 
constraints substantially compared to those obtained when limited to monopole-quadrupole. 

\subsection{Wedges or multipoles?}
\label{multipolesorwedges_section}

In \S\ref{reproducing_hda_section} and \S\ref{practical_section} 
we analyse a particular case where the constraining power on $H$ of the $\Delta\mu=0.5$ 
clustering wedges outperforms the monopole-quadrupole 
pair, and see substantial improvement when including the hexadecapole 
(see Figure \ref{hzdm_hexcorrectionterm}). 
However, in these analyses we limit the range of scales to $40<s<150$\hmpcii. 
Here we generalize these tests by varying the minimum scale included in the analysis  
$s_{\rm min}$ while keeping the maximum scale fixed to $s_{{\rm max}}=150$\hmpcii,
and find interesting trends for the constraining power of the various statistics. 
All results given here are in velocity-space. 

\begin{figure*}
\includegraphics[width=0.9\textwidth]{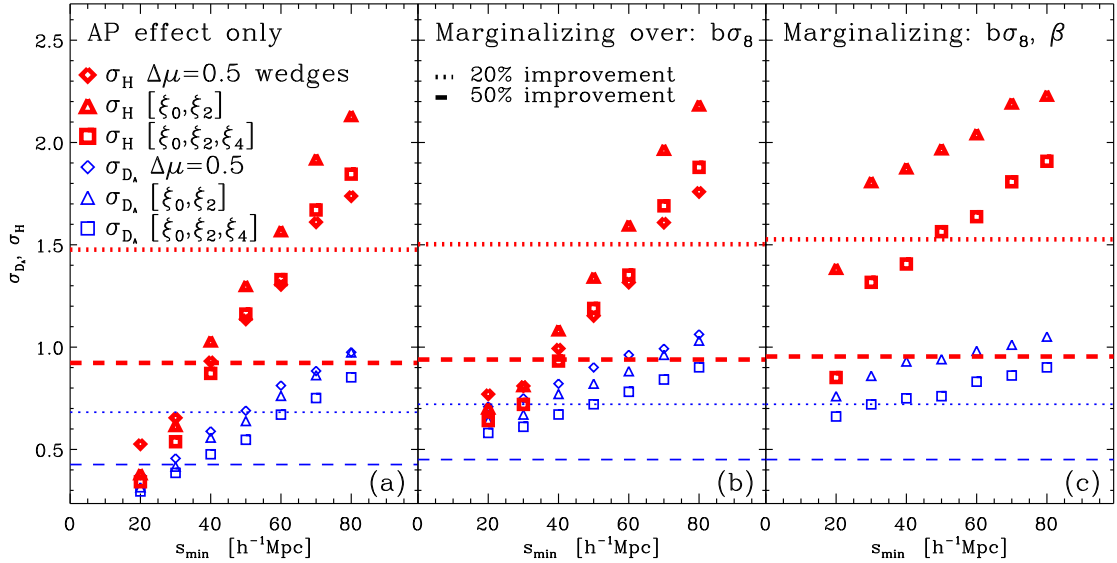}
\caption[$\sigma_H/H,\sigma_{\dA}/\dA$ as a function analysis range]{
Relative uncertainties $\sigma_H/H$ and $\sigma_{D_{\rm A}/\dA}$ 
as a function of the range analysed [$s_{\rm min}$, $150$\hmpcii] 
(given in $[\%]$).  
The different symbols, indicated in the legend, represent 
the various 1D statistics used. 
Thick red symbols are for $\sigma_H$, 
and thin blue for $\sigma_{\dA}$.
(a):  ``\AP only"- $[H,\dA]$ with fixed amplitude. 
(b): $[H,\dA]$ when marginalizing over $b\sigma_8$ without prior.
(c):  $[H,\dA]$ when marginalizing over $b\sigma_8,\beta$ without prior.
Slopes indicate the improved constraints obtained by 
using information from the broad band of $\xi$. 
The dotted lines indicate an improvement of $0.8$ 
from $\sigma_\Phi/\Phi(s_{\rm max}=80 \ h^{-1}{\rm Mpc})$ 
(when fitting for [$\xi_0,\xi_2,\xi_4$]; $\Phi=[H,\dA]$), 
and the dashed lines an improvement of $0.5$.
The \AP distortion applied here is using $w^{\rm FID}=-1.1$ 
instead of  the true value $-1$ 
when converting $z$ to comoving distances.
}
\label{uncertainties_plot}
\end{figure*}


Our findings are summarized in Figure \ref{uncertainties_plot}, 
in which we show the uncertainties on $H$ and $\dM$ as a function of $s_{\rm min}$. 

Results for an ``\AP effect only" are displayed in Figure \ref{uncertainties_plot}a, which
generalize the top right plot in Figure \ref{hzdm_hexcorrectionterm}. ``\AP only"  refers to tests
in which we fix amplitude parameters to their true values and test only for $H$ and $\dA$. 
When marginalizing over $b\sigma_8$ we obtain results displayed in 
Figure \ref{uncertainties_plot}b, which generalizes the top plot of Figure \ref{hzdm_fits_amplitude}. 
Figure \ref{uncertainties_plot}c shows the results obtained when marginalizing over both 
$b\sigma_8$ and $\beta$, and is a generalization of the bottom plot of Figure \ref{hzdm_fits_amplitude}. 
No priors are assumed for $b\sigma_8$ or $\beta$. 

In all cases we find that adding information from $\xi_4$ to the multipole analysis improves the obtained 
constraints on $H$ and $D_{\rm A}$ by a substantial amount. 

The improvement in the constraints as smaller scales are included in the analysis 
emphasizes the fact that although the \baf is essential to perform the \AP test, 
one can extract more information by analyzing the full broad shape of $\xi$ 
\citep{sanchez08, shoji09a}. 

In all parameter spaces tested, the slope of $\sigma_H(s_{\rm min})$ (thick red)
is steeper than that of $\sigma_{\dA}$ (thin blue), indicating that the broad shape 
of these statistics is more sensitive to $H$. The fact that the  ``\AP only"  test 
(Figure  \ref{uncertainties_plot}a) has steeper slopes than the others is expected. 
In this case each data point posses a high constraining power. 
\PWpaper explain that marginalizing over the amplitude is somewhat degenerate with the 
\AP effect and hence leads to a reduction of the slope when 
adding $b\sigma_8$   (Figure \ref{uncertainties_plot}b) 
and even more when adding $\beta$  (Figure \ref{uncertainties_plot}c).
Nevertheless, even when marginalizing over $b\sigma_8$ we notice 
an improvement of factor two in the uncertainty on $H$ and a factor of $\sim 1.25$ in
$\sigma_{\dA}$ when using the full information up to $40$\hmpcii, 
with respect to setting $s_{\rm min}=80$\hmpc (i.e, focusing on the scales
around the \bafii).  

The answer to the question of which statistics should be preferred is not simple 
as it appears to depend on the range of scales used. 
For high values of $s_{\rm min}$ the results obtained by means of the clustering wedges
(diamonds) outperform those obtained with the $[\xi_0,\xi_2]$ combination (triangles). 
However, we notice the different slopes of $\sigma_{\dM}(s_{\rm min})$ and $\sigma_{H}(s_{\rm min})$  
such that the multipole pair should be preferred when a large range of scales is considered
(i.e. low $s_{\rm min}$). 

A puzzling result is the fact that $\sigma_{\dM}(s_{\rm min})$ and $\sigma_{H}(s_{\rm min})$
have a steeper slope for the case of the $[\xi_0,\xi_2,\xi_4]$ combination 
than for the clustering wedges. The $\sigma_{H}(s_{\rm min})$ results indicate 
that the $\Delta\mu=0.5$ wedges are preferred at 
$s_{\rm min}>50$\hmpc (although $[\xi_0,\xi_2,\xi_4]$ is preferred to 
determine $\dA$). 
This might be explained by angular effects causing higher multipoles, 
which we have not corrected for properly with the multipoles, 
but happen not to affect the wedges, which 
are combinations of all multipoles. 
 
 For this analysis we have chosen $w^{\rm fid}=-1.1$ of our fiducial cosmology 
(instead of $w^{\rm true}=-1$). When testing for  $w^{\rm fid}=-0.9$ 
we find similar trends but with varying  crossover points. For example, in the ``\AP only"
case, the clustering wedges are preferred over the  [$\xi_0,\xi_2,\xi_4$] combination 
in determining $H$  at $s_{\rm min}>55$\hmpc  instead of $>45$\hmpcii.

\section{Predicted Constraints on $H$ and $\dA$ from BOSS}
\label{boss_section}

Here we apply both the multipole and  wedge 
techniques to the 32 realizations of the Horizon Run mock galaxy catalogues 
described in \S \ref{mocks_section}, which serve as simulated realizations 
of the final BOSS volume. We investigate the accuracy with which $H$ and $\dA$ can be obtained 
for a survey covering $10,000$ square degrees in the redshift
range $0.16<z<0.6$ with density  $n\sim 3 \cdot 10^{-4} h^3$Mpc$^{-3}$. 
For full details of the mocks please refer to \S \ref{mocks_section}. 
Analyzing $32$ realizations, we obtain results for the full sample, 
and of for two subsamples split at $z=0.45$. This redshift split is motivated by 
the so-called targeted LOWZ and CMASS BOSS galaxies  
(see \citealt{eisenstein11a} and Padmanabhan et al. (in prep.)).

\begin{figure*}
\includegraphics[width=0.45\textwidth]{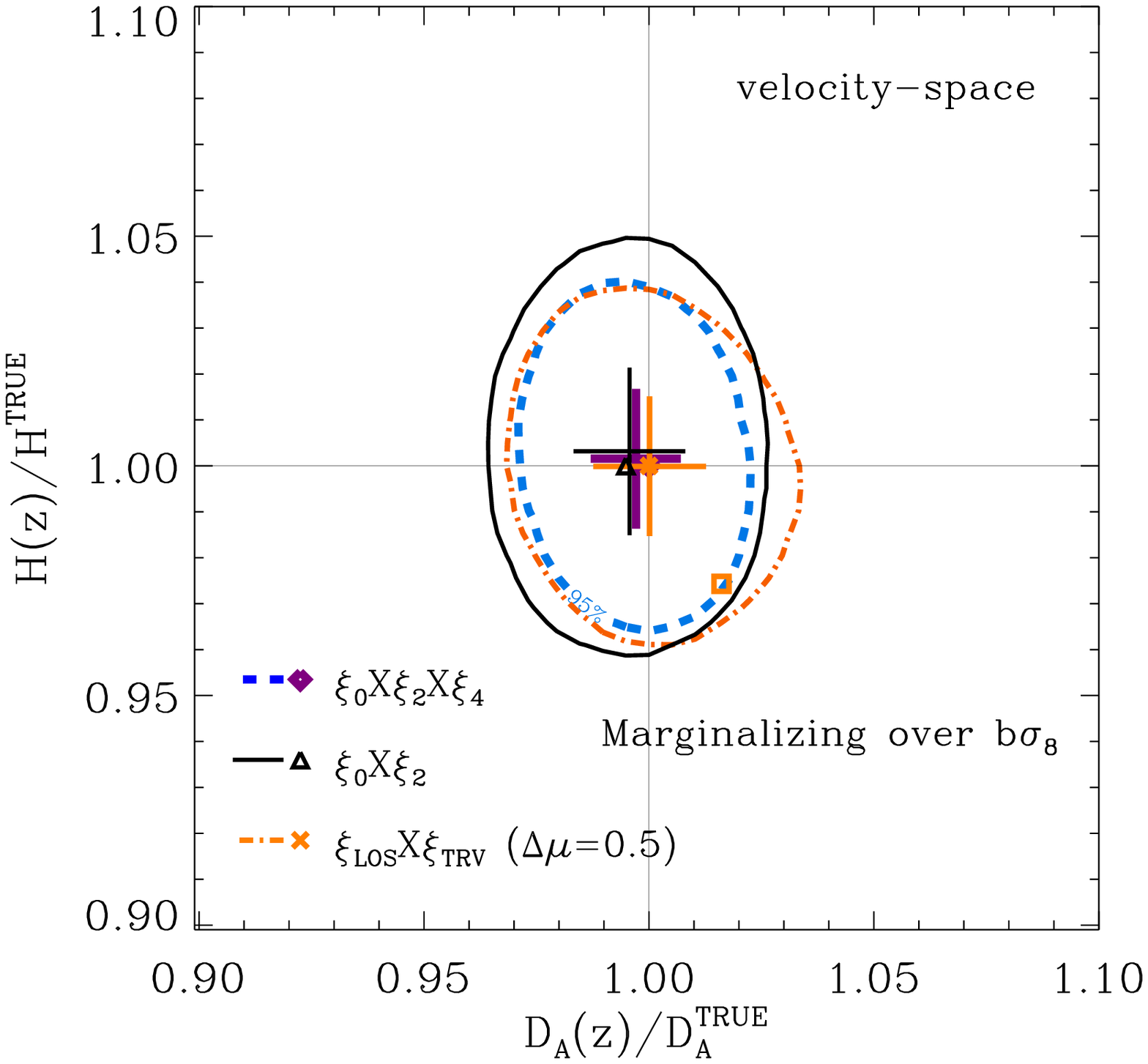}
\includegraphics[width=0.45\textwidth]{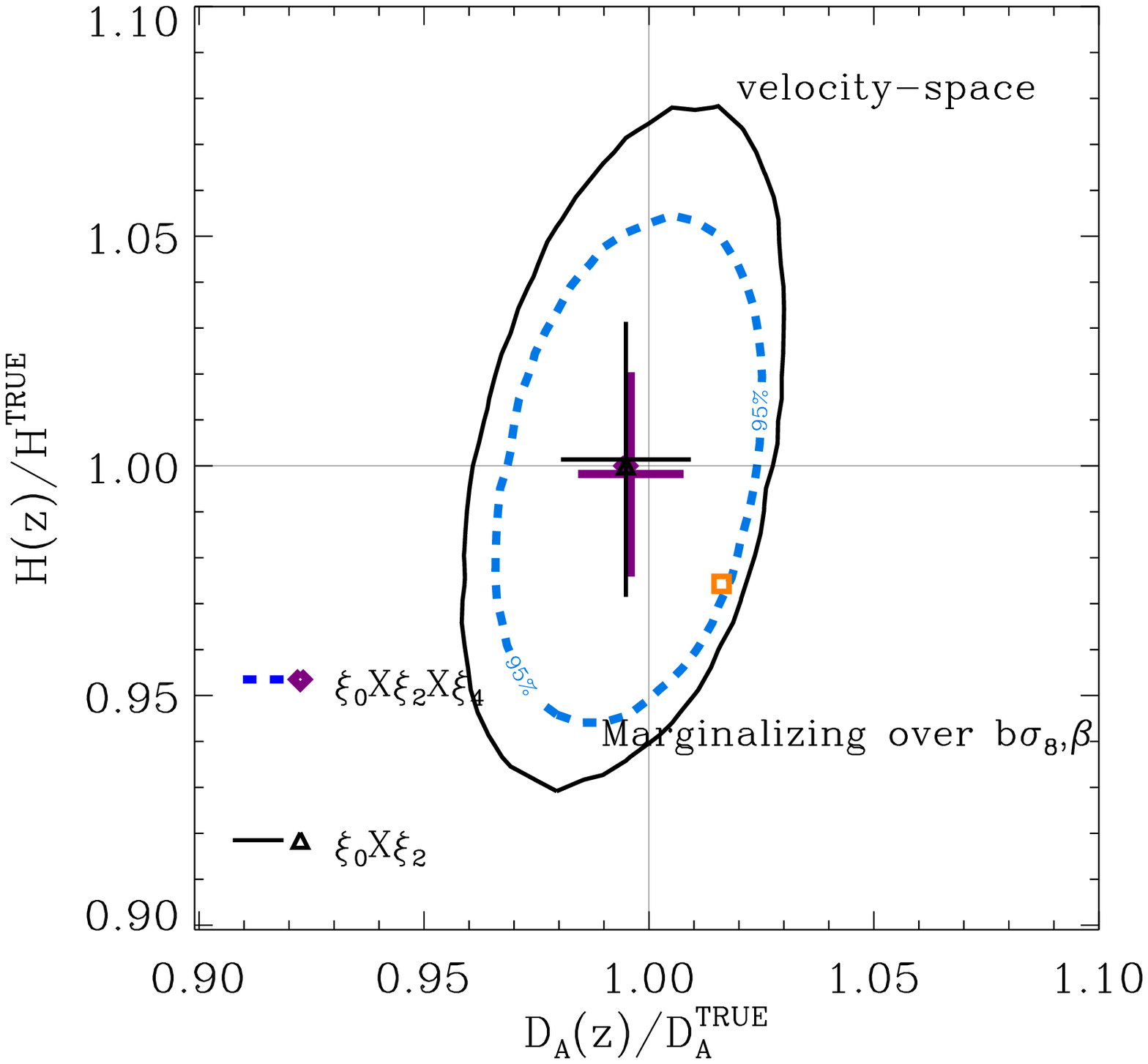}\\
\caption[BOSS predictions: $H-\dA$ joint constraints, a technique comparison]{
BOSS mock galaxy mean  ($0.16<z<0.6$)
$H-\dA$ joint $2\sigma$ constraints from \AP shift tests,  
when analyzing velocity-space clustering 1D statistics.
The parameter space in the left panel is [$\dA$,$H$,$b\sigma_8$].
and  [$\dA$,$H$,$b\sigma_8$,$\beta$] in the right.  
$H-\dM$  joint constraints for the multipole combination [$\xi_0,\xi_2,\xi_4$] 
are shown in dashed blue lines, [$\xi_0,\xi_2$] are in solid black lines, 
and $\Delta\mu=1/2$ wedges are dot-dashed orange lines. 
Crosses indicate marginalized $1\sigma$ results according to color, 
and symbols most likely 2D value.
The \AP distortion applied here is using $w^{\rm FID}=-1.1$ (orange box) 
instead of  the true value $-1$, when converting $z$ to comoving distances.
These constraints correspond to an 
analysis of the broadband $s=[40,150]$\hmpcii, 
without priors, and assumes all 
shape effects are fully understood. 
In both [$\xi_0,\xi_2,\xi_4$] cases the true cosmology is recovered with 
excellent accuracy, and the incorrect input $H,\dA$  
is rejected by $\sim 2\sigma$.
We clearly see a degradation in constraining $H$ 
when including marginalization of $\beta$. 
}
\label{hzdm_boss_plot} 
\end{figure*}
\begin{figure*}
\includegraphics[width=0.45\textwidth]{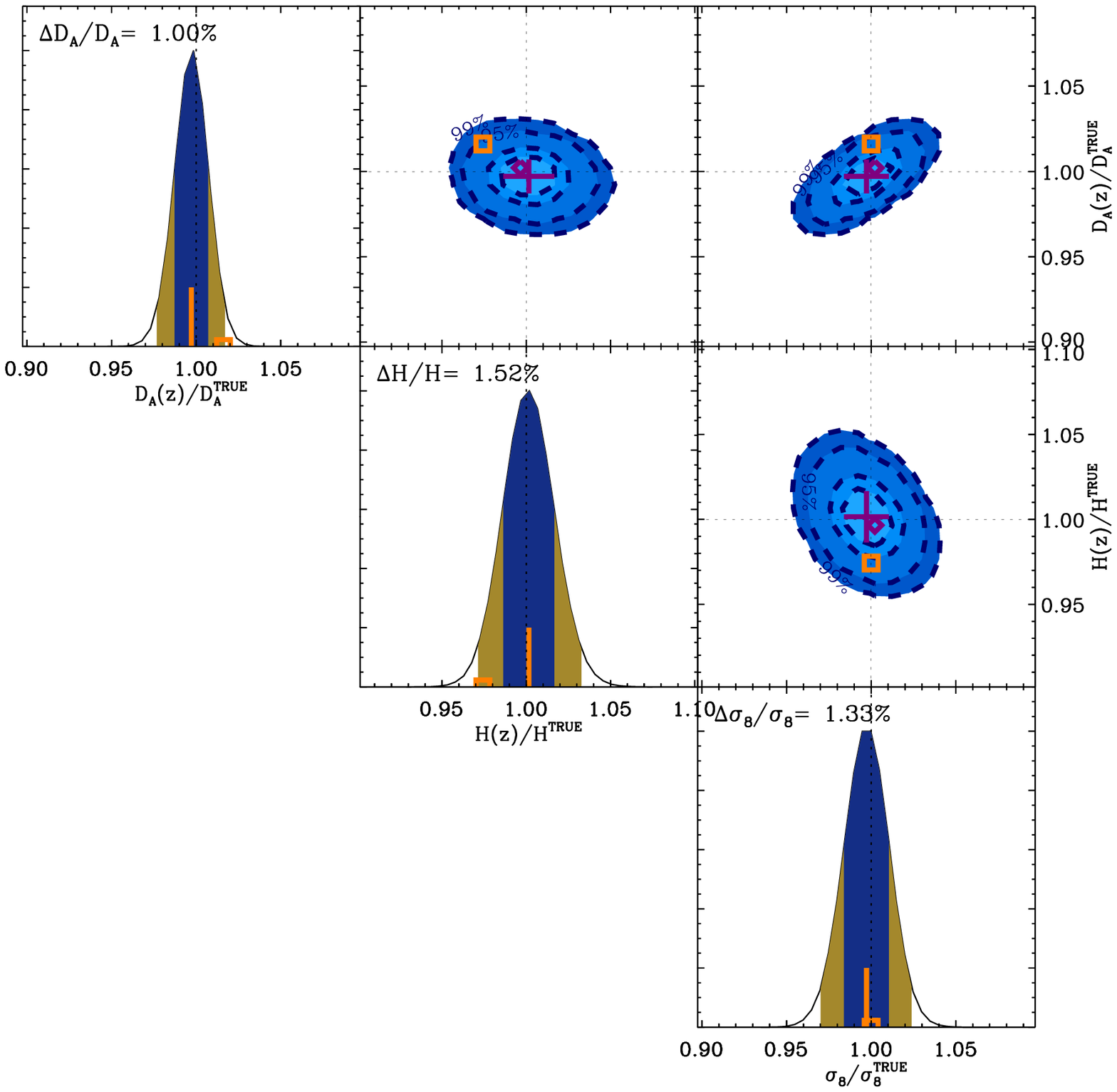}
\includegraphics[width=0.45\textwidth]{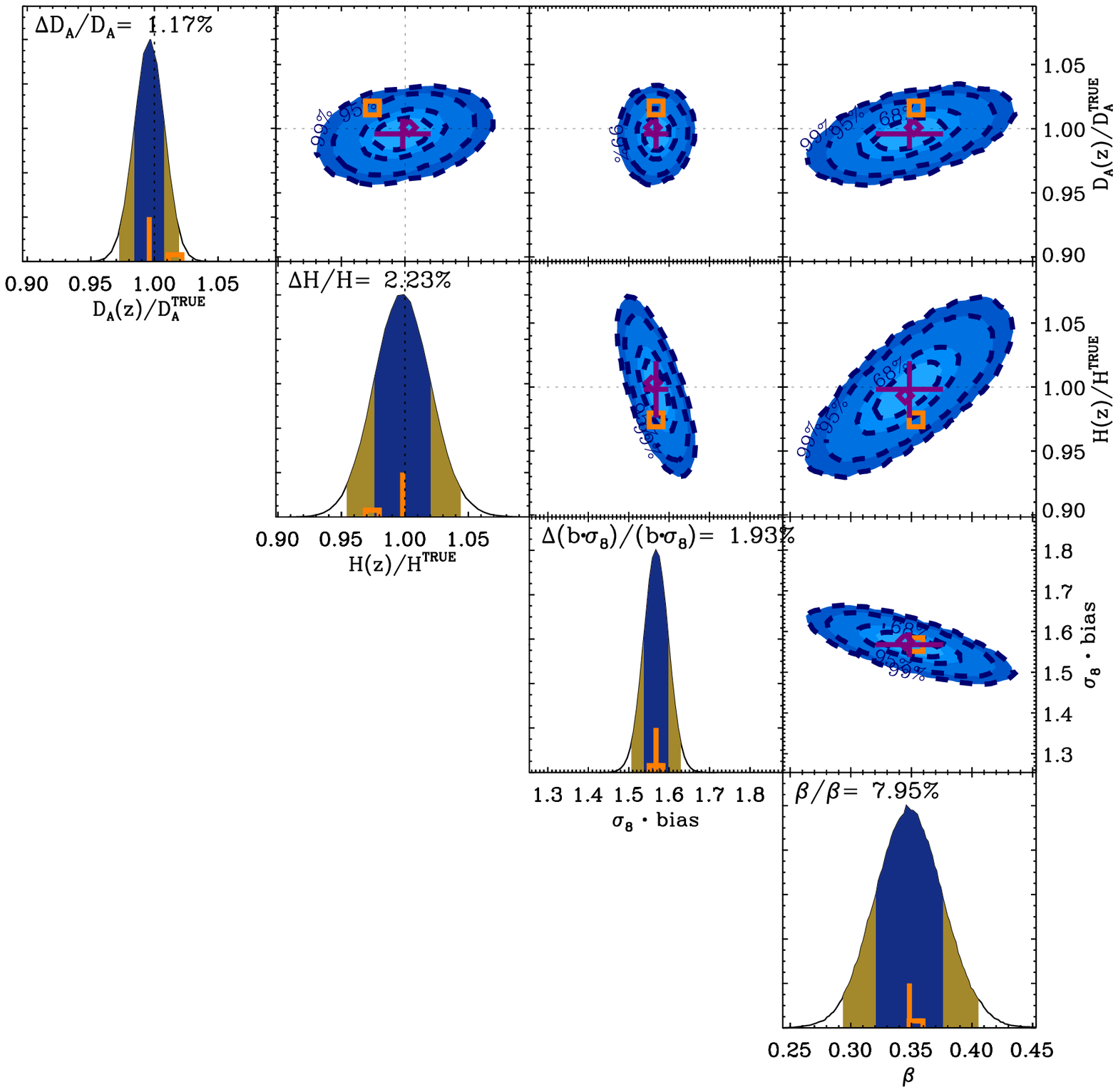}\\
\caption[BOSS predictions: $H-\dA$ joint constraints with $\beta,b\sigma8$]{
BOSS mock galaxy mean  ($0.16<z<0.6$) 
joint and marginalized constraints from  
\AP shift tests,  
when analyzing velocity-space clustering [$\xi_0,\xi_2,\xi_4$].
(These results are the same as those displayed in Figure 
\ref{hzdm_boss_plot}, meaning analyzing region $s=[40,150]$\hmpcii). 
The parameter space in the left panel is [$\dA$,$H$,$b\sigma_8$].
and  [$\dA$,$H$,$b\sigma_8$,$\beta$] in the right.  
The joint constraints are for CL: [$40,68,95,99$]$\%$. 
On the top of each 1D likelihood plot we inscribe for each parameter $\Phi$ 
the marginalized $1\sigma$ result, also summarized in Table \ref{tab:wa}. 
The orange boxes demonstrate the fiducial parameters 
input into the analysis (in the case of $H$ and $\dA$ this is 
the \AP effect using $w^{\rm FID}=-1.1$ instead of the true $-1$ value 
when converting redshifts to comoving distances). 
In all cases the true cosmology is recovered with 
excellent accuracy, and the incorrect input $H,\dA$  
is rejected by $\sim 2\sigma$.
We clearly see a $\sim 30\%$ degradation in constraining $H$ 
when including marginalization of $\beta$, 
and a $\sim 15\%$ degradation in $\dA$.
}
\label{hzdm_boss_plot_2} 
\end{figure*}
 
As in previous sections,  when building our templates, we gradually increase our parameter space, 
starting from solely the \AP effect and adding amplitude parameters. 
For each case we analyse both real- and velocity-space and obtain the corresponding 
results as expected from one BOSS volume.

On a technical note, 
the mock ``data" and templates  are based on the mock mean of 
$32$ Horizon Run mocks. To obtain a stable, invertible  $C_{ij}$ we use the 
$160$ LasDamas mocks,  normalized by the variance of the $32$ Horizon Run mocks. 

Our $0.16<z<0.6$ results are shown in Figures \ref{hzdm_boss_plot} and \ref{hzdm_boss_plot_2} 
for a particular range $40<s<150$\hmpcii. 
The left plots in each figure are results for the parameter space [$H,\dA,b\sigma_8$] 
and the right plots for  [$H,\dA,b\sigma_8,\beta$]. 
Figure \ref{hzdm_boss_plot} compares the statistic combinations
in the $H-\dA$ plane. 
In Figure \ref{hzdm_boss_plot_2} we display one and two-dimensional likelihood functions 
for the above three and four parameter spaces when using  the [$\xi_0,\xi_2,\xi_4$] combination. 
Marginalized $1\sigma$ uncertainty values are indicated on the 1D likelihood panels.
In Table \ref{tab:wa} we summarize the $1\sigma$ uncertainty values obtained for the various
statistic combinations, parameter spaces, choices of $s_{\rm min}$, and sample analysed.

In all cases investigated we find that the true cosmology ($H, \dA$) is recovered to 
 high accuracy (much better than $1\%$). 
 We report that the velocity-space results 
yield a noticeable increase in uncertainty of $H$ 
compared to real-space results. 
Below we will limit our explanations 
to velocity-space results.

When allowing the amplitude parameters to vary, 
we find that 
uncertainties in the $H-\dA$ plane are degraded, as expected. 
When adding $\beta$ in velocity-space, 
we notice an increase in the uncertainty 
of $H$ by a fraction range of  $1.1$ to $1.6$, depending 
on the scales included in the analysis. 
For $\dA$ the fraction changes by $1.01$ to $1.10$.

\begin{table*} 
\begin{minipage}{172mm}
\centering
\begin{center}
\caption[BOSS $H,\dA$ predictions]{
BOSS predictions based on velocity-space Horizon Run mock galaxies$^{a}$
}
\begin{tabular}{@{}cccccc@{}}
\hline
\hline
Parameter Space & 1D Projection & Volume$^{b}$ & analysis range \hmpcii & $\sigma_{H}/H$ $\%$ $^c$ & $\sigma_{\dA}/\dA$ $\%$ $^c$ \\

 \AP only$^d$ &$[\xi_0,\xi_2]$                        &  $0.16<z<0.6$      & $[40,150]$&$1.71$&$0.90$ \\  
  \AP only &$[\xi_0,\xi_2,\xi_4]^{e}$             & $0.16<z<0.6$       & $[40,150]$&$1.41$&$0.72$ \\  
   \AP only &$\Delta\mu=1/2$ wedges  & $0.16<z<0.6$      & $[40,150]$&{\bf 1.33}&$0.99$  \\ \\
  
    \APii, $b\sigma_8$ &$[\xi_0,\xi_2]$                     &  $0.16<z<0.6$      & $[40,150]$&$1.82$&$1.24$ \\  
  \APii, $b\sigma_8$ &$[\xi_0,\xi_2,\xi_4]$             & $0.16<z<0.6$       & $[40,150]$&$1.52$&$1.00$ \\  
   \APii, $b\sigma_8$& $\Delta\mu=1/2$ wedges  & $0.16<z<0.6$      & $[40,150]$&{\bf 1.52 }&$1.25$ \\ 
       \APii, $b\sigma_8$ &$[\xi_0,\xi_2]$                     &  $0.16<z<0.6$      & $[60,150]$&$3.13$&$1.52$ \\  
  \APii, $b\sigma_8$ &$[\xi_0,\xi_2,\xi_4]$             & $0.16<z<0.6$       & $[60,150]$&$2.28$&$1.32$ \\  
   \APii, $b\sigma_8$ &$\Delta\mu=1/2$ wedges  & $0.16<z<0.6$      & $[60,150]$&{\bf 2.44}&$1.48$ \\ \\

       \APii, $b\sigma_8$, $\beta$ &$[\xi_0,\xi_2]$                  &  $0.16<z<0.6$      & $[40,150]$&$2.99$&$1.44$ \\  
  \APii, $b\sigma_8$, $\beta$ &$[\xi_0,\xi_2,\xi_4]$             & $0.16<z<0.6$       & $[40,150]$&$2.24$&$1.17$ \\  
         \APii, $b\sigma_8$, $\beta$ &$[\xi_0,\xi_2]$                &  $0.16<z<0.6$      & $[60,150]$&$3.45$&$1.57$ \\  
  \APii, $b\sigma_8$, $\beta$ &$[\xi_0,\xi_2,\xi_4]$             & $0.16<z<0.6$       & $[60,150]$& $2.53$&$1.35$ \\  \\

       \APii, $b\sigma_8$ &$[\xi_0,\xi_2]$                 &  $0.16<z<0.45$      & $[40,150]$&$2.42$&$1.85$ \\  
         \APii, $b\sigma_8$& $[\xi_0,\xi_2]$              &  $0.45<z<0.60$      & $[40,150]$&$2.43$&$1.67$ \\  
  \APii, $b\sigma_8$ &$[\xi_0,\xi_2,\xi_4]$             & $0.16<z<0.45$       & $[40,150]$&2.13&1.46 \\  
  \APii, $b\sigma_8$ &$[\xi_0,\xi_2,\xi_4]$             & $0.45<z<0.6$       & $[40,150]$&1.85&1.41 \\  
    \APii, $b\sigma_8$& $\Delta\mu=1/2$ wedges  & $0.16<z<0.45$      & $[40,150]$&{\bf 2.22}&$1.81$ \\ 
    \APii, $b\sigma_8$& $\Delta\mu=1/2$ wedges  & $0.45<z<0.6$      & $[40,150]$&{\bf 2.19}&$1.60$ \\ \\
  
         \APii, $b\sigma_8$, $\beta$ &$[\xi_0,\xi_2]$                 &  $0.16<z<0.45$      & $[40,150]$&$4.05$&$2.95$ \\  
         \APii, $b\sigma_8$, $\beta$& $[\xi_0,\xi_2]$              &  $0.45<z<0.60$      & $[40,150]$&$3.84$&$2.00$ \\  
  \APii, $b\sigma_8$, $\beta$& $[\xi_0,\xi_2,\xi_4]$             & $0.16<z<0.45$       & $[40,150]$&3.23&1.61 \\  
  \APii, $b\sigma_8$, $\beta$ &$[\xi_0,\xi_2,\xi_4]$             & $0.45<z<0.6$       & $[40,150]$&2.79&1.70 \\  \\
\hline
\end{tabular}
\label{tab:wa}
\end{center}
$^a$ We use the $C_{ij}$ based on 160 LasDamas (SDSS-II LRG) mocks normalized by the $C_{ii}$ of 32 Horizon Run (BOSS) mocks.\\
$^b$ All BOSS mocks are volume limited and cover 1/4 of the sky with $n\sim 3\cdot 10^{-4}h^{3}$Mpc$^{-3}$. For $0.16<z<0.6$: $V=4$\hgpcii, $0.16<z<0.45$: 1.8 \hgpcii, $0.45<z<0.6$: 2.2 \hgpcii. \\
$^c$ $1\sigma$ meaning $68.4\%$ CL when all other parameters marginalized over. \\
$^d$ ``\AP only" means amplitude is fixed and we test for $H$, $\dA$. \\
$^e$ Analyses of $\xi_4$ assume it can be measured and modeled for. In BOSS, we do, however, expect low S/N at \baf scales.
\end{minipage}
\end{table*}

As in the previous sections, we find in all cases that adding the $\xi_4$ information imporves the 
constraints obtained by means of the multipoles. We see an improvement of 
a factor $1.2$ to $1.35$ in $H$ and between $1.15$ to $1.25$ in $\dA$. This is comparable to results
found in \cite{taruya11a}, who perform a Fisher Matrix analysis. 

As for the clustering wedges, we find that they outperform the monopole-quadrupole 
pair in $H$ while giving similar constraints on $\dA$. Interestingly, in this case the $\Delta\mu=0.5$ 
clustering wedges measure $H$ with a similar accuracy to that of $[\xi_0,\xi_2,\xi_4]$. 

As to values expected from BOSS, we find that, assuming that $\beta$ 
is fixed and the broad shape understood down to $40$\hmpcii, 
our best $0.16<z<0.6$ constraints obtained are: $\Delta H/H\sim 1.52\%$ (similar for wedges
and [$\xi_0,\xi_2,\xi_4$]) and $\Delta\dA/\dA\sim1\%$ ([$\xi_0,\xi_2,\xi_4$]).

When splitting to the two subsamples, using [$\xi_0,\xi_2,\xi_4$] 
at $40<s<150$\hmpcii, NEAR yields
[$\Delta H/H,\Delta\dA/\dA$]$\sim[2.13,1.46]\%$ and FAR [$1.85,1.41\%$].

\cite{schlegel09a} use Fisher-Matrix analysis to obtain much more optimistic 
estimates, than those shown here, even though they focus on the baryonic acoustic 
wiggles in $P(k)$, and not the broad shape and do not use the $\xi_4$.
This is probably due to the fact that they assume the ``reconstruction" of the 
feature which, if applicable without introducing bias, should improve the obtained
constraints (\citealt{eisenstein07}). We have not applied the technique on the mocks.

\section{Discussion}
\label{discussion_section}

The purpose of this study is to investigate possible ways to break   the $H-\dA$ degeneracy 
by including information in the anisotropy in the $\xi(\mu,s)$ plane produced by geometrical 
redshift distortions.

\subsection{Relating our analysis to previous studies}

This concept has been studied in the full 2D $P(\mu,k)$ plane by \cite{hu03a}, \cite{wagner08a},
and \cite{shoji09a}, who investigated the power of using the \baf to determine the equation of state 
of dark energy. In practical terms, however, there are a few difficulties in applying this approach
on real data, namely the low S/N of the measurements in the full $\mu-k$ plane,
the practical problems related to estimating accurate covariance matrices for them,
and the difficulties in constructing realistic models that take non-linearities into account. 

Following \PWpaper and \cite{taruya10a}, we break the $H-\dA$ degeneracy by using projections of the 
$\mu-s$ plane, which have the advantage of a higher S/N, while preserving much of the essential 
information. As near future surveys will provide fairly noisy $P(\mu,k)$ planes, 
we find the projection approach more useful in the short term. 

These last two studies focused on the monopole-quadrupole (or ``multipole") pair 
in $k-$space. We demonstrate similar results for the first time in configuration 
space, and introduce an alternative method in the form of clustering 
wedges $\xi(\Delta\mu,s)$, and compare their constraining power to 
that of the multipoles.  

The projection approach also simplifies covariance issues, as one resorts 
to a much smaller covariance matrix, which is more likely to be invertible 
and stable, when using a reasonable number of mock realizations. 
An alternative method suggested by \cite{taruya10a} is to ignore 
non-Gaussianties by using a linear Cov$_{\ell,\ell'}(k)$ based on an analytic model. 
While this might be a fine approach for simple estimates, when analyzing real data 
one should take into account observational effects, most straightforwardly achieved 
by mock realizations with a similar window function, such 
as the mock catalogues produced
by the LasDamas group (McBride et al.; in prep), and used here. 

In this study we analyse geometric effects (or \AP effects after \citealt{alcock79})
in clustering. We study the effect of using an incorrect value for the 
dark-energy equation of state $w$, when converting redshifts to comoving 
distances, causing slight shifts in the inferred positions of the galaxies in respect 
to the real positions. Instead of the true $w=-1$ value we use $-0.9$ and $-1.1$, 
which is within the allowed region for this parameter according to current
observations  (\citealt{komatsu09a}, \citealt{sanchez09a}, \citealt{percival09b}, \citealt{reid09a}).  
It is interesting to note that the $-0.9$ shift (which causes larger dilations and warps)  
yields slightly, but noticeable, larger uncertainties in both real- and velocity-space 
(to see this compare the corresponding results in Figure   \ref{hzdm_hexcorrectionterm}). 
It would be interesting to see if this is a systematic trend increasing with 
dilation and warping, or if we obtained this by chance. Although we obtain similar absolute 
results  in $H$ and $\dA$ with the different \AP effects, this does point out to a possible systematic 
in the estimated ucertainties. This could be tested by applying different fiducial cosmologies 
on actual data, and comparing final results. 

\PWpaper examine much larger warps ($\epsilon>1$ compared to our more realistic $|\epsilon|\sim$0.003) 
and  show a trend of increasing uncertainty $\sigma_\epsilon$ with increasing $\epsilon$ for  
$\epsilon>2$ (see their Table 1). Their argument for using large $\epsilon$ is 
that warping is degenerate between dynamic and geometric $z$-distortions, where the former clearly
dominates the latter.  \cite{ballinger96a} suggest that  the degeneracy between 
dynamic and geometric distortions may be resolved by using measurements at various redshifts,
as they are affected differently.

We perform  similar analyses as \PWpaper on the multipoles up to a few technical differences, 
which  should not affect the results. 
The first difference is that they warp the box of their simulation, that is,
distort the positions of the particles according to given values of $\alpha$ and $\epsilon$, while 
we  imitate the observer's point of view by assuming an incorrect cosmology when converting
redshifts to comoving distances.
Second, we constrain both dilation $\alpha$ 
and warping $\epsilon$ parameters simultaneously, while they assume $\alpha$ is constrained 
by the monopole independently, and this information is then combined 
with the quadrupole to constrain $\epsilon$.

\subsection{Modelling issues}

In this study we avoid modelling issues, for the most part, by using the true mock-mean signal 
as a template. By doing so we assume all parameters 
and effects are known except for the \AP effect ($H,\dA$), 
and test the  effects of marginalizing over amplitude parameters $b\sigma_8,\beta$. 
As this assumes ideal conditions, and we do not test shape effects and non-linearities,  
we consider the tests performed here merely as proofs of concept. This means that we show  
that Equations (\ref{s_distorted_equation})--(\ref{quad_equation}), initially introduced in 
$k-$space form  by \PWpaperii, and our equivalent version of the clustering wedges,  
Equations (\ref{los_distorted})--(\ref{correct_los}), accurately describe the \AP effect on 
projections of $\xi(\mu,s)$,  and can be used to obtain constraints on the values of $H$
and $\dA$. 

In our tests on mock catalogues we show that the \AP effect does not introduce substantial
amplitude or shape effects. Minute amplitude deviations are seen in the
$\sigma_8/\sigma_8^{\rm TRUE}$ column in the bottom left plot of Figure \ref{hzdm_boss_plot_2}, 
as deviations of the results from unity. This amplitude test shows that 
the \AP effect does not introduce an amplitude bias. 

We find that when marginalizing over $\beta$ without priors, $H$ constraints 
are substantially degraded. 
This is clearly seen in Figures  \ref{hzdm_boss_plot}, \ref{hzdm_boss_plot_2}  
(degradation by factor of $\sim 1.5$) and  Table \ref{tab:wa}. 
(For a generalization of this effect as a function 
of range of analysis see Figure \ref{uncertainties_plot}.)
The uncertainties on $\beta$ are fairly large, too. 
The latter could be decreased 
by performing in parallel the \cite{kaiser87} quadrupole test 
based on the squashing effect 
(see \citealt{tocchini11a} for a newly proposed method to reduce uncertainties on $\beta$
through the quadrupole test). 

In order to use the \AP correction in practice on the broad shape of anisotropic clustering projection, 
or even if focusing only on the \bafii, a realistic model based 
on physical principles should be used as templates for these statistics. Of special concern is
understanding the distortions of the baryonic acoustic feature itself. 
For example, comparing the linear theory for $\xi_2$ of Figure  \ref{poc_ld_matsubara} 
with the results from the mock catalogues of Figure \ref{poc_ld}  we notice that the  
\baf is  distorted.  In this case it appears to change from a dip to a bump.   
 
The only studies that the authors are aware of that attempt to resolve this issue are \cite{taruya09a}
and  \cite{taruya10a}. Following a model that includes velocity-dispersion 
decompression given in \cite{scoccimarro04a}, they 
improve the  standard perturbation theory velocity-space power spectrum.
They show significant improvement compared to linear theory and previous attempts (see references within). 
They conclude that density and velocity terms need to be improved, as well as scale-dependent and
stochastic effects of galaxy bias. 
\cite{samushia11a} also show fairly good fits 
to the LasDamas mock catalogue using a more simplistic approach (see their Figure 11).

\subsection{More practicalities}

An approach that could potentially reduce non-linear effects is the {\it reconstruction}
technique of the \bafii. \cite{eisenstein07} proposed to reconstruct the monopole to its original
linear form by reversing the displacements of galaxies using the Zel'dovic approximation
(\citealt{zel'dovich70a}). \cite{seo10a} demonstrate for dark matter particles 
in real- and velocity-space that the non-linearities can be corrected for to a high 
degree, and \cite{mehta11a} have recently reported that reconstruction should work for biased 
matter tracers in an unbiased manner. Focusing on the feature, and ignoring effects 
of the full shape, \cite{seo10a} show that the $\alpha$ parameter ($\sim \dA^2/H$) can be reproduced 
 accurately in  unbiased fashion. The fact that their velocity-space results at low redshifts 
show a larger scatter than in real-space (see  Figures 3--5 in \citealt{seo10a}) indicates that
there is still information in the higher multipoles, especially the quadrupole. 
Nonetheless, this is encouraging in the matter case, and it would be interesting 
to see if the remaining quadrupole after reconstruction 
would be useful to constrain cosmology. If  the dynamic $\xi_2$ could be eliminated, however,  
Equation  (\ref{quad_equation}) yields the simple relation $\xi^{\mathcal D}_2=  2\epsilon d\xi_0/d\ln s$, 
and hence one needs to model only for the monopole, which in this case would be (very close)
to its original linear form. 

One needs also to consider the fact that dynamical distortions 
are degenerate with geometrical. \cite{ballinger96a} argue that, 
although breaking this degneracy might be impossible at one given redshift, 
analyzing various epochs could break degeneracies as both effects evolve differently. 
It would be interesting to see if this could be done in practice, where in reality one might be using 
different sets of tracers, as well as amplitude bias evolution of the tracers (\citealt{fry96a}).

\section{Conclusions}
\label{conclusion_section}

We demonstrate that by correcting for the geometric effects 
of 1D projections of the clustering $\xi(\mu,s)$, it is possible to 
constrain $H, \dA$ to  high accuracy 
We perform tests on the commonly used 
monopole-quadrupole pair 
as well as an alternative basis in the form of {\it clustering  wedges},  
introduced here for the first time. By doing so, we prove that 
the geometrical effects (\citealt{alcock79})  are accurately described 
by Equations  (\ref{los_distorted})-(\ref{correct_los}), 
which illustrate that even a wide ``radial" wedge $\mu>0.5$ 
is mostly sensitive to $H$ and a ``transverse"  wedge ($\mu<0.5$) 
is sensitive to $\dA$, up to small intermixing corrections  terms ${\mathcal C_{||,\perp}}$. 

Throughout this study we use both analytic formulae and 
realistic mock galaxy catalogues to 
compare the constraining power of 
the wide $\Delta\mu=0.5$ wedges with the previously proposed 
multipole statistics, the monopole-quadrupole pair (\PWpaperalt). 
Our main findings are: 
\begin{enumerate}
\item{The \cite{alcock79} effect in 
$\xi$ is very well described by Equations (\ref{mono_equation}) and (\ref{quad_equation}),  
(\ref{los_distorted})--(\ref{correct_los})  
(e.g, see Figures \ref{poc_ld},  \ref{poc_ld_matsubara}).}
\item{Adding the hexadecapole in the multipole analysis improves 
constraints on $H$ by a factor of $1.2-1.35$ and improves $\dA$ 
measurements by $1.15-1.25$ (Figures \ref{hzdm_hexcorrectionterm}--\ref{uncertainties_plot})}. 
In a recent study \cite{taruya11a} have used 
Fisher matrix analysis and obtained similar results. 
\item{The clustering wedges serve as an alternative 
basis  (see Figure \ref{multipole_wedges}) containing much of the 
same $H$, $\dA$ information as the standard multipole projection (e.g, Figure \ref{hzdm_fitsi}).}
\item{Limiting the analysis to the \baf region,  
the $\Delta\mu=0.5$ clustering wedges can 
slightly outperform the [$\xi_0,\xi_2,\xi_4$] combination 
in constraining $H$ (see Figure \ref{uncertainties_plot}). 
This might be due to the fact that higher 
multipoles need to be taken into account. 
See \S \ref{multipolesorwedges_section} for a detailed discussion.}
\item{Constraints on $H$ and $\dA$ can 
be substantially improved by analyzing the broad shape 
of $\xi$ (see Figure \ref{uncertainties_plot} and discussion in \S\ref{multipolesorwedges_section}). 
}
\end{enumerate} 

\begin{figure}
\includegraphics[width=0.47\textwidth]{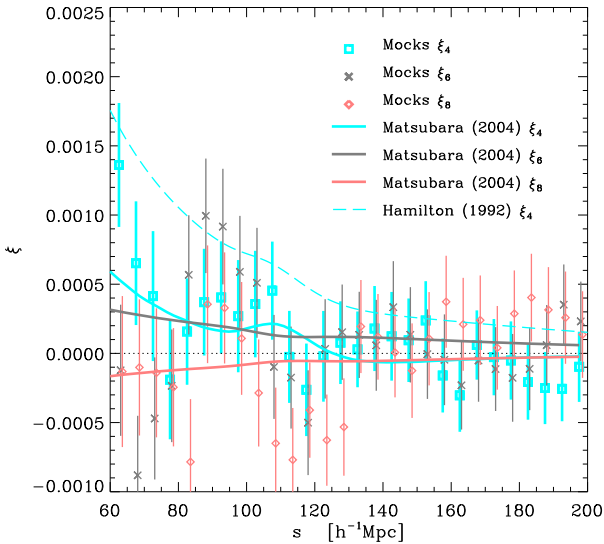}
\caption[High $\ell$ mock and analytic results]{
Mean high order multipoles from the set of 160 LasDamas mock catalogues compared
with linear theory predictions based on \citet{hamilton92} (dashed line) and
\citet{matsubara04} (solid lines). The variances correspond to the uncertainty in the 
mean measurements. Large angle effects \citep[which are included in the formulae of][]{matsubara04}
reduce the amplitude of $\xi_4$ and cause the apparent $\xi_6$, $\xi_8$ results. 
}
\label{highmult_plot}  
\end{figure}

The improved constraining power of clustering wedges might be explained 
by the fact that they contain information from higher order multipoles 
(see Equation  \ref{wedge_definition}). 
We also argue that at ``low" redshift (e.g, $\avg{z}$=0.33 as tested here) 
angular effects introduce even higher order multipole contributions which might 
have an effect at  \baf scales, when correcting 
for the \AP effect (see Figure \ref{highmult_plot} 
and Appendix \ref{projectionspractice_section} for details). 
 
To predict how well the ongoing BOSS 
survey will do, we analyse mock galaxies of the volume and number density expected for this sample. 
Although we predict improvements in constraining $H$ by using information 
on $\xi_4$, this assumes both measuring a signal and being able to model it. 
In practice, BOSS should yield  low S/N at \baf scales, 
meaning that the use of $\xi_4$ would only be possible at smaller scales, 
where velocity-dispersion effects which are difficult to model are more important.   

Because we test for only a few types of 1D projections, there is room for further investigation 
for an optimal statistic for extracting cosmological information from clustering 
through the \AP test. Nonetheless, having various types of statistics yielding consistent
results provides an important tool to test for systematics. 

We suggest that a Fisher Matrix analysis could serve as an analytic method to  
test for an optimal method, to be followed up with mock catalogue tests. 

Once the above are achieved, Figure \ref{hzplane_parameters} 
demonstrates that degeneracies between $\Omega_{\rm M0}$, $\Omega_{\rm K}$ and 
the dark energy equation of state $w$ within the $H-\dA$ plane 
may be resolved by using clustering at high redshifts 
($z>2$; e.g, through Lyman-$\alpha$ forest, and 21 cm measurements) 
in a complementary fashion to other observations (e.g, the 
temperature fluctuations of the CMB). At lower redshifts the degeneracies 
between these parameters is quite large
at a single redshift, but might be resolved by  
measuring $H(z)$ at various redshifts. 

In the coming years a variety of new large volume 
galaxy surveys will measure the large-scale clustering pattern
of the Universe with unprecedented precision. 
Investigations of techniques such as the reconstruction of the \baf 
(\citealt{eisenstein07, seo10a, mehta11a}, Padmanabhan et al (in prep.)) 
suggest possible substantial improvements on constraining the 
cosmic evolution out of the information from these surveys and hence on our understanding of dark energy.
That said, these measurements will always be bound to the \AP effect. 
The 1D clustering projection techniques discussed here 
will be essential to ensure that the full potential 
of the information contained in the \AP effect will be realised.

\section*{Acknowledgements}

We thank Nikhil Padmanabhan for his insight.
We thank Chris Blake and Bob Nichol for commenting on our manuscripts and fruitful input.
We thank 
David Hogg, 
Daniel Eisenstein, 
Takahiko  Matsubara, 
Will Percival, 
Atsushi Taruya 
and Martin White for helpful conversations.
We thank the LasDamas project for making their mock catalogues publicly available. 
In particular we are much obliged to Cameron McBride for supplying mocks on demand 
as well for his insights on S/N issues in particular and statistics in general. 
We thank the Horizon team for making their mocks public,
and in particular  Changbom Park and Juhan Kim for discussions on usage.
E.K thanks Guang-Tun Zhu for his technical help.
E.K thanks the NYU Horizon Fellowship for summer travel 
support which contributed to this collaboration. 
E.K was partially supported by a Google Research Award and NASA Award
NNX09AC85G.  M.B was supported by Spitzer G05-AR-50443 and NASA Award.

\appendix

\section{Hexadecapole terms}
\label{quad_terms_appendix}

In \S \ref{dialtionwarping_section} we give analytic 
expressions for the \AP effect on the monopole and quadrupole. 
Here we extend this treatment to take into account the hexadecapole $\xi_4$ contribution. 
In this case Equation (\ref{quad_equation}) becomes: 
\begin{dmath}
\label{full_quad_equation}
\xi^{\mathcal D}_2(s) = \xi_2(\alpha s)  +  \epsilon \left(  2\frac{d\xi_0(s)}{d\ln(s)} + \frac{6}{7}\xi_2(\alpha s) + \frac{4}{7} \frac{d\xi_2(s)}{d\ln(s)} 
 +  \frac{20}{7}\xi_4(\alpha s) +  \frac{4}{7}\frac{d\xi_4(s)}{d\ln(s)} \right). 
\end{dmath}
In our analyses of the \AP effect on the $\xi_2$ measurement from the mock catalogues, 
we find the $\xi_4$ and d$\xi_4/$d$\ln s$ corrections negligible. 
Limiting the hexadecapole to $\ell<=4$ contributions we obtain:
\begin{dmath}
\label{hex_equation}
\xi^{\mathcal D}_4(s)=\xi_4(\alpha s)  + \epsilon \left(\frac{36}{35}\frac{d\xi_2(s)}{d\ln(s)} -  \frac{10}{77}\xi_4(\alpha s) + \frac{115}{154}\frac{d\xi_4(s)}{d\ln(s)}\right), 
\end{dmath} 
where a higher $\ell$ would be required for completeness. 

\section{Testing wedge widths and intermixing terms}
\label{wedgescorrestion_section}

\begin{figure}
\includegraphics[width=0.47\textwidth]{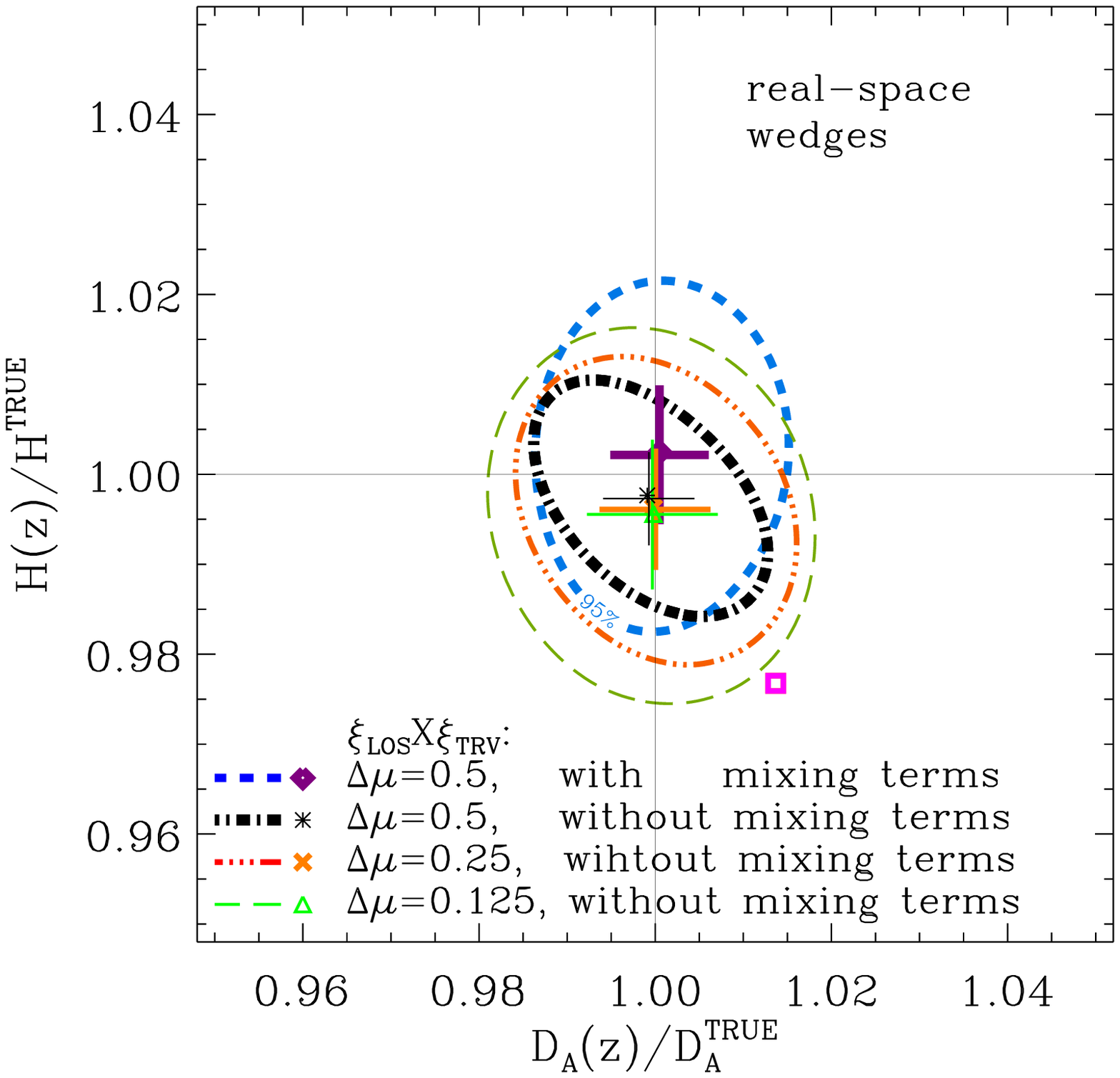}
\includegraphics[width=0.47\textwidth]{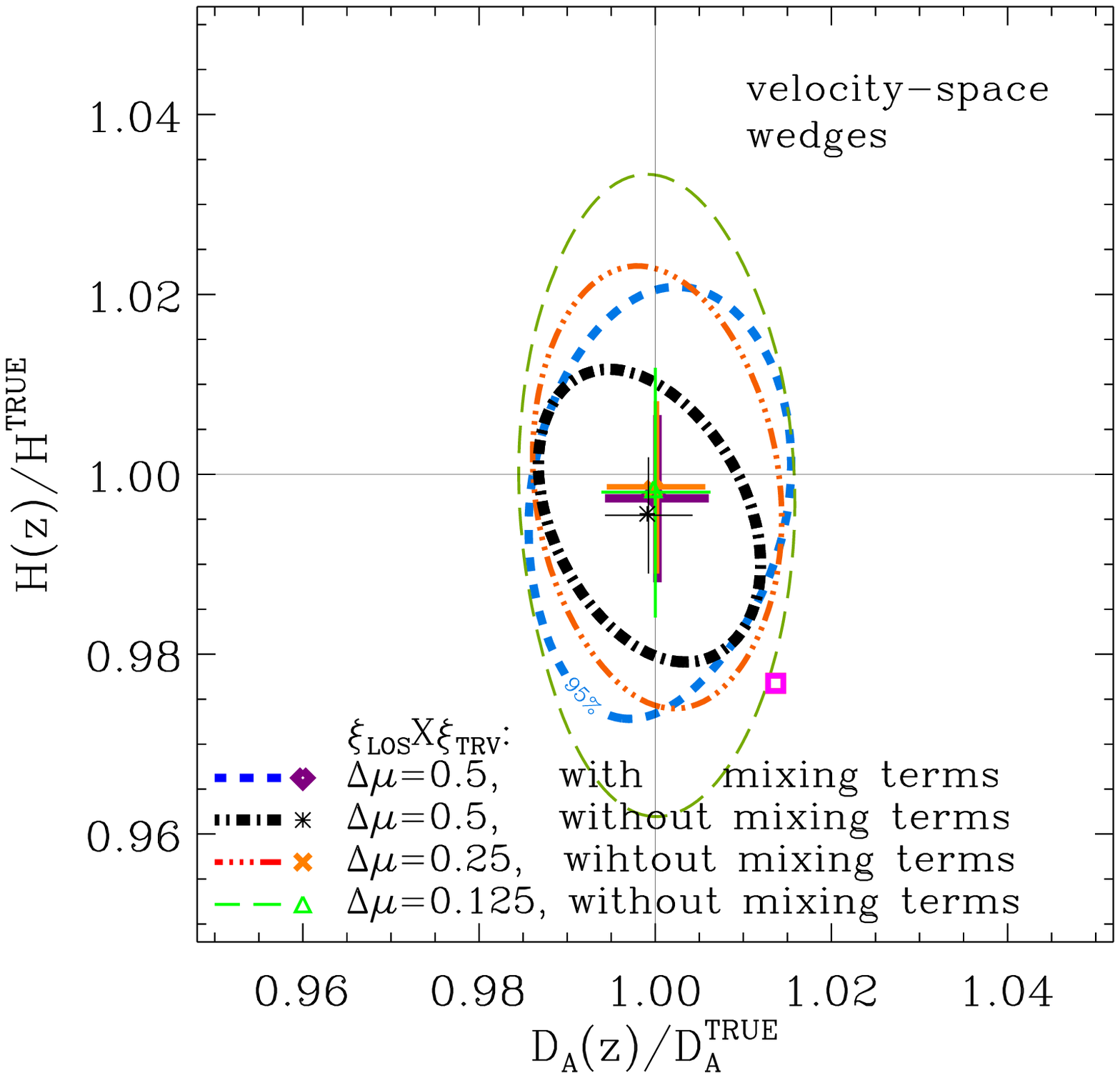}
\caption[Varying $\Delta\mu$ and testing ${\mathcal C_{||,\perp}}$]{
Joint $2\sigma$ constraints on the $H-\dA$ plane obtained 
when analyzing the \AP effect for various clustering widths $\Delta\mu$ 
in real-space (upper panel) and velocity-space (lower panel). 
For $\Delta\mu=0.5$ we test the effect of the correciton terms 
${\mathcal C}_{||,\perp}$ (short-dashed blue; same results as in Figure \ref{hzdm_fitsi}) 
and without (thick dot-dashed black; ${\mathcal C}_{||,\perp}=0$). 
(Ignoring intermixing terms 
means $\xi_{||}$ depends 
solely on $H$ and $\xi_\perp$ on $\dM$.)  
As indicated in the legend for each statistic, 
the symbols are the most likely 2D values, and the crosses 
show the marginalized $1\sigma$ results. 
We see that not including  ${\mathcal C}_{||,\perp}$ biases the results, 
and underestimates uncertainties. 
 We also show
results for analyses of thinner wedges of 
$\Delta\mu=0.25, \ 0.125$ (as indicated in legend). 
For these results we do not use correction terms in the analysis. 
We notice a clear trend of less bias as $\Delta \mu$ decreases, 
which is expected as the intermixing terms are less important.
The increase in uncertainty reflects the fact that we are using 
less information.
In velocity-space this is seen very sharply by the elongation 
along the $H$ direction. 
}
\label{hzdm_correctionterm}
\end{figure}

In Figure  \ref{hzdm_correctionterm} we explore the effectiveness of 
the wedge correction terms ${\mathcal C}_{||,\perp}$ (Equation \ref{correct_los}). 
These results should be compared to those shown in the bottom plots of Figure  \ref{hzdm_fitsi},
The short-dashed blue lines are the $2\sigma$ results shown in Figure \ref{hzdm_fitsi}, 
meaning when including  ${\mathcal C}_{||,\perp}$ terms.
The thick single-dot-dashed black lines are the 
$2\sigma$ results from the same tests where we set 
 ${\mathcal C}_{||,\perp}=0$. Two interesting differences are apparent. 
The most obvious one is the small bias relative to the true cosmology 
which produces a shift in the contour lines (although within the $1\sigma$ region). 
Interestingly, our second observation is 
that it yields apparently tighter constraints than 
the corrected method (colored contour). 
These observations are noticeable in both spaces. 


In Figure \ref{hzdm_correctionterm} we also investigate other choices of radial and transverse wedges 
and give $2\sigma$ results for: $\Delta\mu=0.125$ and  $0.25$, where
we do not use correction terms (i.e, ${\mathcal C}_{||,\perp}$=0). 

Our results clearly show, as expected, that using more information, 
i.e., through wider wedges 
yields tighter constraints. We also notice that decreasing 
$\Delta\mu$ reduces the bias in the obtained constraints on $\dA$ 
both in real- and velocity-space. In velocity-space the bias is improved
also in the $H$ direction. 
We have not investigated the constraining 
power of multiple thin wedges, 
but rather that of the two extermes $\mu>\Delta\mu$ and $\mu<\Delta\mu$. 

In both spaces we notice that the narrower clustering wedges 
yield contours with reduced correlation coefficients  
between parameters $H$ and $\dM$. This is expected due to weaker 
intermixing terms with decreasing $\Delta\mu$. 
This is most prominent in velocity-space where the contours 
sharpen towards the line-of-sight. In real-space the 
ellipticity appears to decreas with $\Delta \mu$. 

\section{Proof of Concept: using Analytic Predictions}
\label{matsubara_section_ap}

Here we demonstrate the applicability of Equations (\ref{mono_equation})--(\ref{correct_los})
to reproduce the distortions in the various multipoles and clustering wedges introduced by the use
of an incorrect cosmology when transforming the observed redshifts to distances. For this we 
use the formulae of  \cite{matsubara04} for the velocity-space $\xi(\mu,s)$. 
We use their equation 1 which yields the two-point correlation function $\xi(z_1,z_2,\gamma)$ 
for pairs of objects located at redshifts $z_1$ and $z_2$ separated by 
observer angle $\gamma$. Applying Equation (\ref{comoving_equation}),
and basic Euclidean comoving geometry, we project this result around $z_1$ 
to obtain a $\xi(\mu,s)$ plane, where the line-of-sight 
direction is defined as the bisecting vector 
from the observer to $\vec{s}$ (i.e., the same as we do with the mock catalogues). 
We then project $\xi(\mu,s)$ into the various multipole and wedge components. 

\begin{figure*}
\includegraphics[width=0.45\textwidth]{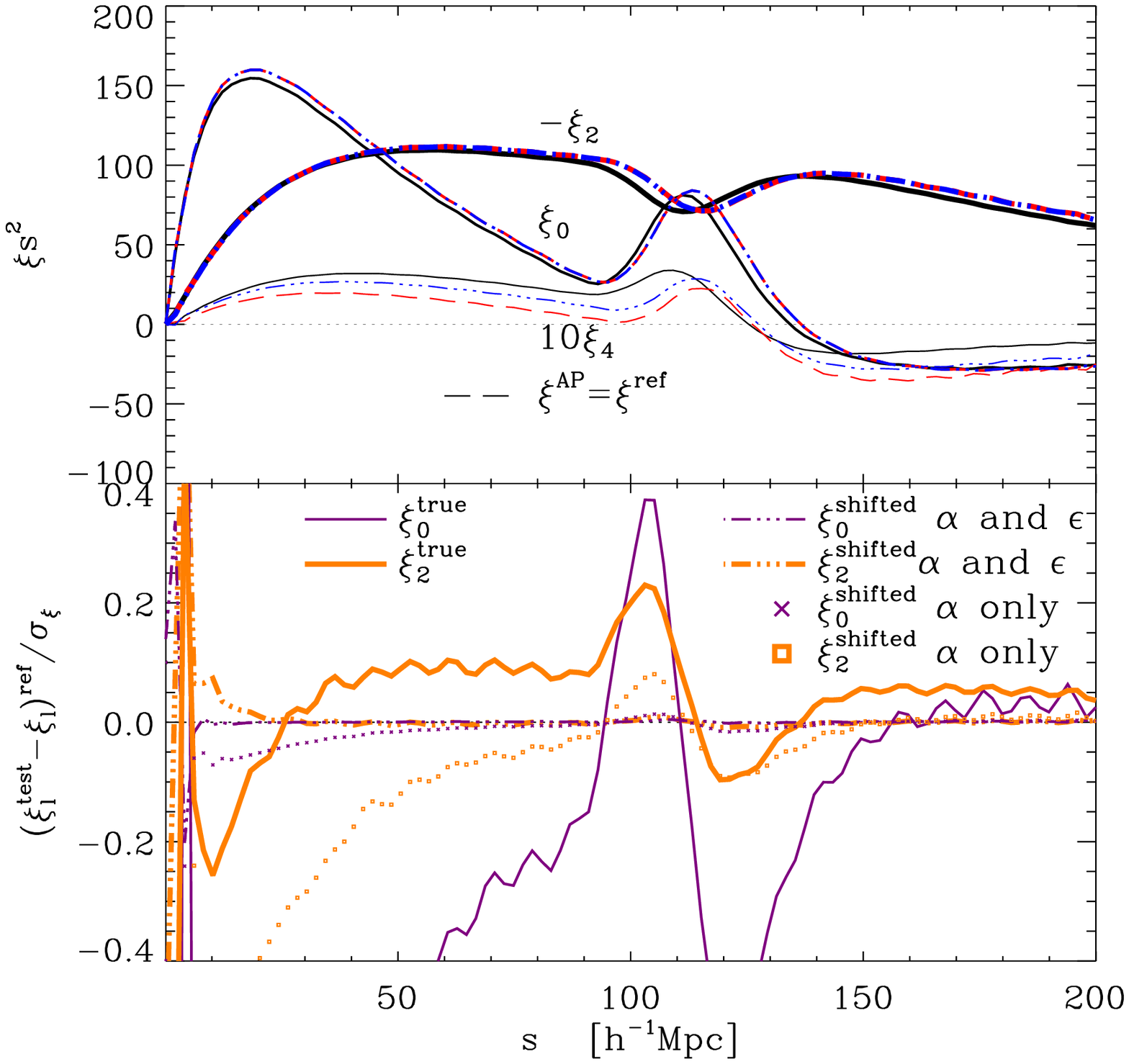}
\includegraphics[width=0.45\textwidth]{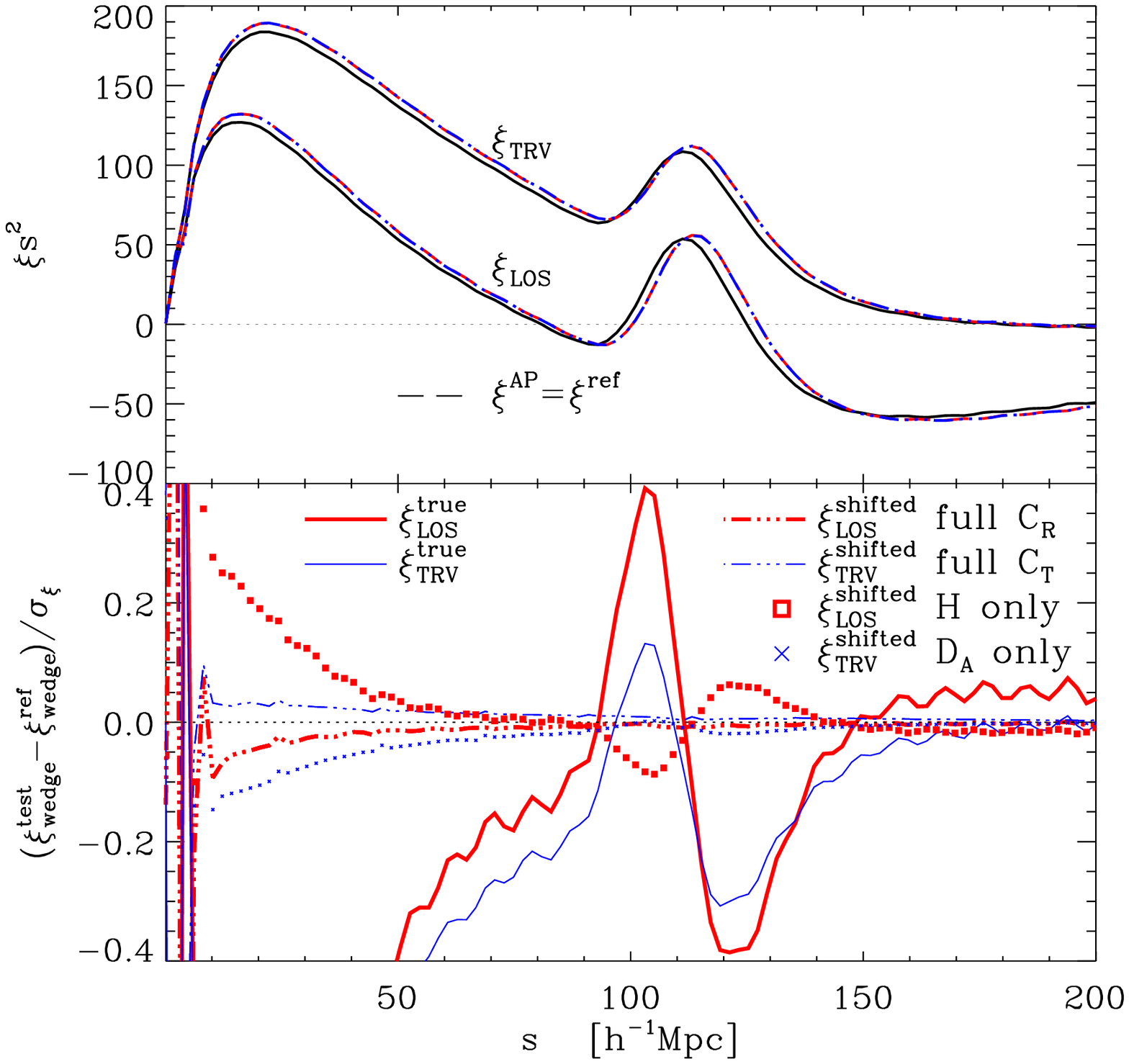}
\caption[Geometrical effects on analytical 1D $\xi$ projections]{
Analytical velocity-space clustering projections (based on \citealt{matsubara04}) 
with and without the \AP effect. 
Left: multipoles  (corresponding to Equations \ref{mono_equation}, \ref{quad_equation}).
Right: $\Delta\mu=1/2$ wedges  (Equations \ref{los_distorted}-\ref{correct_los}). 
In the top panel of each plot are the 1D projections, 
and in the bottom panels the difference 
of each result with the DISTORTED signal (labeled as AP) 
normalized by the uncertainty of one realization. 
As indicated in legend, solid lines 
are the true projection signals ($\TRUE$), the long-dashed lines 
are the AP signals ($\DIST$). 
We apply the AP shift on true signals to obtain 
the triple-dot-dashed lines ($\SH$).
 A perfect shift would yield a null result in the bottom panels. 
The symbols represent applying only zeroth order corrections. 
For multipoles this means applying only isotropic dilation correction ($\alpha$) 
and no warping terms ($\epsilon$). We see that the warping is crucial 
for correcting $\xi_2$ but negligible (although visible here) for the monopole. 
For the  wedges zeroth order means applying only $H$ ratio corrections 
for radial wedge, and applying $\dA$ ratio for transverse (meaning no ${\mathcal C}_{||,\perp}$ corrections). 
The \AP distortion applied here is using $w^{\rm FID}=-1.1$ instead of  the true value $-1$ 
when converting $z$ to comoving distances.
}
\label{poc_ld_matsubara} 
\end{figure*}


For the true signal $\xi^{\TRUE}$ $\xi$ we use the true input $P(k)$ cosmology in Equation (\ref{comoving_equation}). 
For the distorted signal $\xi^{\DIST}$ we apply the \AP effect in which we apply the same exact procedure, 
except that in  Equation (\ref{comoving_equation}) we 
assume an equation of state $w=-1.1$ instead of the true $-1$ value. 
For the shifted signal $\xi^{\SH}$ we apply the scaling given by Equations (\ref{mono_equation})
and (\ref{quad_equation}) on the $\xi^{\TRUE}$ signal of the monopole, quadrupole pair
(and also on $\xi_4$), and Equations (\ref{los_distorted}), (\ref{trv_distorted}), and
(\ref{correct_los}) for the $\Delta\mu=0.5$ wedges. 

Our results are shown in the upper panels of Figure \ref{poc_ld_matsubara}
for the multipoles (left) and clustering wedges (panels), rescaled by $s^2$ for clarity.
The bottom panels compare the predictions of Equations (\ref{mono_equation})--(\ref{correct_los})
with the $\xi^{\DIST}$ (or $\xi^{\rm AP}$) results. A perfect description of the \AP effect would result in 
zero values along the solid line. By comparing results of 
$(\xi^{\SH}_{\rm stat}-\xi^{\rm AP}_{\rm stat})/\sigma_{\xi_{\rm stat}}$ 
and $(\xi^{\TRUE}_{\rm stat}-\xi^{\rm AP}_{\rm stat})/\sigma_{\xi_{\rm stat}}$  we verify 
that Equations (\ref{mono_equation})--(\ref{correct_los}) accuratedly describe the \AP effect and can be 
used to correct for it. Here ``stat" means the various 1D statistics investigated: $\xi_{0,2,||,\perp}$. 

We also examine the importance  of the first order $\epsilon$ correction in each projection relative 
to the zeroth order. In the multipoles the zeroth order term 
is defined as applying only the isotropic dilation parameter $\alpha$, 
meaning assuming $\xi_\ell^{\mathcal D}(s)=\xi_\ell(\alpha s)$. 
In the wedges the zeroth order means applying only 
the $H$ ratio in the radial, and $\dM$ in the transverse. 
This means assuming ${\mathcal C}_{||,\perp}=0$ in Equations  (\ref{los_distorted}) and (\ref{trv_distorted}) 
for $\Delta\mu=1/2$.

The first order correction in all projections is defined here by adding the warping term 
$\epsilon$ which include derivatives of the projections. 

The improvement in the monopole is subtle, and negligible for any practical case. 
In $\xi_2$ the $\epsilon$ term is, of course, essential for describing the \AP shift. 
In the $\Delta\mu=0.5$ wedges, interestingly, the first order corrections yield slightly better results,
meaning that the radial wedge is most sensitive to $H$ and the transverse on $\dM$. 
In Appendix  \ref{wedgescorrestion_section} we show, using mock catalogues, that the
${\mathcal C}(\epsilon)$ intermixing terms are essential to use in the wedge \AP correction technique 
to yield unbiassed $H, \ \dM$ results, and that without them the uncertainties in these parameters 
are underestimated.

\section{Projections in practice: relating wedges to multipoles}
\label{projectionspractice_section}

\begin{figure*}
\includegraphics[width=0.9\textwidth]{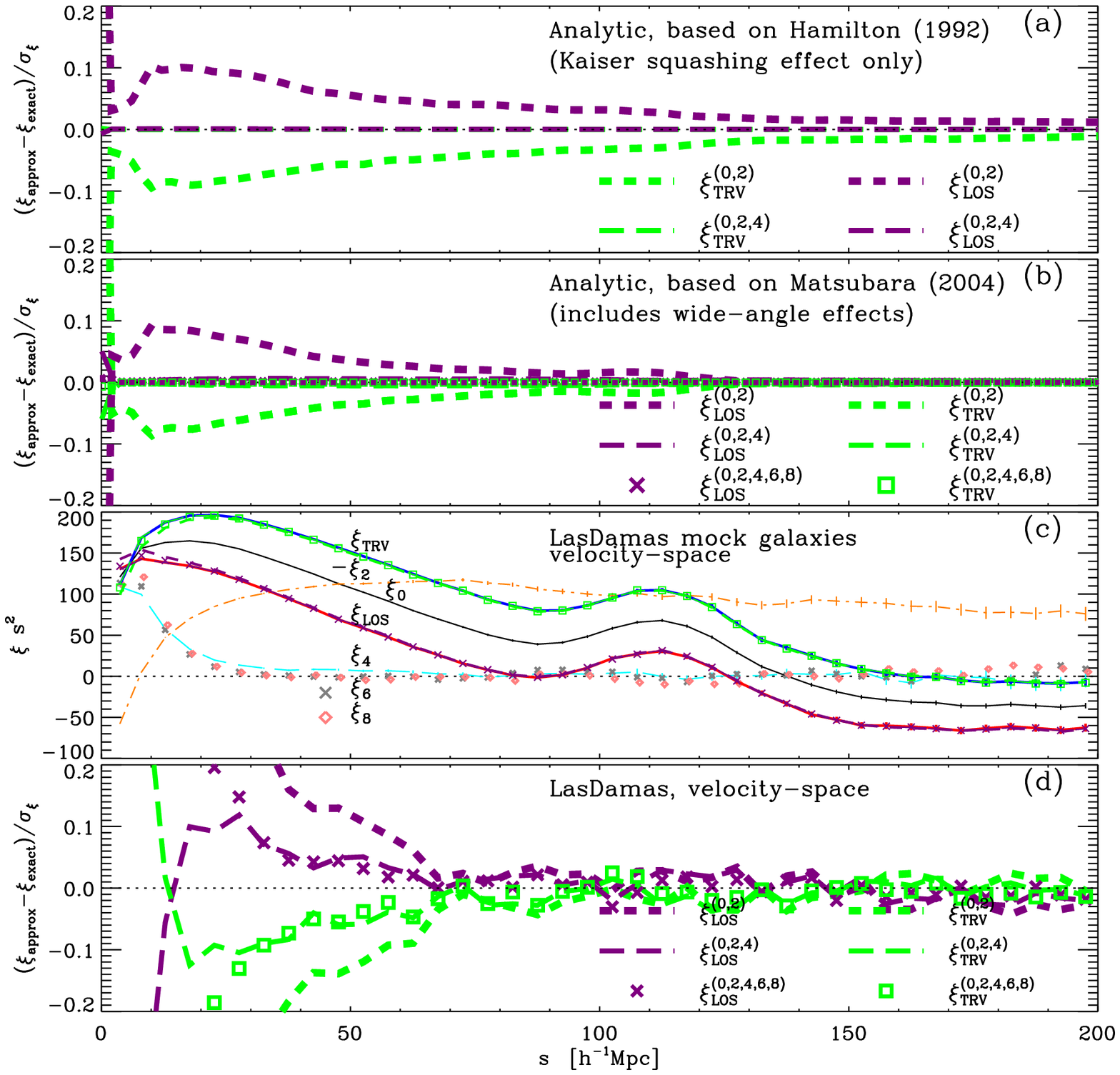}
\caption[Relating clustering wedges to multipoles]{
Panels (a),(b),(d) show the difference between clustering 
wedges inferred from multipoles 
($\xi^{([\ell])}$), $\xi_{\rm wedge}^{\rm approx}$, and 
those calculated directly, $\xi_{\rm wedge}^{\rm exact}$, 
normalized by the uncertainty of one mock realization. 
$\xi_{\rm wedge}^{([\ell])}$ indicates 
the multipoles [$\ell$] used in the approximation.   
Panel (a) show results for analytic formulae in \citet{hamilton92},
and (b) \citet{matsubara04}, respectively. 
Including the contribution from higher order multipoles improves the results
with respect to those obtained when only the monopole-quadrupole pair is used. 
Panels (c),(d) show similar results for the LasDamas mock catalogues, where 
(c) shows the actual statistics. At large scales, $\xi_0$ and $\xi_2$ are sufficient
to accurately describe clustering wedges obtained from the mock catalogues, 
and higher order multipoles are required for $s<30$\hmpcii.
}
\label{multipole_wedges}  
\end{figure*}

In Figure  \ref{multipole_wedges}  we put Equation (\ref{wedgemonoquad_equality}) 
to the test both analytically and using mock galaxy catalogues by comparing $\Delta\mu=0.5$ wedges 
which are calculated directly to those approximated by the multipoles. 

All uncertainty bars in the mocks are for the mock mean. All results (analytical and mocks) assume 
measurements of galaxies at $\avg{z}\sim0.33$ with a bias of $b\sim 2$ with the same flat
$\Lambda$CDM cosmology.


\subsection{Relating multipoles and wedges: analytic treatment}

In this section we test the relation between the clustering wedges and the multipoles
by using theoretical predictions for $\xi(\mu,s)$. For this we use the 
prescriptions of  \cite{hamilton92} (Figure  \ref{multipole_wedges}a) 
and \cite{matsubara04} (Figure \ref{multipole_wedges}b). 
Both of these formalisms take into account only the Kaiser squashing effect 
in the plane-parallel approximation (and no velocity-dispersion), 
but the latter also includes wide-angle effects as 
well as linear growth evolution between points $z_1$ and $z_2$ 
(see Appendix \ref{matsubara_section_ap} for a brief explanation 
on manipulations of Equation 1 of \citealt{matsubara04}). 
For the \cite{hamilton92} prescription we calculate 
$\xi_0$, $\xi_2$, and $\xi_4$ directly by using their Equations 6-8, 
where the matter correlation function is calculated from the output of 
CAMB (\citealt{lewis02a}). For the formulae of \cite{matsubara04} we
use the same matter $\xi(r)$ to compute $\xi(\mu,s)$,\footnote{
In practice, $\xi(\mu,s)$ requires very fine $\mu$ bins (actually, in
$\theta$). Here we show results for $\delta\theta\sim {5 \cdot 10^{-3}}^\circ$, where for
e.g. $\delta\theta=1^\circ$, there are binning effects.}
and integrate to obtain multipoles using Equation  (\ref{eq_extract_multipole}).\footnote{
As a consistency check we verify that these integrated multipoles yield
the same result as when calculating directly in the  \cite{hamilton92}  algorithm. 
For discussion on binning issues with actual data see Appendix \ref{xi_estimators}}
In each case, the ``directly" calculated wedges are 
computed by integrating  $\xi(\mu,s)$, 
and defining the radial $\xi_{\rm \parallel}$ as $\mu>0.5$ 
and transverse $\xi_{\rm \perp}$ as $\mu<0.5$.  

To calculate the ``approximated" clustering wedges $\xi_{\rm Wedge}^{(0,2)}$, that is, 
only taking into account the monopole and quadrupole contributions, we use Equation (\ref{wedgemonoquad_equality}). 
When adding the term of the hexadecapole $\xi_{\rm Wedge}^{(0,2,4)}$, 
we add the contribution of Equation (\ref{hex_term}). 
When using the formulae of \cite{matsubara04} (panel b) 
we also calculate $\xi_{\rm Wedge}^{(0,2,4,6,8)}$ by following 
a similar procedure using $\xi_6$ and $\xi_8$.  Results are given in terms of $(\xi_{\rm wedge}^{\rm true}-\xi_{\rm wedge}^{\rm approximated})/\sigma_\xi$, 
where $\sigma_\xi$ 
corresponds to the $\sqrt{C_{ii}}$ obained from the $0.16<z<0.44$ LasDamas mocks.

Due to the fact that according to the prescription of \cite{hamilton92} $\xi(\mu,s)$ 
contains contributions only from even multipoles up to the hexadecapole, this $\ell=4$
contribution is critical for an absolute definition of the wedges.
By neglecting this term the wedge approximations $\xi_{\rm Wedge}^{(0,2)}$ are inaccurate at the 
 $\sim 1.5\%$ level (in terms of fractions $\xi^{\rm approx}/\xi^{\rm exact}$) at the barionic acoustic feature,  
which corresponds to $4\%$ of $\sigma_\xi$.

For the formulae of \cite{matsubara04} we follow a similar procedure,  
where we also note contribution due to higher multipoles. 
We notice improvement from $\xi_{\rm Wedge}^{(0,2)}$
to $\xi_{\rm Wedge}^{(0,2,4)}$
and even further improvement with $\xi_{\rm Wedge}^{(0,2,4,6,8)}$. 
The $\xi_6, \ \xi_8$ contributions are not expected from the linear squashing effect,
but rather are a result of wide observer angle effects. 
We verify that $\xi_6$ and $\xi_8$ vanish when running the 
algorithm at $z=3$. 

Comparing the results for $\xi_{\rm wedges}^{(0,2)}$ obtained using the 
prescription of \cite{matsubara04} to those from the recipe of \cite{hamilton92}, 
we note that the former asymptots to zero faster. 
This can be explained by the fact that the 
$\xi_4$ term in  \cite{matsubara04} is suppressed 
due to the rise of higher order terms. 
This is clearly seen in Figure \ref{highmult_plot} 
where  $\xi_6$ is comparable to $\xi_4$ at the \baf 
and surpasses it at larger scales and $\xi_8$ is comparable to $\xi_4$ at $s>130$\hmpcii. 

To conclude, these tests show that the clustering wedges can be accuratedly described in terms of the
multipoles, and hence can be used as an alternative basis. 


\subsection{Analysis of mock catalogues}

Here we perform a similar analysis as in the previous section but using the 
results for the mean multipoles and wedges obtained from the $160$ LasDamas
mock catalogues described in \S\ref{mocks_section} in real-space and velocity-space.
As these include velocity-dispersion effects, as well as the expected wide-angle effects, we
study even multipoles up to the eighth order. In practice, here all clustering multipoles and 
wedges are estimated by integrating $\xi(\mu,s)$ which is calculated with the 
\cite{landy93a} estimator. In  Appendix \ref{xi_estimators} we discuss an alternative approach 
to calculating these 1D statistics. 

We find that the monopole and quadrupole approximations  $\xi_{\rm Wedge}^{(0,2)}$  
appear to be sufficient to describe the clustering wedges 
on all scales in real-space. 
In velocity-space they appear to be sufficient for scales $s>70$\hmpcii, 
as seen as thick short dashed lines in Figure \ref{multipole_wedges}c,d. 
In real-space, of course, the only term is the monopole, 
but we do obtain higher order multipoles resulting from noise, 
and on large scales from angular effects. In velocity-space adding even multipoles $\ell>2$ 
improves results at $s<70$\hmpcii. These $\ell>2$ terms appear due to velocity-dispersion effects
on large scales and dominate at $s>30$\hmpcii  \citep{scoccimarro04a}. 

In Figure \ref{highmult_plot} we focus on the mock $\ell=4,6,8$ multipoles, where uncertainties 
are for the mock mean of $160$ realizations. These are compared to those analytically obtained from 
\citet{hamilton92} ($\xi_4$ only: dashed line) and \citet{matsubara04} (solid lines). 
The mock $\xi_4$ result appears to have structure  similar to that expected according to  
\citet{matsubara04} (that is, inclulding wide-angle effects). We might be seeing a \baf at
$\sim 100$\hmpcii.  Although the large uncertainties indicate 
low significance, we note that similar real-space analysis yields a negative $\xi_4$ (but consistant 
with zero at a level of $2\sigma$), indicating that the velocity-space result is not due to angular effects 
only. Figure 11 in \cite{samushia11a} shows similar trends in the mock $\xi_4$. 

We conclude that the two wide clustering wedges ($\Delta\mu=0.5$) are defined fairly well by
the monopole and quadrupole in velocity-space (and monopole only in real-space), and hence may
be used as a alternative basis to these multipoles to project most of the information
contained in $\xi(\mu,s)$.

\section{$\xi$ Estimators}\label{xi_estimators}

\begin{figure*}
\includegraphics[width=0.47\textwidth]{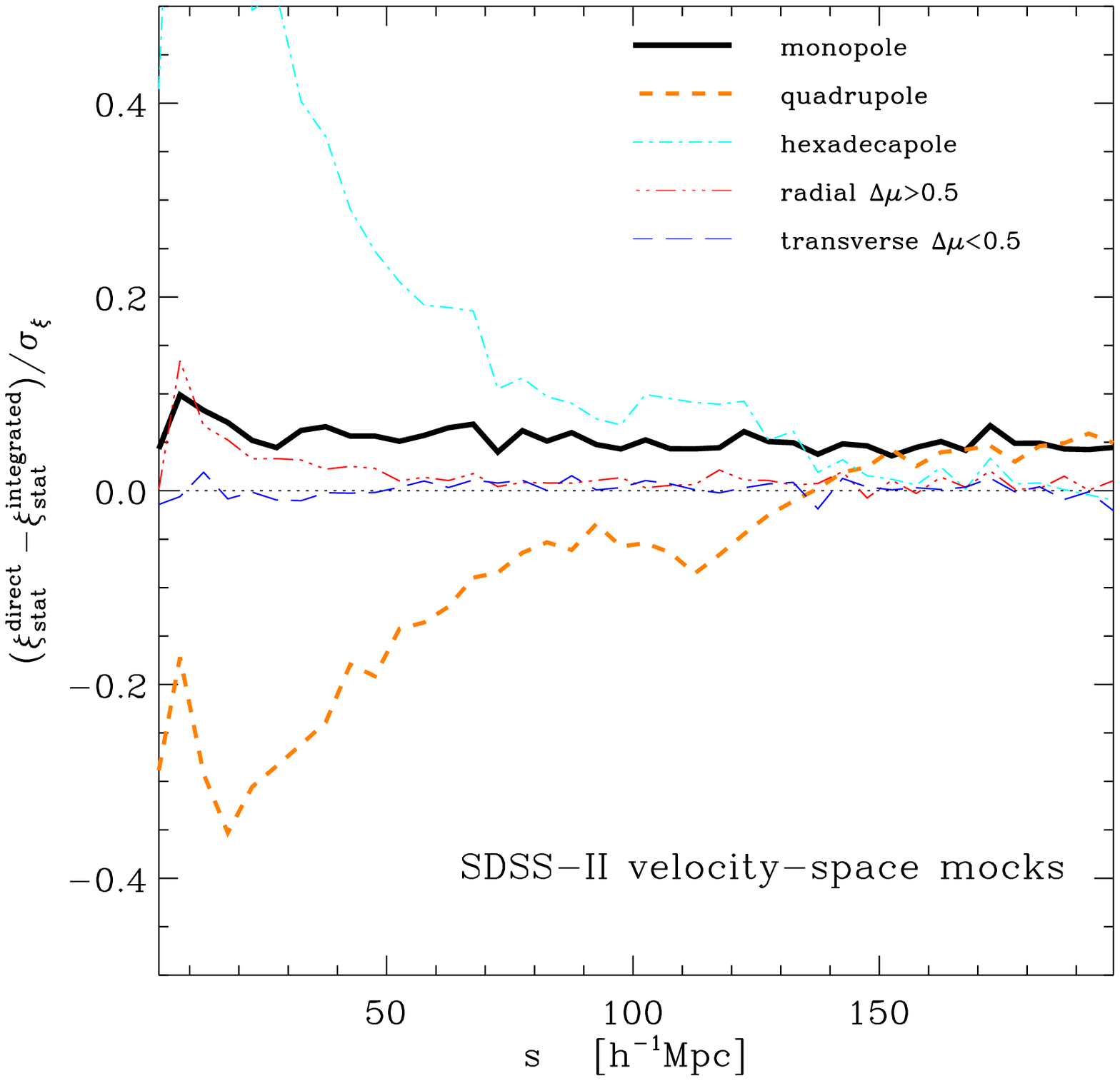}
\includegraphics[width=0.47\textwidth]{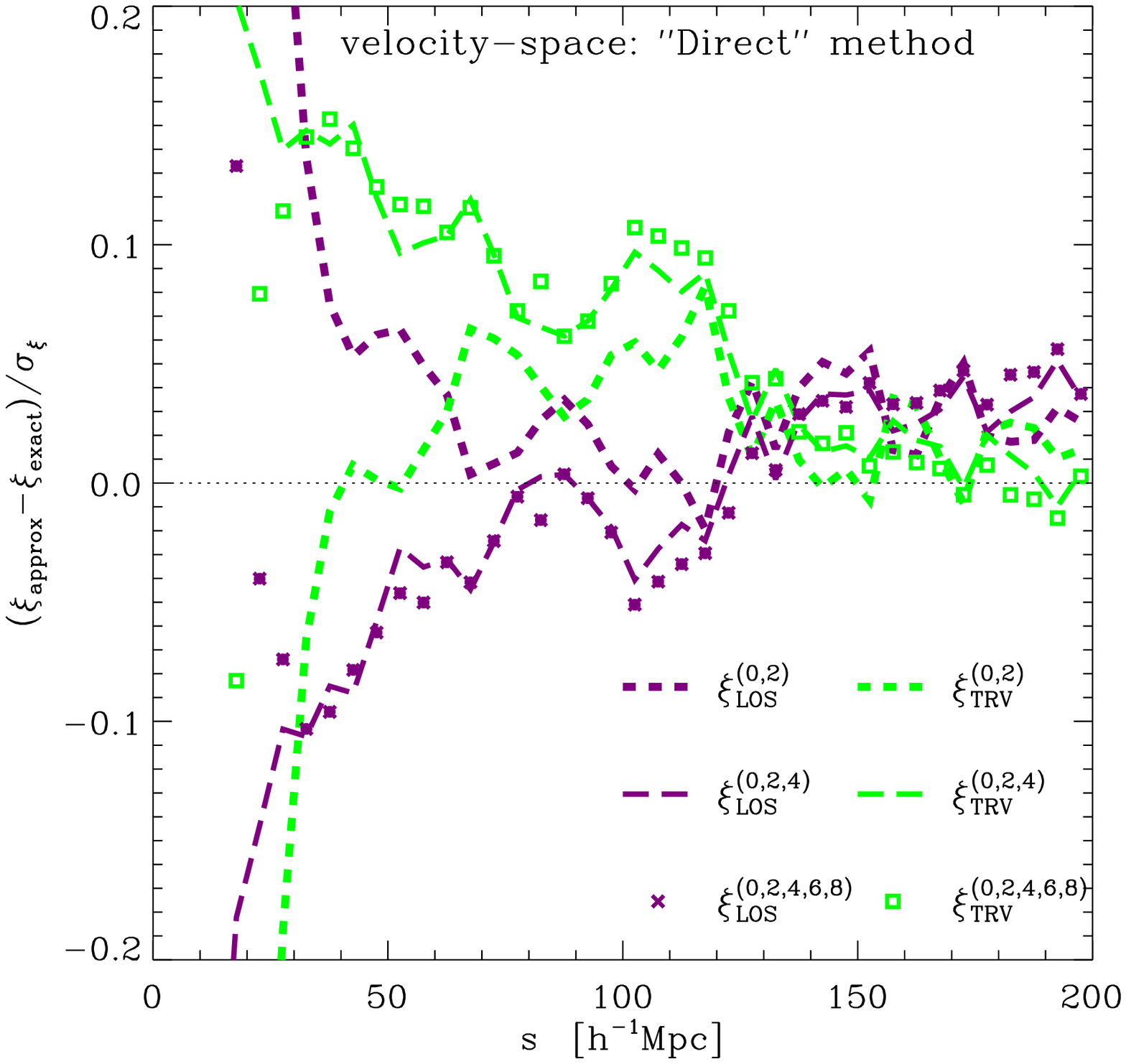}
\caption[$\xi$ systematics: comparing ``integrated" to ``direct"]{
The left panel shows differences between the statistics obtained when using the 
``direct" method ($\xi^{\rm direct}$; 1D binning in pair-counter) and the 
``integrated" method ($\xi^{\rm integrated}$; integrating over 2D binning) 
in units of $\sigma_\xi$. Differences arrise due to wide angle effects 
for these clustering scales at $\avg{z}=0.33$. 
We clearly see that the wide wedges are the least affected. 
The right panel shows the relationship between the difference 
between the actual  $\Delta\mu=0.5$ clustering wedges to 
those statistics approximated when using 
multipole combinations $\xi^{[\ell]}$ (Equation \ref{wedge_definition}) in 
units of $\sigma_\xi$.
In all cases shown here we use the ``direct'' measurements, 
while in Figure \ref{multipole_wedges}d we show 
a simliar comparison for $\xi^{\rm integrated}$. 
We clearly see that the latter test is more successful, 
indicating the ``integrated" method to be a more natural technique.
}
\label{directint_plot}  
\end{figure*}

For our $\xi$ analysis of the mocks we have used the \cite{landy93a} estimator which has been 
shown in various studies as the lowest variance known (e.g, \citealt{kerscher00a,kazin10a}). 
We examine two  methods in which the 1D $\xi_\ell$ multipole and wedge statistics can be calculated:
\begin{enumerate}
\item{
``Integrated": calculating $\xi(\mu,s)$ in 2D bins and integrating using Equation  \ref{eq_expansion} 
for the multipoles and Equation \ref{wedge_definition} for the wedges.
}
\item{``Direct": calculating the statistics in 1D bins directly in the pair-counter.}
\end{enumerate}
For the ``direct" method
the 1D estimators can be generically written as:
\beq\label{ls_direct_equation}
\hat{\xi}^{\rm direct}_\ell(s)=
\frac{DD_{\rm stat}(s)+RR_{\rm stat}(s)-2\cdot DR_{\rm stat}(s)}{RR_{\rm projection \ type}(s)}.
\eeq
The $DD, \ RR, \ DR$ are the usual normalized data-data, random-random, data-random 
pair counts. By normalized we refer to the fact that it is common to reduce shot-noise effects 
of the random points by assigning many more random points $N_{\rm R}$ than data $N_{\rm D}$:  
$r\equiv N_{\rm R}/N_{\rm D}\gg 1$.\footnote{
For LasDamas catalogues we use
$r\sim 30$, and for the Horizon Run mocks we use $r\sim 2$.} 
This means, that when one counts number of data-data points $N_{\rm dd}$ 
the estimator requires it to be normalized to the random-random 
$N_{\rm rr}$ by $DD\equiv N_{\rm dd}\cdot r^2$, while
for data-random this is  $DR\equiv N_{\rm dr}\cdot r$. 

The ``stat" subscript indicates the required weighting for each statistic. 
For the multipoles stat$=\ell$, meaning the weight is the $\ell$ 
Legendre polynomials ${\mathcal P}_\ell(\mu)$. 
Technically this means when counting data-random pairs of the multipole
$N^\ell_{\rm dr}(s)$, e.g, it is increased by: 
$N_{\rm dr}^{\ell}+= w_{\rm d}\cdot w_{\rm r} \cdot {\mathcal P}_\ell(\mu)$, 
if a pair is within range of $[s-0.5\Delta s, s+0.5\Delta s]$ 
and has a $\mu$ 
 within $[\mu-0.5\Delta\mu , \mu+0.5\Delta\mu]$.
The weights 
$w_i$ could be due to incompleteness in uniformity (angular or radial), 
otherwise they are unity. 
For the multipoles the denominator term 
is that of the monopole, meaning $RR_{\rm multipoles}=RR_0$. 

The wedge estimators are similar in the nominator, but the denominator is different. 
For example,  data-random pairs $N_{\rm dr}^{\rm wedge}$ 
for clustering wedge $\mu_{\rm min}<\mu<\mu_{\rm max}$ 
is increased by $w_{\rm d}\cdot w_{\rm r}$ given a pair in the separation 
range  $[s-\Delta/2 s, s+\Delta s/2]$  and within the wedge $\mu$ range, 
and zero contribution otherwise. 
Because the nominator ``weighting" is that of a top hat, 
the denominator term $RR_{\rm wedge}$ 
in the same as in the nominator.

In this study we focused on two wide $\Delta\mu=0.5$ wedges, 
but this can, of course, be generalized to finer widths. 
One simple sanity check for $\Delta\mu=0.5$ wedges is to verify that the average of both 
wedges yields the monopole. 

These two methods (``direct" and ``integrated") should yield a similar result 
given $RR$ is not a function of $\mu$, but only 
of separation $s$. If this is the case, 
the ``direct" method should be preferred because  
the ``integrated" yields lower S/N  due 
to binning effects. 

\cite{samushia11a} point out, however, both wide angle 
effects as well as observer angle effects on the 
current SDSS-II geometry. 
When analyzing $RR(\mu,s)$ of the LasDamas SDSS-II mocks,  
we find a $\sim 10\%$ of $1\sigma$ difference between 
$RR(1,s)$ to $RR(0,s)$ for the \baf scale, 
indicating the survey is wider than it is broader. 
In the BOSS $0.16<z<0.6$ mocks we obtain similar conclusions 
only this time a $\sim 5\%$ of $1\sigma$ effect. 
In this case, by using the ``direct" method (Equation \ref{ls_direct_equation}), 
one is introducing an angular weighting of  
$RR(\mu,s)/RR_0(s)$ such that the $RR(\mu,s)$ 
is disregarded.
In other words, Equation (\ref{ls_direct_equation})
is not equal to Equation (\ref{eq_extract_multipole}). 
For example, the ``direct" multipoles actually take the form:

\begin{dmath}
{\hat{\xi}_\ell(s)= \frac{DD_\ell(s)+RR_\ell(s)-2DR_\ell(s)}{RR_0(s)}}
=\frac{2\ell+1}{2}
\int_{-1}^{+1}{\rm d}\mu\frac{DD(\mu,s)+RR(\mu,s)-2DR(\mu,s)}{RR(\mu,s)}{\mathcal P}_\ell(\mu)\cdot\frac{RR(\mu,s)}{RR(s)}. 
\end{dmath}

In the left panel of Figure \ref{directint_plot} we 
show the difference between the two methods 
for clustering $\xi_{0,2,4}$ and $\Delta\mu=0.5$ wedges. 
The difference between ``direct" and ``integrated" is given 
in units of $\sigma_\xi$ for the SDSS-II mocks. 

We clearly see an over estimation of the ``direct" results  
in most scales for the multipoles. At the \baf scale this effect 
in units of $1\sigma$ is $\sim 5\%$ for $\xi_0$, 
$\sim 7\%$ for $\xi_2$ and $\sim 8\%$ for $\xi_4$, 
with the differences for the last two cases increasing at smaller scales. 
We note that the uncertainties of $\xi_2$ and $\xi_4$ 
are much larger than that of $\xi_0$ making the ratios 
$\xi^{\rm direct}_\ell/\xi^{\rm integrated}_\ell$  
larger than that of the monopole. 

The $\Delta\mu=0.5$ wedges, on the other hand, 
are shown to yield much smaller differences. 
This is probably due to the fact that the wide 
wedges do not correlate pairs at $\mu=0$ with 
those at $\mu=1$. 

Another comparison is shown on the 
right panel of Figure \ref{directint_plot} 
which is similar to that in Figure \ref{multipole_wedges}d. 
As described in Appendix \ref{projectionspractice_section}, both test
the relation between clustering wedges and multipoles
(Equation \ref{wedgemonoquad_equality}), where 
the right panel of  Figure \ref{directint_plot}  
corresponds to the ``direct" method, 
and Figure \ref{multipole_wedges}d shows the ``integrated" measurements. 
We clearly see that the ``integrated" method performs better. 

Although the ``integrated" estimator method appears to be closer 
to the natural definition of multipoles and wedges, 
there is a tradeoff of increase in uncertainties due to binning effects 
(one is integrating of more noisy $DD$). Also, 
by using the ``integrated" method explained here, 
one might be adding observer angle effects 
discussed in \cite{samushia11a}. 
For this reason, we conclude that the user 
should be aware of the two methods, and choose 
accordingly. 
We recommend applying both and comparing the obtained results. 
The ``direct" method might be preferable 
if the user ads the $RR(\mu,s)/RR_0(s)$ weighting 
into the multipole $\xi_\ell$ model. 
For the $\Delta\mu=0.5$ wedges we show 
no substantial difference in the actual $\xi$, 
so one would choose the one that yields  lower 
uncertainties. 


\begin{thebibliography}{76}


\bibitem[{{Alcock} \& {Paczynski}(1979)}]{alcock79}
{Alcock}, C. \& {Paczynski}, B. 1979, \nat, 281, 358

\bibitem[{{Arnalte-Mur} {et~al.}(2011){Arnalte-Mur}, {Labatie}, {Clerc},
  {Mart{\'{\i}}nez}, {Starck}, {Lachi{\`e}ze-Rey}, {Saar}, \&
  {Paredes}}]{arnalte11a}
{Arnalte-Mur}, P., {Labatie}, A., {Clerc}, N., {Mart{\'{\i}}nez}, V.~J.,
  {Starck}, J., {Lachi{\`e}ze-Rey}, M., {Saar}, E., \& {Paredes}, S. 2011,
  ArXiv e-prints

\bibitem[{{Ballinger} {et~al.}(1996){Ballinger}, {Peacock}, \&
  {Heavens}}]{ballinger96a}
{Ballinger}, W.~E., {Peacock}, J.~A., \& {Heavens}, A.~F. 1996, \mnras, 282,
  877

\bibitem[{{Berlind} \& {Weinberg}(2002)}]{berlind02a}
{Berlind}, A.~A. \& {Weinberg}, D.~H. 2002, \apj, 575, 587

\bibitem[{{Blake} {et~al.}(2011){Blake}, {Brough}, {Colless}, {Contreras},
  {Couch}, {Croom}, {Davis}, {Drinkwater}, {Forster}, {Gilbank}, {Gladders},
  {Glazebrook}, {Jelliffe}, {Jurek}, {Li}, {Madore}, {Martin}, {Pimbblet},
  {Poole}, {Pracy}, {Sharp}, {Wisnioski}, {Woods}, {Wyder}, \&
  {Yee}}]{blake11a}
{Blake}, C., {Brough}, S., {Colless}, M., {Contreras}, C., {Couch}, W.,
  {Croom}, S., {Davis}, T., {Drinkwater}, M.~J., {Forster}, K., {Gilbank}, D.,
  {Gladders}, M., {Glazebrook}, K., {Jelliffe}, B., {Jurek}, R.~J., {Li}, I.,
  {Madore}, B., {Martin}, C., {Pimbblet}, K., {Poole}, G., {Pracy}, M.,
  {Sharp}, R., {Wisnioski}, E., {Woods}, D., {Wyder}, T., \& {Yee}, H. 2011,
  ArXiv e-prints

\bibitem[{{Blake} {et~al.}(2007){Blake}, {Collister}, {Bridle}, \&
  {Lahav}}]{blake07}
{Blake}, C., {Collister}, A., {Bridle}, S., \& {Lahav}, O. 2007, \mnras, 374,
  1527

\bibitem[{{Blake} \& {Glazebrook}(2003)}]{blake03}
{Blake}, C. \& {Glazebrook}, K. 2003, \apj, 594, 665

\bibitem[{{Cabr{\'e}} \& {Gazta{\~n}aga}(2009)}]{cabre09i}
{Cabr{\'e}}, A. \& {Gazta{\~n}aga}, E. 2009, \mnras, 393, 1183

\bibitem[{{Cole} {et~al.}(2005){Cole}, {Percival}, {Peacock}, {Norberg},
  {Baugh}, {Frenk}, {Baldry}, {Bland-Hawthorn}, {Bridges}, {Cannon}, {Colless},
  {Collins}, {Couch}, {Cross}, {Dalton}, {Eke}, {De Propris}, {Driver},
  {Efstathiou}, {Ellis}, {Glazebrook}, {Jackson}, {Jenkins}, {Lahav}, {Lewis},
  {Lumsden}, {Maddox}, {Madgwick}, {Peterson}, {Sutherland}, \&
  {Taylor}}]{cole05a}
{Cole}, S., {Percival}, W.~J., {Peacock}, J.~A., {Norberg}, P., {Baugh}, C.~M.,
  {Frenk}, C.~S., {Baldry}, I., {Bland-Hawthorn}, J., {Bridges}, T., {Cannon},
  R., {Colless}, M., {Collins}, C., {Couch}, W., {Cross}, N.~J.~G., {Dalton},
  G., {Eke}, V.~R., {De Propris}, R., {Driver}, S.~P., {Efstathiou}, G.,
  {Ellis}, R.~S., {Glazebrook}, K., {Jackson}, C., {Jenkins}, A., {Lahav}, O.,
  {Lewis}, I., {Lumsden}, S., {Maddox}, S., {Madgwick}, D., {Peterson}, B.~A.,
  {Sutherland}, W., \& {Taylor}, K. 2005, \mnras, 362, 505

\bibitem[{{Colless} {et~al.}(2003){Colless}, {Peterson}, {Jackson}, {Peacock},
  {Cole}, {Norberg}, {Baldry}, {Baugh}, {Bland-Hawthorn}, {Bridges}, {Cannon},
  {Collins}, {Couch}, {Cross}, {Dalton}, {De Propris}, {Driver}, {Efstathiou},
  {Ellis}, {Frenk}, {Glazebrook}, {Lahav}, {Lewis}, {Lumsden}, {Maddox},
  {Madgwick}, {Sutherland}, \& {Taylor}}]{colless03a}
{Colless}, M., {Peterson}, B.~A., {Jackson}, C., {Peacock}, J.~A., {Cole}, S.,
  {Norberg}, P., {Baldry}, I.~K., {Baugh}, C.~M., {Bland-Hawthorn}, J.,
  {Bridges}, T., {Cannon}, R., {Collins}, C., {Couch}, W., {Cross}, N.,
  {Dalton}, G., {De Propris}, R., {Driver}, S.~P., {Efstathiou}, G., {Ellis},
  R.~S., {Frenk}, C.~S., {Glazebrook}, K., {Lahav}, O., {Lewis}, I., {Lumsden},
  S., {Maddox}, S., {Madgwick}, D., {Sutherland}, W., \& {Taylor}, K. 2003,
  ArXiv Astrophysics e-prints

\bibitem[{{Crocce} \& {Scoccimarro}(2008)}]{crocce08}
{Crocce}, M. \& {Scoccimarro}, R. 2008, \prd, 77, 023533

\bibitem[{{Crocce} {et~al.}(2011){Crocce}, {Gaztanaga}, {Cabre}, {Carnero}, \&
  {Sanchez}}]{crocce11a}
{Crocce}, M., {Gaztanaga}, E., {Cabre}, A., {Carnero}, A., \& {Sanchez}, E.
  2011, ArXiv e-prints


\bibitem[{{Dalal} {et~al.}(2008){Dalal}, {Dor{\'e}}, {Huterer}, \&
  {Shirokov}}]{dalal08a}
{Dalal}, N., {Dor{\'e}}, O., {Huterer}, D., \& {Shirokov}, A. 2008, \prd, 77,
  123514

\bibitem[{{Drinkwater} {et~al.}(2010){Drinkwater}, {Jurek}, {Blake}, {Woods},
  {Pimbblet}, {Glazebrook}, {Sharp}, {Pracy}, {Brough}, {Colless}, {Couch},
  {Croom}, {Davis}, {Forbes}, {Forster}, {Gilbank}, {Gladders}, {Jelliffe},
  {Jones}, {Li}, {Madore}, {Martin}, {Poole}, {Small}, {Wisnioski}, {Wyder}, \&
  {Yee}}]{drinkwater10a}
{Drinkwater}, M.~J., {Jurek}, R.~J., {Blake}, C., {Woods}, D., {Pimbblet},
  K.~A., {Glazebrook}, K., {Sharp}, R., {Pracy}, M.~B., {Brough}, S.,
  {Colless}, M., {Couch}, W.~J., {Croom}, S.~M., {Davis}, T.~M., {Forbes}, D.,
  {Forster}, K., {Gilbank}, D.~G., {Gladders}, M., {Jelliffe}, B., {Jones}, N.,
  {Li}, I., {Madore}, B., {Martin}, D.~C., {Poole}, G.~B., {Small}, T.,
  {Wisnioski}, E., {Wyder}, T., \& {Yee}, H.~K.~C. 2010, \mnras, 401, 1429

\bibitem[{{Eisenstein} \& {Hu}(1998)}]{eisenstein98a}
{Eisenstein}, D.~J. \& {Hu}, W. 1998, \apj, 496, 605

\bibitem[{{Eisenstein} {et~al.}(1998){Eisenstein}, {Hu}, \&
  {Tegmark}}]{eisenstein98b}
{Eisenstein}, D.~J., {Hu}, W., \& {Tegmark}, M. 1998, \apjl, 504, L57+

\bibitem[{Eisenstein {et~al.}(1999)Eisenstein, Hu, \& Tegmark}]{eisenstein99a}
Eisenstein, D.~J., Hu, W., \& Tegmark, M. 1999, \apj, 518, 2

\bibitem[{Eisenstein {et~al.}(2005)}]{eisenstein05b}
Eisenstein, D.~J. {et~al.} 2005, \apj, 633, 560

\bibitem[{{Eisenstein} {et~al.}(2007){Eisenstein}, {Seo}, {Sirko}, \&
  {Spergel}}]{eisenstein07}
{Eisenstein}, D.~J., {Seo}, H.-J., {Sirko}, E., \& {Spergel}, D.~N. 2007, \apj,
  664, 675

\bibitem[{{Eisenstein} {et~al.}(2011){Eisenstein}, {Weinberg}, {Agol},
  {Aihara}, {Allende Prieto}, {Anderson}, {Arns}, {Aubourg}, {Bailey},
  {Balbinot}, \& et~al.}]{eisenstein11a}
{Eisenstein}, D.~J., {Weinberg}, D.~H., {Agol}, E., {Aihara}, H., {Allende
  Prieto}, C., {Anderson}, S.~F., {Arns}, J.~A., {Aubourg}, E., {Bailey}, S.,
  {Balbinot}, E., \& et~al. 2011, ArXiv e-prints

\bibitem[{{Estrada} {et~al.}(2009){Estrada}, {Sefusatti}, \&
  {Frieman}}]{estrada09a}
{Estrada}, J., {Sefusatti}, E., \& {Frieman}, J.~A. 2009, \apj, 692, 265




\bibitem[{{Friedman}(1922)}]{friedman22a}
{Friedman}, A. 1922, Zeitschrift fur Physik, 10, 377

\bibitem[{{Fry}(1996)}]{fry96a}
{Fry}, J.~N. 1996, \apjl, 461, L65+

\bibitem[{{Gazta{\~n}aga} {et~al.}(2009){Gazta{\~n}aga}, {Cabr{\'e}}, \&
  {Hui}}]{gaztanaga08iv}
{Gazta{\~n}aga}, E., {Cabr{\'e}}, A., \& {Hui}, L. 2009, \mnras, 399, 1663

\bibitem[{{Glazebrook} \& {Blake}(2005)}]{glazebrook05a}
{Glazebrook}, K. \& {Blake}, C. 2005, \apj, 631, 1

\bibitem[{Hamilton(1992)}]{hamilton92}
Hamilton, A. 1992, \apj, 385, L5

\bibitem[{{Hamilton}(1998)}]{hamilton98a}
{Hamilton}, A.~J.~S. 1998, in Astrophysics and Space Science Library, Vol. 231,
  The Evolving Universe, ed. {D.~Hamilton}, 185--+

\bibitem[{Hogg(1999)}]{hogg99cosm}
Hogg, D.~W. 1999, astro-ph/9905116

\bibitem[{{Hu} \& {Haiman}(2003)}]{hu03a}
{Hu}, W. \& {Haiman}, Z. 2003, \prd, 68, 063004

\bibitem[{{Hubble} \& {Humason}(1931)}]{hubble31}
{Hubble}, E. \& {Humason}, M.~L. 1931, \apj, 74, 43

\bibitem[{{Hui} {et~al.}(2007){Hui}, {Gazta{\~n}aga}, \& {Loverde}}]{hui07a}
{Hui}, L., {Gazta{\~n}aga}, E., \& {Loverde}, M. 2007, \prd, 76, 103502

\bibitem[{{H{\"u}tsi}(2006)}]{hutsi06a}
{H{\"u}tsi}, G. 2006, \aap, 449, 891

\bibitem[{{Jackson}(1972)}]{jackson72}
{Jackson}, J.~C. 1972, \mnras, 156, 1P

\bibitem[{{Jones} {et~al.}(2009){Jones}, {Read}, {Saunders}, {Colless},
  {Jarrett}, {Parker}, {Fairall}, {Mauch}, {Sadler}, {Watson}, {Burton},
  {Campbell}, {Cass}, {Croom}, {Dawe}, {Fiegert}, {Frankcombe}, {Hartley},
  {Huchra}, {James}, {Kirby}, {Lahav}, {Lucey}, {Mamon}, {Moore}, {Peterson},
  {Prior}, {Proust}, {Russell}, {Safouris}, {Wakamatsu}, {Westra}, \&
  {Williams}}]{jones09a}
{Jones}, D.~H., {Read}, M.~A., {Saunders}, W., {Colless}, M., {Jarrett}, T.,
  {Parker}, Q.~A., {Fairall}, A.~P., {Mauch}, T., {Sadler}, E.~M., {Watson},
  F.~G., {Burton}, D., {Campbell}, L.~A., {Cass}, P., {Croom}, S.~M., {Dawe},
  J., {Fiegert}, K., {Frankcombe}, L., {Hartley}, M., {Huchra}, J., {James},
  D., {Kirby}, E., {Lahav}, O., {Lucey}, J., {Mamon}, G.~A., {Moore}, L.,
  {Peterson}, B.~A., {Prior}, S., {Proust}, D., {Russell}, K., {Safouris}, V.,
  {Wakamatsu}, K.-I., {Westra}, E., \& {Williams}, M. 2009, \mnras, 399, 683

\bibitem[{Kaiser(1987)}]{kaiser87}
Kaiser, N. 1987, \mnras, 227, 1

\bibitem[{{Kazin} {et~al.}(2010{\natexlab{a}}){Kazin}, {Blanton},
  {Scoccimarro}, {McBride}, {Berlind}, {Bahcall}, {Brinkmann}, {Czarapata},
  {Frieman}, {Kent}, {Schneider}, \& {Szalay}}]{kazin10a}
{Kazin}, E.~A., {Blanton}, M.~R., {Scoccimarro}, R., {McBride}, C.~K.,
  {Berlind}, A.~A., {Bahcall}, N.~A., {Brinkmann}, J., {Czarapata}, P.,
  {Frieman}, J.~A., {Kent}, S.~M., {Schneider}, D.~P., \& {Szalay}, A.~S.
  2010{\natexlab{a}}, \apj, 710, 1444

\bibitem[{{Kazin} {et~al.}(2010{\natexlab{b}}){Kazin}, {Blanton},
  {Scoccimarro}, {McBride}, \& {Berlind}}]{kazin10b}
{Kazin}, E.~A., {Blanton}, M.~R., {Scoccimarro}, R., {McBride}, C.~K., \&
  {Berlind}, A.~A. 2010{\natexlab{b}}, \apj, 719, 1032

\bibitem[{{Kerscher} {et~al.}(2000){Kerscher}, {Szapudi}, \&
  {Szalay}}]{kerscher00a}
{Kerscher}, M., {Szapudi}, I., \& {Szalay}, A.~S. 2000, \apjl, 535, L13


\bibitem[{{Kim} {et~al.}(2009){Kim}, {Park}, {Gott}, \& {Dubinski}}]{kim09a}
{Kim}, J., {Park}, C., {Gott}, J.~R., \& {Dubinski}, J. 2009, \apj, 701, 1547

\bibitem[{Komatsu {et~al.}(2009)Komatsu, Dunkley, Nolta, Bennett, Gold,
  Hinshaw, Jarosik, Larson, Limon, Page, Spergel, Halpern, Hill, Kogut, Meyer,
  Tucker, Weiland, Wollack, \& Wright}]{komatsu09a}
Komatsu, E., Dunkley, J., Nolta, M.~R., Bennett, C.~L., Gold, B., Hinshaw, G.,
  Jarosik, N., Larson, D., Limon, M., Page, L., Spergel, D.~N., Halpern, M.,
  Hill, R.~S., Kogut, A., Meyer, S.~S., Tucker, G.~S., Weiland, J.~L., Wollack,
  E., \& Wright, E.~L. 2009, The Astrophysical Journal Supplement, 180, 330

\bibitem[{{Labini} {et~al.}(2009){Labini}, {Vasilyev}, \&
  {Baryshev}}]{labini09}
{Labini}, F.~S., {Vasilyev}, N.~L., \& {Baryshev}. 2009, ArXiv e-prints

\bibitem[{{Landy} \& {Szalay}(1993)}]{landy93a}
{Landy}, S.~D. \& {Szalay}, A.~S. 1993, \apj, 412, 64

\bibitem[{Lewis {et~al.}(2002)}]{lewis02a}
Lewis, I. {et~al.} 2002, \mnras, 334, 673

\bibitem[{{Linder}(2003)}]{linder03a}
{Linder}, E.~V. 2003, \prd, 68, 083504

\bibitem[{{Martinez} {et~al.}(2009){Martinez}, {Arnalte-Mur}, {Saar}, {de la
  Cruz}, {Pons-Borderia}, {Paredes}, {Fernandez-Soto}, \&
  {Tempel}}]{martinez08}
{Martinez}, V.~J., {Arnalte-Mur}, P., {Saar}, E., {de la Cruz}, P.,
  {Pons-Borderia}, M.~J., {Paredes}, S., {Fernandez-Soto}, A., \& {Tempel}, E.
  2009, \apjl, 696

\bibitem[{{Matsubara}(2004)}]{matsubara04}
{Matsubara}, T. 2004, \apj, 615, 573

\bibitem[{{Mehta} {et~al.}(2011){Mehta}, {Seo}, {Eckel}, {Eisenstein},
  {Metchnik}, {Pinto}, \& {Xu}}]{mehta11a}
{Mehta}, K.~T., {Seo}, H., {Eckel}, J., {Eisenstein}, D.~J., {Metchnik}, M.,
  {Pinto}, P., \& {Xu}, X. 2011, ArXiv e-prints

\bibitem[{{Meiksin} {et~al.}(1999){Meiksin}, {White}, \&
  {Peacock}}]{meiksin99a}
{Meiksin}, A., {White}, M., \& {Peacock}, J.~A. 1999, \mnras, 304, 851

\bibitem[{{Norberg} {et~al.}(2009){Norberg}, {Baugh}, {Gazta{\~n}aga}, \&
  {Croton}}]{norberg09a}
{Norberg}, P., {Baugh}, C.~M., {Gazta{\~n}aga}, E., \& {Croton}, D.~J. 2009,
  \mnras, 396, 19

\bibitem[{{Okumura} {et~al.}(2008){Okumura}, {Matsubara}, {Eisenstein}, {Kayo},
  {Hikage}, {Szalay}, \& {Schneider}}]{okumura08}
{Okumura}, T., {Matsubara}, T., {Eisenstein}, D.~J., {Kayo}, I., {Hikage}, C.,
  {Szalay}, A.~S., \& {Schneider}, D.~P. 2008, \apj, 676, 889

\bibitem[{Padmanabhan \& White(2008)}]{padmanabhan08a}
Padmanabhan, N. \& White, M. 2008, Physical Review D, 77, 123540, (c) 2008: The
  American Physical Society

\bibitem[{{Padmanabhan} {et~al.}(2007)}]{padmanabhan07b}
{Padmanabhan}, N. {et~al.} 2007, \mnras, 378, 852

\bibitem[{{Peebles} \& {Yu}(1970)}]{peebles70a}
{Peebles}, P.~J.~E. \& {Yu}, J.~T. 1970, \apj, 162, 815

\bibitem[{Percival {et~al.}(2007)Percival, Cole, Eisenstein, Nichol, Peacock,
  Pope, \& Szalay}]{percival07a}
Percival, W.~J., Cole, S., Eisenstein, D.~J., Nichol, R.~C., Peacock, J.~A.,
  Pope, A.~C., \& Szalay, A.~S. 2007, Monthly Notices of the Royal Astronomical
  Society, 381, 1053

\bibitem[{{Percival} {et~al.}(2010){Percival}, {Reid}, {Eisenstein}, {Bahcall},
  {Budavari}, {Fukugita}, {Gunn}, {Ivezic}, {Knapp}, {Kron}, {Loveday},
  {Lupton}, {McKay}, {Meiksin}, {Nichol}, {Pope}, {Schlegel}, {Schneider},
  {Spergel}, {Stoughton}, {Strauss}, {Szalay}, {Tegmark}, {Weinberg}, {York},
  \& {Zehavi}}]{percival09b}
{Percival}, W.~J., {Reid}, B.~A., {Eisenstein}, D.~J., {Bahcall}, N.~A.,
  {Budavari}, T., {Fukugita}, M., {Gunn}, J.~E., {Ivezic}, Z., {Knapp}, G.~R.,
  {Kron}, R.~G., {Loveday}, J., {Lupton}, R.~H., {McKay}, T.~A., {Meiksin}, A.,
  {Nichol}, R.~C., {Pope}, A.~C., {Schlegel}, D.~J., {Schneider}, D.~P.,
  {Spergel}, D.~N., {Stoughton}, C., {Strauss}, M.~A., {Szalay}, A.~S.,
  {Tegmark}, M., {Weinberg}, D.~H., {York}, D.~G., \& {Zehavi}, I. 2010,
  \mnras, 401, 2148

\bibitem[{{Perlmutter} {et~al.}(1999)}]{perlmutter99a}
{Perlmutter}, S. {et~al.} 1999, \apj, 517, 565

\bibitem[{{Reid} {et~al.}(2010){Reid}, {Percival}, {Eisenstein}, {Verde},
  {Spergel}, {Skibba}, {Bahcall}, {Budavari}, {Fukugita}, {Gott}, {Gunn},
  {Ivezic}, {Knapp}, {Kron}, {Lupton}, {McKay}, {Meiksin}, {Nichol}, {Pope},
  {Schlegel}, {Schneider}, {Strauss}, {Stoughton}, {Szalay}, {Tegmark},
  {Weinberg}, {York}, \& {Zehavi}}]{reid09a}
{Reid}, B.~A., {Percival}, W.~J., {Eisenstein}, D.~J., {Verde}, L., {Spergel},
  D.~N., {Skibba}, R.~A., {Bahcall}, N.~A., {Budavari}, T., {Fukugita}, M.,
  {Gott}, J.~R., {Gunn}, J.~E., {Ivezic}, Z., {Knapp}, G.~R., {Kron}, R.~G.,
  {Lupton}, R.~H., {McKay}, T.~A., {Meiksin}, A., {Nichol}, R.~C., {Pope},
  A.~C., {Schlegel}, D.~J., {Schneider}, D.~P., {Strauss}, M.~A., {Stoughton},
  C., {Szalay}, A.~S., {Tegmark}, M., {Weinberg}, D.~H., {York}, D.~G., \&
  {Zehavi}, I. 2010, \mnras, 404, 60

\bibitem[{{Riess} {et~al.}(1998){Riess}, {Filippenko}, {Challis},
  {Clocchiatti}, {Diercks}, {Garnavich}, {Gilliland}, {Hogan}, {Jha},
  {Kirshner}, {Leibundgut}, {Phillips}, {Reiss}, {Schmidt}, {Schommer},
  {Smith}, {Spyromilio}, {Stubbs}, {Suntzeff}, \& {Tonry}}]{riess98}
{Riess}, A.~G., {Filippenko}, A.~V., {Challis}, P., {Clocchiatti}, A.,
  {Diercks}, A., {Garnavich}, P.~M., {Gilliland}, R.~L., {Hogan}, C.~J., {Jha},
  S., {Kirshner}, R.~P., {Leibundgut}, B., {Phillips}, M.~M., {Reiss}, D.,
  {Schmidt}, B.~P., {Schommer}, R.~A., {Smith}, R.~C., {Spyromilio}, J.,
  {Stubbs}, C., {Suntzeff}, N.~B., \& {Tonry}, J. 1998, \aj, 116, 1009

\bibitem[{{Samushia} {et~al.}(2011){Samushia}, {Percival}, \&
  {Raccanelli}}]{samushia11a}
{Samushia}, L., {Percival}, W.~J., \& {Raccanelli}, A. 2011, ArXiv e-prints

\bibitem[{{S{\'a}nchez} {et~al.}(2008){S{\'a}nchez}, {Baugh}, \&
  {Angulo}}]{sanchez08}
{S{\'a}nchez}, A.~G., {Baugh}, C.~M., \& {Angulo}, R. 2008, \mnras, 390, 1470

\bibitem[{{S{\'a}nchez} {et~al.}(2009){S{\'a}nchez}, {Crocce}, {Cabre},
  {Baugh}, \& {Gaztanaga}}]{sanchez09a}
{S{\'a}nchez}, A.~G., {Crocce}, M., {Cabre}, A., {Baugh}, C.~M., \&
  {Gaztanaga}, E. 2009, \mnras, 400, 1643

\bibitem[{{Schlegel} {et~al.}(2009){Schlegel}, {White}, \&
  {Eisenstein}}]{schlegel09a}
{Schlegel}, D., {White}, M., \& {Eisenstein}, D. 2009, ArXiv e-prints

\bibitem[{{Scoccimarro}(2004)}]{scoccimarro04a}
{Scoccimarro}, R. 2004, \prd, 70, 083007

\bibitem[{{Seo} {et~al.}(2010){Seo}, {Eckel}, {Eisenstein}, {Mehta},
  {Metchnik}, {Padmanabhan}, {Pinto}, {Takahashi}, {White}, \& {Xu}}]{seo10a}
{Seo}, H., {Eckel}, J., {Eisenstein}, D.~J., {Mehta}, K., {Metchnik}, M.,
  {Padmanabhan}, N., {Pinto}, P., {Takahashi}, R., {White}, M., \& {Xu}, X.
  2010, \apj, 720, 1650

\bibitem[{{Seo} \& {Eisenstein}(2003)}]{seo03}
{Seo}, H.-J. \& {Eisenstein}, D.~J. 2003, \apj, 598, 720

\bibitem[{{Shoji} {et~al.}(2009){Shoji}, {Jeong}, \& {Komatsu}}]{shoji09a}
{Shoji}, M., {Jeong}, D., \& {Komatsu}, E. 2009, \apj, 693, 1404

\bibitem[{{Smith} {et~al.}(2008){Smith}, {Scoccimarro}, \& {Sheth}}]{smith08}
{Smith}, R.~E., {Scoccimarro}, R., \& {Sheth}, R.~K. 2008, \prd, 77, 043525

\bibitem[{{Taruya} {et~al.}(2010){Taruya}, {Nishimichi}, \&
  {Saito}}]{taruya10a}
{Taruya}, A., {Nishimichi}, T., \& {Saito}, S. 2010, \prd, 82, 063522

\bibitem[{{Taruya} {et~al.}(2009){Taruya}, {Nishimichi}, {Saito}, \&
  {Hiramatsu}}]{taruya09a}
{Taruya}, A., {Nishimichi}, T., {Saito}, S., \& {Hiramatsu}, T. 2009, \prd, 80,
  123503

\bibitem[{{Taruya} {et~al.}(2011){Taruya}, {Saito}, \&
  {Nishimichi}}]{taruya11a}
{Taruya}, A., {Saito}, S., \& {Nishimichi}, T. 2011, ArXiv e-prints

\bibitem[{Tegmark {et~al.}(1998)Tegmark, Hamilton, Strauss, Vogeley, \&
  Szalay}]{tegmark98b}
Tegmark, M., Hamilton, A. J.~S., Strauss, M.~A., Vogeley, M.~S., \& Szalay,
  A.~S. 1998, \apj, 499, 555

\bibitem[{{Tegmark} {et~al.}(2006)}]{tegmark06a}
{Tegmark}, M. {et~al.} 2006, \prd, 74, 123507

\bibitem[{{Tian} {et~al.}(2010){Tian}, {Neyrinck}, {Budav{\'a}ri}, \&
  {Szalay}}]{tian10a}
{Tian}, H.~J., {Neyrinck}, M.~C., {Budav{\'a}ri}, T., \& {Szalay}, A.~S. 2010,
  ArXiv e-prints

\bibitem[{{Tocchini-Valentini} {et~al.}(2011){Tocchini-Valentini}, {Barnard},
  {Bennett}, \& {Szalay}}]{tocchini11a}
{Tocchini-Valentini}, D., {Barnard}, M., {Bennett}, C.~L., \& {Szalay}, A.~S.
  2011, ArXiv e-prints

\bibitem[{{Wagner} {et~al.}(2008){Wagner}, {M{\"u}ller}, \&
  {Steinmetz}}]{wagner08a}
{Wagner}, C., {M{\"u}ller}, V., \& {Steinmetz}, M. 2008, \aap, 487, 63

\bibitem[{{White} {et~al.}(2010){White}, {Blanton}, {Bolton}, {Schlegel},
  {Tinker}, {Berlind}, {da Costa}, {Kazin}, {Lin}, {Maia}, {McBride},
  {Padmanabhan}, {Parejko}, {Percival}, {Prada}, {Ramos}, {Sheldon}, {de
  Simoni}, {Skibba}, {Thomas}, {Wake}, {Zehavi}, {Zheng}, {Nichol},
  {Schneider}, {Strauss}, {Weaver}, \& {Weinberg}}]{white10a}
{White}, M., {Blanton}, M., {Bolton}, A., {Schlegel}, D., {Tinker}, J.,
  {Berlind}, A., {da Costa}, L., {Kazin}, E., {Lin}, Y., {Maia}, M., {McBride},
  C., {Padmanabhan}, N., {Parejko}, J., {Percival}, W., {Prada}, F., {Ramos},
  B., {Sheldon}, E., {de Simoni}, F., {Skibba}, R., {Thomas}, D., {Wake}, D.,
  {Zehavi}, I., {Zheng}, Z., {Nichol}, R., {Schneider}, D., {Strauss}, M.~A.,
  {Weaver}, B.~A., \& {Weinberg}, D.~H. 2010, ArXiv e-prints

\bibitem[{{York} {et~al.}(2000)}]{york00a}
{York}, D.~G. {et~al.} 2000, \aj, 120, 1579

\bibitem[{{Zel'Dovich}(1970)}]{zel'dovich70a}
{Zel'Dovich}, Y.~B. 1970, \aap, 5, 84

\end{thebibliography}
\end{document}